\documentclass[a4paper,11pt]{article}
\usepackage{indentfirst}
\usepackage{hyperref}
\usepackage{multirow}
\usepackage{latexsym,bm,amsmath,amssymb}
\usepackage{graphicx}
\usepackage{epsfig}
\usepackage{epstopdf}
\usepackage{subfigure}
\usepackage{cite}
\usepackage{color}
\usepackage[top=1in, bottom=1in, left=1.in, right=1.in]{geometry}
\usepackage{appendix}
\usepackage{ulem}
\usepackage{slashed}
\usepackage{enumitem}

\newcommand{\bea}{\begin{eqnarray}}
\newcommand{\eea}{\end{eqnarray}}

\begin{document}
\title{\bf\Large New Insights on Low Energy $\pi N$ Scattering Amplitudes: Comprehensive Analyses at $\mathcal{O}(p^3)$ Level}
\author
{\sc
Yu-Fei Wang$^{1}$, De-Liang~Yao$^{2,3}$, Han-Qing~Zheng$^{1,4}$
\vspace*{0.5cm} \\
$^{1}${\it Department of Physics and State Key Laboratory of Nuclear Physics and Technology,} \\
{\it Peking University,  Beijing 100871, China}\\
$^{2}${\it School of Physics and Electronics, Hunan University, Changsha 410082, China }\\
$^{3}${\it Instituto de F\'{\i}sica Corpuscular (centro mixto CSIC-UV), }\\
{\it Institutos de Investigaci\'{o}n de Paterna,  Apartado 22085, 46071, Valencia, Spain}\\
$^{4}${\it  Collaborative Innovation Center of Quantum Matter, Beijing 100871, China}
}
\maketitle
\begin{abstract}
A production representation of partial-wave $S$ matrix is utilized to construct low-energy elastic pion-nucleon scattering amplitudes from cuts and poles on complex Riemann sheets. Among them, the contribution of left-hand cuts is estimated using the $\mathcal{O}(p^3)$ results obtained in covariant baryon chiral perturbation theory within the extended-on-nass-shell scheme. By fitting to data on partial-wave phase shifts, it is indicated that the existences of hidden poles in $S_{11}$ and $P_{11}$ channels, as conjectured in our previous paper~[Eur. Phys. J. C. 78(7): 543 (2018)], are firmly established. Specifically, the pole mass of the $S_{11}$ hidden resonance is determined to be $(895\pm81)-(164\pm23)i$ MeV, whereas, the virtual pole in the $P_{11}$ channel locates at $(966\pm18)$ MeV. It is found that analyses at the $\mathcal{O}(p^3)$ level improves significantly the fit quality, comparing with the previous $\mathcal{O}(p^2)$ one. Quantitative studies with cautious physical discussions are also conducted for the other $S$- and $P$-wave channels.
\end{abstract}
\section{Introduction}\label{intro}
Pion-nucleon ($\pi N$) scattering, as one of the fundamental processes of low-energy quantum chromodynamics (QCD), has been extensively studied since the middle of the last century, see e.g. Ref.~\cite{Bransden:1973}. Up to now, there exists a wealth of long-time accumulated experimental data on differential cross section and polarizations~\cite{Hohler:1983}. To decode the inherent physics, efforts have been devoted to performing partial wave analysis (PWA) of the $\pi N$ scattering amplitudes by groups: Karlsruhe~\cite{Koch:1980ay,Koch:1985bn}, Matsinos~\cite{Matsinos:2006sw}, GWU~\cite{SAID}, etc. Pole information about excitations, e.g. the Roper resonance $N^\ast(1440)$~\cite{Roper:1964zza}, can be extracted from PWA, usually relying on various phenomenological models used in the analyses.
However, a modern study of $\pi N$ scatterings is no longer restricted to the extraction of the nucleon excitations, but aims at a precision description of the $\pi N$ amplitudes, not only in the physical region but also in the subthreshold region, and offers reliable inputs for any other related physics, for instance, $\pi N$ sigma term serving as a crucial input for the interpretation of dark-matter searches~\cite{Bottino:1999ei,Bottino:2001dj}. As such, it is necessary to perform investigations in a model-independent way so as to gain an overall understanding of the underlying dynamics of $\pi N$ physics.

At low energies, baryon chiral perturbation theory (BChPT) is one of such model-independent methods in the study of $\pi N$ scatterings. As an expansion in powers of external momenta and light-quark masses, BChPT relies on a hierarchy of the contributions (Feynman diagrams) known as power counting. The presence of baryon fields as explicit degrees of freedom introduces new scales that do not vanish in the chiral limit, leading to the occurrence of power counting breaking (PCB) terms~\cite{Gasser:1987rb} in loop diagrams. In the past thirty years, to remedy this issue various approaches have been proposed: e.g., the heavy baryon (HB) formalism~\cite{Jenkins:1990jv,Bernard:1992qa}, the infrared regularization (IR) prescription~\cite{Ellis:1997kc,Becher:1999he}, and the extended-on-mass-shell (EOMS) scheme~\cite{Gegelia:1999qt,Gegelia:1999gf,Fuchs:2003qc} (see also Ref.~\cite{Epelbaum:2015vea} for extension beyond low-energy region). {Consequently, analyses on low-energy as well as resonance region of $\pi N$ scatterings have been done in the various approaches, see, for example, Refs.~\cite{Fettes:1998ud,Fettes:2000xg,Fettes:2000bb,Becher:2001hv,Mai:2009ce,Bruns:2010sv,Alarcon:2011kh,Chen:2012nx,Yao:2016vbz,Siemens:2016hdi,Lu:2018zof}}. In particular, results obtained in the EOMS scheme has proven to be more suitable for amplitude analysis in the sense that, compared to the other schemes, the analytic structure of the calculated physical quantities is properly kept~\cite{Pascalutsa:2004ga,Pascalutsa:2011fp,Alarcon:2012kn}. Though great achievements have been gained using BChPT with EOMS scheme, one should still keep in mind that those results only fulfill the unitarity perturbatively. To restore exact unitarity, a plenty of unitarization methods have been developed, but with shortcomings like violation of analyticity, breaking of crossing symmetry, etc.. Thus, more rigorous model-independent tools are required.

A rigorous manner is to construct partial-wave $\pi N$ amplitudes under the guidance of the axiomatic principles of $S$-matrix theory, such as unitarity, crossing symmetry, analyticity and so on. Attempts in this direction have been performed in Refs~\cite{Chew:1957zz,Hamilton:1963zz,Steiner:1970mh, Gasparyan:2010xz} by imposing dispersion relations. Likewise, the recent Roy-Steiner (RS) equation analysis in Refs~\cite{Ditsche:2012fv,Hoferichter:2015hva} is an alternative and has gained great achievements in the description of $\pi N$ physics. More recently, in Ref.~\cite{Wang:2017agd}, a fresh look at low energy $\pi N$ scatterings has been established with the help of the Peking University (PKU) representation~\cite{Xiao:2000kx,Zheng:2003rw,Zhou:2006wm,Zhou:2004ms}, a production parametrization of partial wave $S$-matrix on the whole complex plane for two-body elastic scattering amplitudes. In this paper, we intend to carry out a more comprehensive study of the partial wave $\pi N$ scattering amplitudes using PKU representation.

The advantage of the use of the PKU representation is two-fold. On the one hand, it is suitable for pole analysis. The PKU representation separates partial waves into various terms contributing either from poles or branch cuts. The corresponding phase shifts extracted from PKU representation are sensitive to subthreshold poles, enabling one to determine pole positions rather accurately. Furthermore, each phase shift contribution has a definite sign, which makes possible the disentanglement of hidden poles from a background. On the other hand, it respects causality honestly. In the PKU representation, the pole contributions are regarded either as hidden poles or as known poles fixed by experiments, while the cut contribution is estimated from perturbative BChPT amplitudes and uncertainties from such an estimation is known to be severely suppressed. Importantly, the philosophy of the PKU representation is not to directly unitarize the amplitude itself, instead, it unitarizes the left-hand (and inelastic) cuts of the perturbative amplitude and hence hazardous spurious poles, violating causality, can be avoided~\cite{Qin:2002hk}. {In addition, the consistency of that representation with crossing symmetry is already examined in Refs.~\cite{Guo:2007ff,Guo:2007hm}. }

In Ref.~\cite{Wang:2017agd}, the application of the PKU representation to pion-nucleon scatterings has led to very interesting findings: a resonance below $\pi N$ threshold (denoted as crazy resonance for short) in the $S_{11}$ channel is discovered for the first time, and a companionate virtual state of the nucleon in the $P_{11}$ channel is found.
Those findings are drawn with the left-hand cuts ({\it l.h.c.}s) estimated by tree-level BChPT amplitudes. That is, only the kinematical cut $(-\infty,(M_N-m_\pi)^2]$ and the segment cut $[(M_N^2-m_\pi^2)^2/M_N^2,M_N^2+2m_\pi^2]$ due to the $u$-channel nucleon exchange are taken into account, where $M_N$ and $m_\pi$ are physical masses of the nucleon and the pion, respectively. In fact, as demonstrated in Refs~\cite{MacDowell:1959zza,Kennedy:1961}, the full structure of the {\it l.h.c.}s for $\pi N$ scattering contains a circular cut as well. Thus the motivation of this work boils down to two aspects as follows. Firstly, we intend to consider the whole structure of the {\it l.h.c.}s and verify that the $S_{11}$ hidden pole still exists in $\mathcal{O}(p^3)$ analysis. Meanwhile, the $\mathcal{O}(p^3)$ calculation also provides cross-check of the $\mathcal{O}(p^2)$ result in the $P_{11}$ channel. Secondly, since the $\mathcal{O}(p^3)$ calculation is expected to determine the contributions from left-hand cuts more precisely, we try to scrutinize the minor discrepancies in the other channels, which were not able to be fitted well in $\mathcal{O}(p^2)$ case. Note that the one-loop BChPT amplitudes are renormalized within the EOMS scheme mentioned above. The utility of EOMS-renormalized amplitudes guarantees the correct analyticity behaviour as required by the foundation of PKU representation, i.e., the principles of $S$-matrix theory.

In Sect.~\ref{cal} basic formalisms of BChPT, partial wave projection and the PKU representation are introduced. Sect.~\ref{known} collects the numerical results of the known contributions used in the PKU-representation analysis at $\mathcal{O}(p^3)$ level, then in Sect.~\ref{subsec:PSana} the two hidden states, pointed out in Ref.~\cite{Wang:2017agd}, are examined and, meanwhile, the other four $S$- and $P$-wave channels are also studied. The scattering length and effective range parameters for all the $S$- and $P$-waves are estimated in Sect.~\ref{subsec:thrpara}. Finally, conclusions and outlook are made in Sect.~\ref{con}. Appendices.~\ref{app:fftree} and \ref{app:ffloop} exhibit the calculations of the chiral amplitudes for tree diagrams and one-loop diagrams respectively, while Appendix.~\ref{app:renorm} shows the procedure of renormalization and EOMS subtraction. In Appendix.~\ref{app:uncer} the uncertainties of the contributions from \textit{l.h.c.}s are discussed.
\section{Theoretical framework\label{cal}}

\subsection{Formal aspects of $\pi N$ scattering amplitudes}

The isospin structure of the $\pi N$ amplitude can be decomposed as
\begin{equation}\label{isode}
T(\pi^a+N_{\text{i}}\to \pi^{a'}+N_{\text{f}})=\chi_{\text{f}}^\dagger\left(\delta^{a'a}T^++\frac{1}{2}\big[\tau^{a'},\tau^a\big]T^-\right)\chi_{\text{i}}\ \mbox{, }
\end{equation}
where $\chi_{\text{i}}$ and $\chi_{\text{f}}$ correspond to the isospinors of initial and final nucleon states, respectively. The amplitudes with isospins $I=\frac{1}{2},\frac{3}{2}$ can be obtained by
\begin{align}
&T^{I=1/2}=T^++2T^-\ ,\\
&T^{I=3/2}=T^+-T^-\ \mbox{. }
\end{align}
As for the Lorentz structure, for an isospin index $I\in\{\frac{1}{2},\frac{3}{2}\}$ or $I\in\{+,-\}$,
\begin{equation}\label{ampTI}
T^I=\bar u(p',s')\Big[A^I(s,t)+\frac{1}{2}(\slashed{q}+\slashed{q}')B^I(s,t)\Big]u(p,s)\ \mbox{, }
\end{equation}
where $s=W^2\equiv(p+q)^2$, $t\equiv(p^\prime-p)^2$ are Mandelstam variables, and $q$ ($p$) and $q'$ ($p^\prime$) are the $4$-momenta of initial and final state pions (nucleons), respectively. $A$ and $B$ are scalar functions of $s$ and $t$.

One can substitute the nucleon spinors $u(p,s)$ and $\bar u(p',s')$ in Eq.~\eqref{ampTI} by helicity eigenstates in the centre of mass frame to obtain the following helicity amplitudes
\begin{equation}\label{TppTpm}
\begin{split}
&T_{++}^I=(\frac{1+z_s}{2})^{\frac{1}{2}}[2M_N A^I(s,t)+(s-m_\pi^2-M_N^2)B^I(s,t)]\ \mbox{, }\\
&T_{+-}^I=-(\frac{1-z_s}{2})^{\frac{1}{2}}s^{-\frac{1}{2}}[(s-m_\pi^2+M_N^2)A^I(s,t)+M_N(s+m_\pi^2-M_N^2)B^I(s,t)]\ \mbox{; }
\end{split}
\end{equation}
the first and second subscripts refer to the helicities of the initial and final nucleon respectively, and subscripts ``$\pm$'' are shorthands for helicity $\pm 1/2$; $z_s=\cos\theta$ with $\theta$ the scattering angle. The partial wave amplitudes for total angular momentum $J$ can be written as
\begin{equation}\label{TppTpmpw}
\begin{split}
&T_{++}^{I,J}=\frac{1}{32\pi}\int_{-1}^1 dz_s T_{++}^I(s,t(s,z_s)) d^J_{-1/2,-1/2}(z_s)\ \mbox{, }\\
&T_{+-}^{I,J}=\frac{1}{32\pi}\int_{-1}^1 dz_s T_{+-}^I(s,t(s,z_s)) d^J_{1/2,-1/2}(z_s)\ \mbox{, }
\end{split}
\end{equation}
where $d^J$ is the standard Wigner $d$-function. Finally one can get the six $S$- and $P$- wave amplitudes (in $L_{2I\ 2J}$ convention) as follows:
\begin{equation}\label{PWamps}
\begin{split}
&T(S_{11})=T_{++}^{1/2,1/2}+T_{+-}^{1/2,1/2}\ \mbox{, }\\
&T(S_{31})=T_{++}^{3/2,1/2}+T_{+-}^{3/2,1/2}\ \mbox{, }\\
&T(P_{11})=T_{++}^{1/2,1/2}-T_{+-}^{1/2,1/2}\ \mbox{, }\\
&T(P_{31})=T_{++}^{3/2,1/2}-T_{+-}^{3/2,1/2}\ \mbox{, }\\
&T(P_{13})=T_{++}^{1/2,3/2}+T_{+-}^{1/2,3/2}\ \mbox{, }\\
&T(P_{33})=T_{++}^{3/2,3/2}+T_{+-}^{3/2,3/2}\ \mbox{. }
\end{split}
\end{equation}
Throughout this work, the partial-wave label $L_{2I,2J}$ is always suppressed if no confusion is caused.

\begin{figure}[htbp]
\centering
\includegraphics[width=0.8\textwidth]{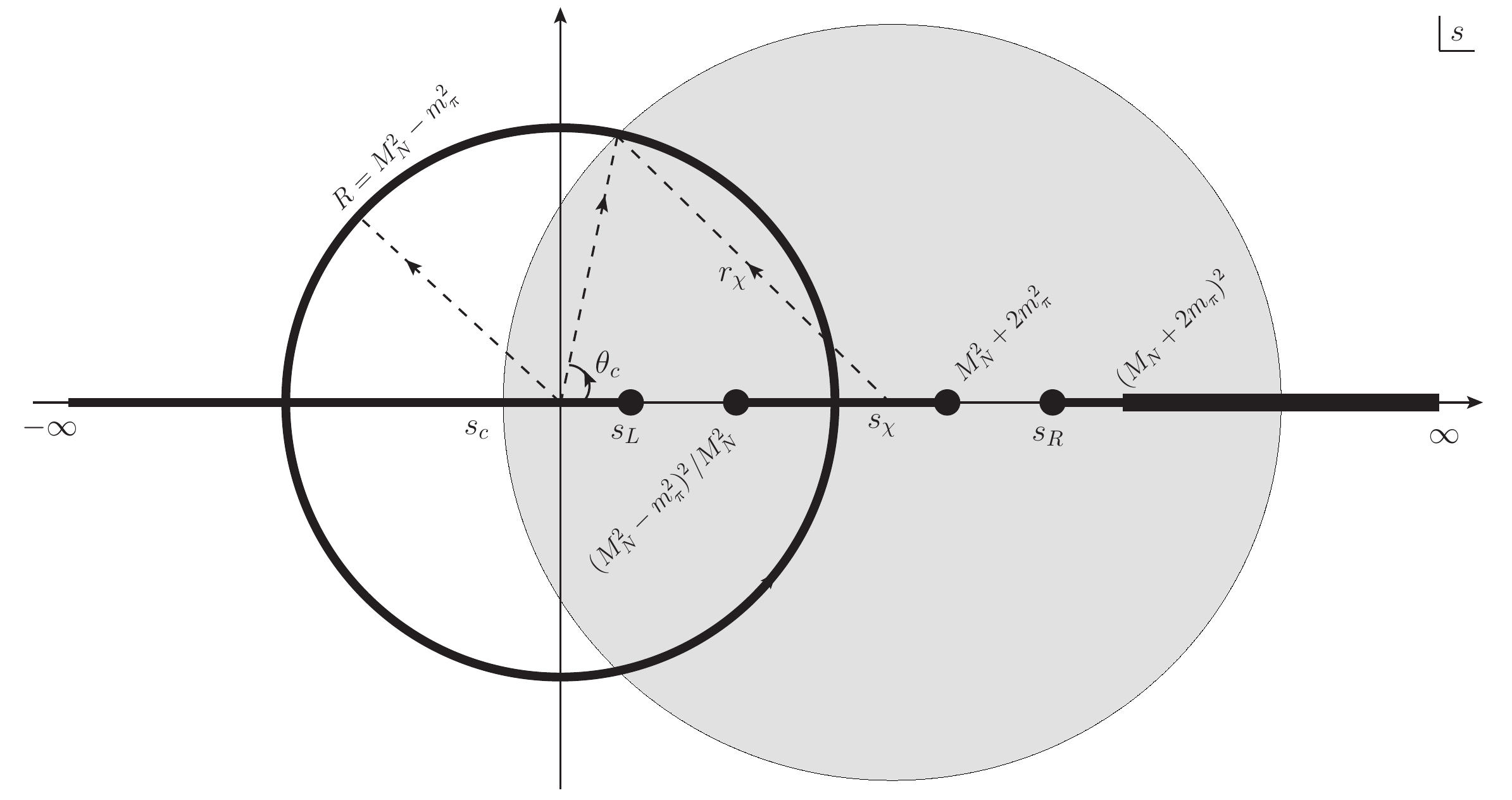}
\caption{The branch cuts of the partial wave $\pi N$ elastic scattering $S$ matrix on the $s$ plane (schematic drawing). The cuts are represented by thick lines and circle. The right hand cut corresponds to the line $[s_R,+\infty)$, and all the others are left hand cuts. The following abbreviations are used: $s_L=(M_N-m_\pi)^2$, $s_R=(M_N+m_\pi)^2$ and $s_\chi=M_N^2+m_\pi^2$. The gray disk indicates the BChPT valid region  characterized by the convergence radius $r_\chi$.  The intersection of the BChPT convergence circle and the on-axis left-hand cuts is denoted by $s_c$.}
\label{fig:cuttotal}
\end{figure}

The branch-cut structure of the partial-wave pion-nucleon scattering amplitudes is generally discussed in Refs.~\cite{MacDowell:1959zza,Kennedy:1961} with details,  which can be shown schematically in Fig.~\ref{fig:cuttotal}. It is worth stressing that the circular cut stems from the $t$ channel continuum $[4m_\pi^2,4M_N^2]$, which is absent at $\mathcal{O}(p^2)$ level but starts to appear at $\mathcal{O}(p^3)$ level.

\subsection{Perturbative BChPT description\label{sec:Lag}}
The $A$ and $B$ scalar functions in Eq.~\eqref{ampTI} can be calculated perturbatively by invoking BChPT. In the framework of covariant BChPT of $SU(2)$ case, the nucleon field is notated as $N$, and the pion as $u$-field
\begin{equation}\label{pif}
u\equiv\exp\Big(\frac{\text{i}\vec{\tau}\cdot \vec{\pi}}{2F}\Big)\ \mbox{, }
\end{equation}
where $F$ is the pion decay constant in the chiral limit and $\vec{\tau}$ stands for Pauli matrices in the flavour space. Relevant pieces of chiral effective Lagrangians for a calculation up to the leading one-loop order, i.e., $\mathcal{O}(p^3)$ level,  are shown as follows~\cite{Fettes:2000gb}:
\begin{equation}\label{p1la}
\mathcal{L}_{\pi N}^{(1)}=\bar N\left(\text{i}\slashed{D}-M+\frac{1}{2}g\slashed{u}\gamma^5\right)N\ \mbox{, }
\end{equation}
\begin{equation}\label{p2la}
\begin{split}
\mathcal{L}_{\pi N}^{(2)}&=c_1\langle\chi_+\rangle\bar NN-\frac{c_2}{4M^2}\langle u^\mu u^\nu\rangle(\bar N D_\mu D_\nu N+\text{h.c.})\\
&+\frac{c_3}{2}\langle u^\mu u_\mu\rangle\bar NN-\frac{c_4}{4}\bar N \gamma^\mu\gamma^\nu \big[u^\mu,u^\nu\big]N\ \mbox{, }
\end{split}
\end{equation}
\begin{equation}\label{p3la}
\begin{split}
\mathcal{L}_{\pi N}^{(3)}&=-\frac{d_1+d_2}{4M}\Big\{\bar N\big[u_\mu,h_\nu^{\ \mu}\big](D^\nu N)+\text{h.c.}\Big\}
+\frac{d_3}{12M^3}\Big\{\bar N \Big[u_\mu,\big[D_\nu,u_\rho\big]\Big](D^{\mu\nu\rho}N)+\text{h.c.}\Big\}\\
&+\frac{d_5}{2M}\Big\{\text{i}\bar N\big[\chi_-,u_\mu\big](D^\mu N)+\text{h.c.}\Big\}\\
&+\frac{d_{14}-d_{15}}{8M}\Big\{\text{i}\bar N \sigma^{\mu\nu}\big\langle\big[D_\rho,u_\mu\big]u_\nu-u_\mu\big[D_\nu,u_\rho\big]\big\rangle (D^\rho N)+\text{h.c.}\Big\}\\
&+\frac{d_{16}}{2}\bar N\gamma^\mu\gamma^5\langle\chi_+\rangle u_\mu N+\frac{\text{i}d_{18}}{2}\bar N\gamma^\mu\gamma^5\big[D_\mu,\chi_-\big] N\ \mbox{. }
\end{split}
\end{equation}
Here $M$ and $g$ are the mass of the nucleon and the axial current coupling constant in the chiral limit, while $c_i$s and $d_j$s are low-energy coupling constants (LECs); $\langle\cdots\rangle$ means matrix tracing in the flavour space. Furthermore, the relevant building blocks are~\cite{Fettes:2000gb}
\begin{equation}
\begin{split}
&D_\mu=\partial_\mu+\Gamma_\mu\ \mbox{, }\\
&\Gamma_\mu=\frac{1}{2}\big[u^\dagger(\partial_\mu-\text{i}r_\mu)u+u(\partial_\mu-\text{i}l_\mu)u^\dagger\big]\ \mbox{, }\\
&u_\mu=\text{i}\big[u^\dagger(\partial_\mu-\text{i}r_\mu)u-u(\partial_\mu-\text{i}l_\mu)u^\dagger\big]\ \mbox{, }\\
&\chi_{\pm}\equiv u^\dagger\chi u^\dagger\pm u\chi^\dagger u\ \mbox{, }\\
&\chi=2B_0(s+i p)\ \mbox{, }\\
&h_\nu^{\ \mu}=\big[D_\nu,u^\mu\big]+\big[D^\mu,u_\nu\big]\ \mbox{, }\\
&D^{\mu\nu\rho}=D^\mu D^\nu D^\rho+\cdots\mbox{(all permutations of $\{\mu,\nu,\rho\}$)}\ \mbox{, }
\end{split}
\end{equation}
with $l_\mu,r_\mu,s,p$ being external sources and $B_0$ a constant related to quark condensation. Finally, the interaction vertex of four pions is also relevant to the $\mathcal{O}(p^3)$ calculation, and the following pion-pion interaction Lagrangian~\cite{Gasser:1983yg} in pure meson ChPT is needed
\begin{equation}\label{piLa}
\mathcal{L}^{(2)}_{\pi\pi}=\frac{F^2}{4}\langle u^\mu u_\mu+\chi_+\rangle\ \mbox{. }
\end{equation}

With the above Lagrangians, it is readily to derive the $\pi N$ scattering amplitudes up to $\mathcal{O}(p^3)$, which are necessary for estimating the contribution of the {\it l.h.c.}s in the PKU representation to be discussed in the next subsection. Though the leading one-loop BChPT amplitudes can be found in Refs~\cite{Alarcon:2012kn,Chen:2012nx}, for the sake of easy reference, here we demonstrate in Appendix~\ref{sec:BChPTcalc} the details of the leading one-loop calculation within the framework of BChPT using EOMS scheme.

The tree and loop amplitudes are collected in Appendices.~\ref{app:fftree} and \ref{app:ffloop} respectively. As pointed out in Ref.~\cite{Gasser:1987rb}, the loop amplitudes contain both ultraviolet (UV) divergences and PCB terms, i.e., the terms in lower chiral order than the naively expected order. For the UV divergence, the dimensional regularization method is employed to pick out the relevant pieces (see Appendix.~\ref{app:ffloop}). We use on-shell renormalization scheme to handle the self-energy of the nucleon, which maintains the correct analytical behavior at the point $s=M_N^2$, see Appendix.~\ref{app:renormdiv}. On the other hand, to remedy the power counting breaking problem, we use EOMS scheme to separate the PCB polynomials, which are then absorbed by the LECs appearing in the chiral effective Lagrangians. Those procedures are shown in Appendix.~\ref{app:EOMS}.

The analytical expressions for all the involved loop integrals are compiled in Appendix~\ref{app:exploop}. Those expressions are obtained by means of dispersion relations with the spectral functions calculated using Cutkosky rule. Therefore, in principle, the BChPT amplitudes shown in this paper are calculable on the whole complex Mandelstam plane.

\subsection{Unitary PKU representation\label{sec:PKU}}
The PKU representation of the partial wave $\pi N$ elastic scattering $S$ matrix can be written as
\begin{equation}\label{eq:Spku}
S(s)=\prod_bS_b(s)\cdot\prod_vS_v(s)\cdot\prod_rS_r(s)\cdot e^{2\text{i}\rho(s)f(s)}\ \mbox{, }
\end{equation}
where $S_b$, $S_v$ and $S_r$ represent the individual contributions of bound states, virtual states and resonances respectively (see Ref.~\cite{Wang:2017agd} for their explicit expressions). The exponential term amounts to the background that carries the information of {\it l.h.c.}s and right-hand inelastic cut ({\it r.h.i.c.}) above inelastic thresholds. The kinematic factor $\rho(s)$ is given by
\begin{equation}\label{rhodef}
\rho(s)=\frac{\sqrt{s-s_L}\sqrt{s-s_R}}{s}\ \mbox{, }
\end{equation}
with $s_L=(M_N-m_\pi)^2$ and $s_R=(M_N+m_\pi)^2$. The function $f(s)$ satisfies a dispersion relation,
\begin{equation}\label{fdisper}
f(s)=\frac{s}{2\pi\text{i}}\int_\text{L}{\rm d}s^\prime\frac{\text{disc}f(s^\prime)}{(s^\prime-s)s^\prime}+\frac{s}{2\pi\text{i}}\int_{\text{R}'}{\rm d}s^\prime\frac{\text{disc}f(s^\prime)}{(s^\prime-s)s^\prime} \ \mbox{, }
\end{equation}
where $\text{L}$ and $\text{R}^\prime$ abbreviate {\it l.h.c.}s and {\it r.h.i.c.}, respectively.
The discontinuities of $f(s)$ along the various cuts can be deduced from the partial-wave BChPT amplitudes $T(L_{2I,2J})$ [c.f. Eq.~\eqref{PWamps}] through the following relations:
\begin{equation}
\begin{split}
&\text{disc}f(s^\prime)=\text{disc}\left[\frac{\ln S(s^\prime)}{2i\rho(s^\prime)}\right]\ \mbox{, }\\
&S(s^\prime)=1+2i\rho(s^\prime)T(s^\prime)\ \mbox{. }
\end{split}
\end{equation}
Specifically, the partial-wave $\pi N$ elastic scattering amplitudes $T(s^\prime)$ are obtained with Eq.~\eqref{PWamps} by inserting the BChPT amplitudes calculated up to $\mathcal{O}(p^3)$ level in the previous subsection.

Taking into account the realistic {\it l.h.c.}s and {\it r.h.i.c.}, as depicted in Figure~\ref{fig:cuttotal}, Eq.~\eqref{fdisper} is expanded to the form
\begin{equation}\label{fdisperpiNp3}
\begin{split}
f(s)&=-\frac{s}{\pi}\int_{s_{c}}^{(M_N-m_\pi)^2} \frac{\ln|S(s^\prime)|{\rm d}s^\prime}{2\rho(s^\prime)s^\prime(s^\prime-s)}+\frac{s}{\pi}\int_{(M_N^2-m_\pi^2)^2/M_N^2}^{2m_\pi^2+M_N^2} \frac{\text{Arg}[S(s^\prime)]{\rm d}s^\prime}{2is^\prime\rho(s^\prime)(s^\prime-s)}\\
&+\frac{s}{\pi}\int_0^{\theta_c}\frac{\ln[S_{in}(\theta)/S_{out}(\theta)]}{2i\rho(s^\prime)(s^\prime-s)\mid_{s^\prime=(M_N^2-m_\pi^2)e^{i\theta}}}{\rm d}\theta
+\frac{s}{\pi}\int_{(2m_\pi+M_N)^2}^{\Lambda_R^2} \frac{\ln[1/\eta(s^\prime)]{\rm d}s^\prime}{2\rho(s^\prime)s^\prime(s^\prime-s)}\ \mbox{, }
\end{split}
\end{equation}
where $S_{in}$ and $S_{out}$ are the $S$ matrices along the circular cut from inside and outside, respectively, and $0<\eta<1$ is the inelasticity along the {\it r.h.i.c.}. Furthermore, three cut-off parameters, $s_c$, $\theta_c$ and $\Lambda_R$, have been taken for the kinematical cut $(-\infty,(M_N-m_\pi)^2]$, the circular cut and the {\it r.h.i.c.} $[(2m_\pi+M_N)^2,\infty)$, in order. The first term of Eq.~\eqref{fdisperpiNp3} is negative definite and dominates the contribution to $f(s)$. The other three terms are quite small numerically. The second term originates from the $u$-channel nucleon exchange which can be approximated as contact interaction at low energies. The third piece only contributes at $\mathcal{O}(p^3)$ level, while the fourth term is positive definite but has been shown to be of insignificant impact~\cite{Wang:2017agd}. Eventually, the resulting function $f(s)$ provides a definitely negative contribution to the partial-wave phase shift, given that the cut-off parameters are chosen inside reasonable domains.

The fact that $f(s)$ contributes negatively is of crucial importance in exposing hidden poles on the second Riemann sheet. Apart from $f(s)$, the other source of negative contribution is thoroughly owing to bound states, which are clearly seen experimentally and hence their contribution can be fixed {\it a priori}. As a result, the disentanglement of the hidden poles  with positive contributions from a negative definite background can be easily achieved, which will be explored in the next section.

\section{Partial wave phase shifts and pole analyses}
\subsection{Pole-hunting strategy}\label{known}
The production presentation of the $S$ matrix results in additive phase shifts with definite signs that are implied by the contributing sources. In general, a bound state contributes negatively, while a pole (virtual or resonance state) on the second Riemann sheets gives positive contribution. The sum of the cut contributions,  identical to $\rho(s)f(s)$, is negative.  Therefore, it is rather convenient to carry out exploration of hidden poles by using partial wave phase shifts, denoted by $\delta(s)$ henceforth.

For easy explanation of the pole-hunting strategy, the $S$-matrix in Eq.~\eqref{eq:Spku} can be rewritten as
\bea\label{eq:Spku2}
S(s)=\overbrace{S_h(s)}^{\rm hidden~poles}\cdot \overbrace{\prod_{b\in \mathcal{B}}S_{b}(s)\prod_{r\in \mathcal{R}} S_r(s)}^{\rm known~poles}\cdot \overbrace{e^{2i\rho(s)f(s)}}^{\rm cuts}=e^{2i\delta(s)}\ .
\eea
In above, the term $S_h(s)$ accounts for the contribution from all possible hidden poles, e.g. virtual states, resonances below the threshold or with extremely large widths, or shadow poles. They cannot be observed directly by experiments, but their existence (and pole positions) can be examined through PKU representation once the last two terms in the first equality are known. To that end, in what follows, we will specify the known-pole and cut contributions by discussing the necessary inputs.

The known-pole contribution can be fixed with their corresponding experimental information. The known poles contain bound states and above-threshold resonances, both of which are well determined experimentally. In our case, the $S$- and $P$-wave known poles below $2$~GeV are complied in Table~\ref{tab:poles}. The bound state in the $P_{11}$ channel is the nucleon itself. For the known resonances, some of them are above the inelastic $\pi\pi N$ threshold and hence are actually located on the third and higher Riemann sheets. Nevertheless, under narrow width approximation, their corresponding shadow poles on the second Riemann sheet may be estimated by
\bea\label{eq:shpole}
{z_r}^{\rm II}=M_{\rm pole} +\frac{i}{2}(\Gamma_{\rm inelastic}-\Gamma_{\pi N})\ ,
\eea
which contribute to $S$ matrix in the same way as the standard resonances on 2nd Riemann sheet, like $\Delta(1232)$.
\begin{table}[htbp]
\begin{center}
 \begin{tabular}  { c | c |  c }
  \hline
 $L_{2I,2J}$    &$\mathcal{B}$ (bound state)& $\mathcal{R}$ (resonance)\\
  \hline
  $S_{11}$  &-& $\{~N^*(1535), N^*(1650), N^*(1895)~\}$\\
  $S_{31}$  &-& $\{~\Delta(1620), \Delta(1900)~\}$\\
  $P_{11}$  &$\{~N(938)~\}$& $\{~ N^*(1440), N^*(1710), N^*(1880)~\}$\\
  $P_{31}$  &-& $\{~\Delta(1910)~\}$\\
  $P_{13}$  &-& $\{~N^*(1720), N^*(1900)~\}$\\
  $P_{33}$  &-& $\{~\Delta(1232),\Delta(1600),\Delta(1920)~\}$\\
  \hline
 \end{tabular}\\
 \caption{Known poles in the two $S$-wave and four $P$-wave channels.\label{tab:poles}}
\end{center}
\end{table}

The cut contribution can be estimated from BChPT amplitudes as demonstrated in subsection~\ref{sec:PKU}. There are three free cut-off parameters, i.e. $s_c$, $\theta_c$ and $\Lambda_R$ in Eq.~\eqref{fdisperpiNp3}. It is emphasized that the cut-off parameters for left-hand cuts, i.e. $s_c$ and $\theta_c$, are in principle unknown parameters. However, to get a first glimpse of the physics, one can assign the values of $s_c$ and $\theta_c$ in accordance with the validity region of BChPT. Similar to the previous work, we assume that the results of {\it l.h.c.}s from the perturbative BChPT calculation remain correct till meeting the $N^*(1440)$ shadow pole, above which complicated coupled-channel dynamics takes place. The distance between the shadow pole and the center point of chiral expansion is denoted as $r_{\chi}$. The boundary of BChPT valid region should be a circle centering at $s_\chi=M_N^2+m_\pi^2$ with the radius $r_{\chi}$. The intersections of this circle with the {\it l.h.c.}s give $s_c=-0.08$ GeV$^2$ and $\theta_c=1.18$ radian, see Figure~\ref{fig:cuttotal}. The two parameters are actually related by
\begin{equation}\label{thetac}
\theta_c=
\begin{cases}
\arccos\left[\frac{2(M_N^4+m_\pi^4)-|s_c-M_N^2-m_\pi^2|^2}{2(M_N^4-m_\pi^4)}\right]&\mbox{($s_c\geq m_\pi^2-M_N^2$), }\\
\pi &\mbox{(\text{otherwise}) }
\end{cases}
\end{equation}
throughout this work: $\theta_c$ will always be obtained from $s_c$, which sometimes takes values other than the chiral estimation $-0.08$~GeV$^2$, by using the above relation. As for $\Lambda_R$, following Ref.~\cite{Wang:2017agd}, it is set to $\Lambda_R=4$~GeV. In addition, the experimental data of inelasticity $\eta$ taken from~\cite{SAID} is used in the calculation of the {\it r.h.i.c.} contribution.

In above we discussed the validity region of BChPT. However, in no way one should expect that the contribution beyond the validity region of BChPT should be negligible. The two parameters $s_c$ and $\theta_c$ are actually taken as free parameters in the fit. Their deviation from the boundary of validity region signals the contribution from high energy region\footnote{From the discussions of the constant $R$ in Sect.~\ref{subsec:thrpara} it is found that even though $s_c$ becomes large, the magnitude of the background contribution is reasonable. }.

Furthermore, to compute the cut contribution, the masses, the nucleon axial charge and the pion decay constant in the BChPT amplitudes take the following values:
\bea
m_\pi = 139.6~{\rm MeV}\ ,\quad M_N=938.3~{\rm MeV}\ ,\quad F_\pi =92.4~{\rm MeV}\ ,\quad g_A=1.27\ ,
\eea
while the relevant LECs, $c_i$ and $d_j$ in Eqs.~\eqref{p2la} and \eqref{p3la}, are set to the fit values of Fit I in Table.~I of Ref.~\cite{Yao:2016vbz}\footnote{Fit I is in absence of explicit $\Delta(1232)$ degree of freedom, i.e. the $\Delta$ field is integrated out.}. Other recent determinations, e.g. Refs.~\cite{Siemens:2016hdi,Siemens:2016jwj}, yield very similar LECs values, and it has already been illuminated in Ref.~\cite{Wang:2017agd} that the change of the LECs has inconsiderable impact on the {\it l.h.c.}s. Moreover, the values of the LECs are determined in Ref.~\cite{Yao:2016vbz} by performing fit to the phase shifts generated by the recent RS equation analysis of the $\pi N$ scatterings in Ref.~\cite{Hoferichter:2015hva}, where both central values and error bars are given. To be consistent, for the pole analyses we will employ the recent RS partial wave phase shifts as well, rather than the ones from PWA by GWU group without errors~\cite{SAID} which are used in Ref.~\cite{Wang:2017agd}.

With the above preparations, we are now in the position to perform phase shift analyses, and explore possible hidden poles with the equipment of the PKU representation.
\subsection{Phase shift analyses}\label{subsec:PSana}
In Ref.~\cite{Wang:2017agd}, among the six $S$- and $P$-wave channels,  the $S_{11}$ and $P_{11}$ waves have been paid special attention to in the study at $\mathcal{O}(p^2)$ level. It is important and necessary to verify if the hidden poles, discovered therein for these two channels, still keep their presence in an $\mathcal{O}(p^3)$ analysis. The $\mathcal{O}(p^3)$ analysis will take into account the full structure of {\it l.h.c.}s, as mentioned in the Introduction, and hence is adequate to verify the very existence of those poles. Besides, higher-order contributions in the chiral expansion will not introduce any other new branch-cut structures, and they should have little impact on the $\mathcal{O}(p^3)$ results obtained here. One refers to Appendix.~\ref{app:uncer} to see the uncertainties caused by numerical details such as the cut-off parameters and the unitarization methods.

In what follows, it will be shown that the $\mathcal{O}(p^3)$ results indeed provide further and stronger evidences indicating that the hidden poles in $S_{11}$ and $P_{11}$ channels definitely exist. In particular, the extra hidden poles in $S_{11}$ and $P_{11}$ waves do play an essential role in establishing a meaningful numerical fit -- without them the chi-squares would be unacceptably large. On the other hand, we will also carry out careful investigations for the other $S$- and $P$-wave channels, which were less studied in Ref.~\cite{Wang:2017agd}.

\subsubsection{$S_{11}$ channel}
\begin{figure}[htbp]
\centering
\includegraphics[width=0.48\textwidth]{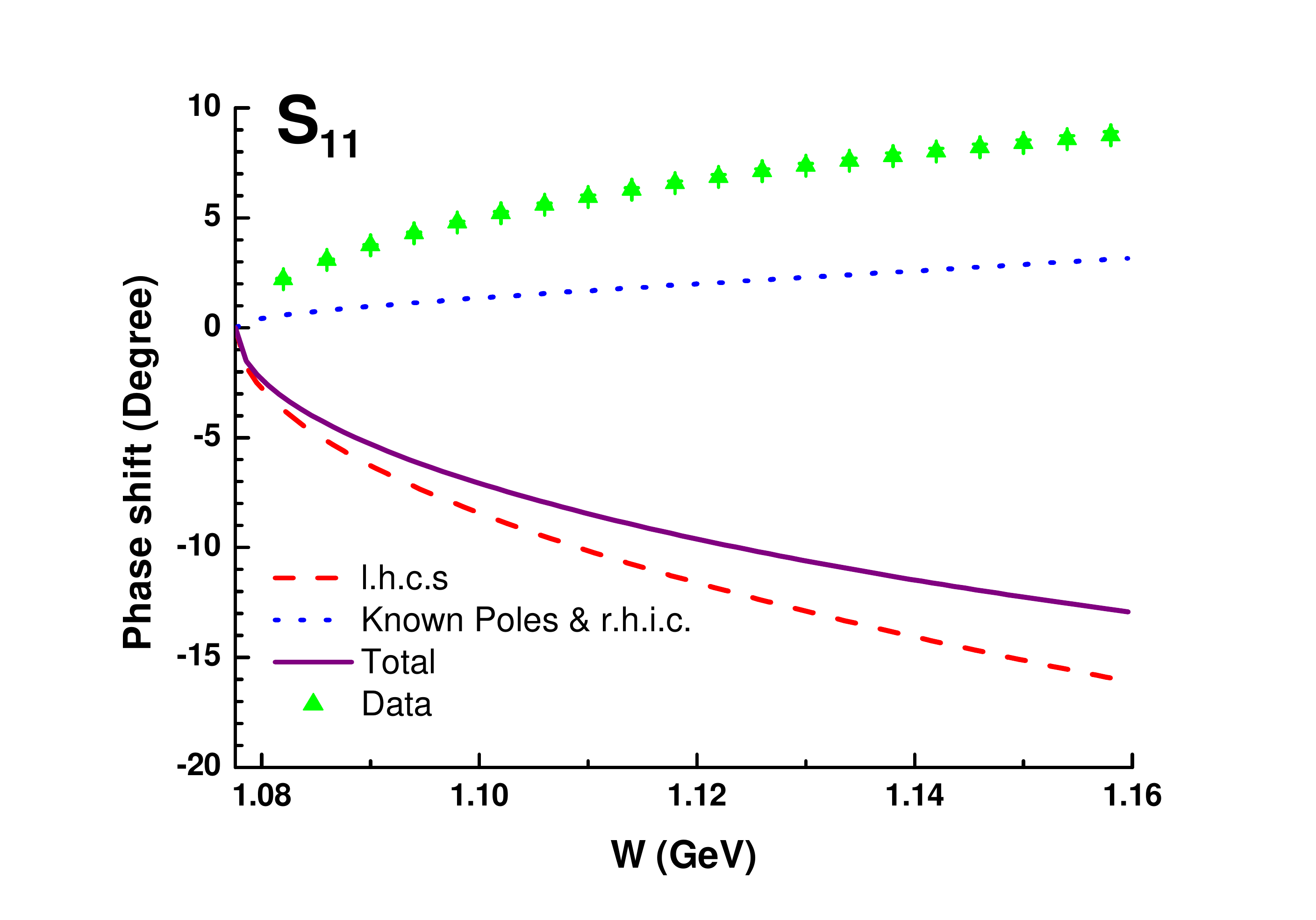}
\includegraphics[width=0.48\textwidth]{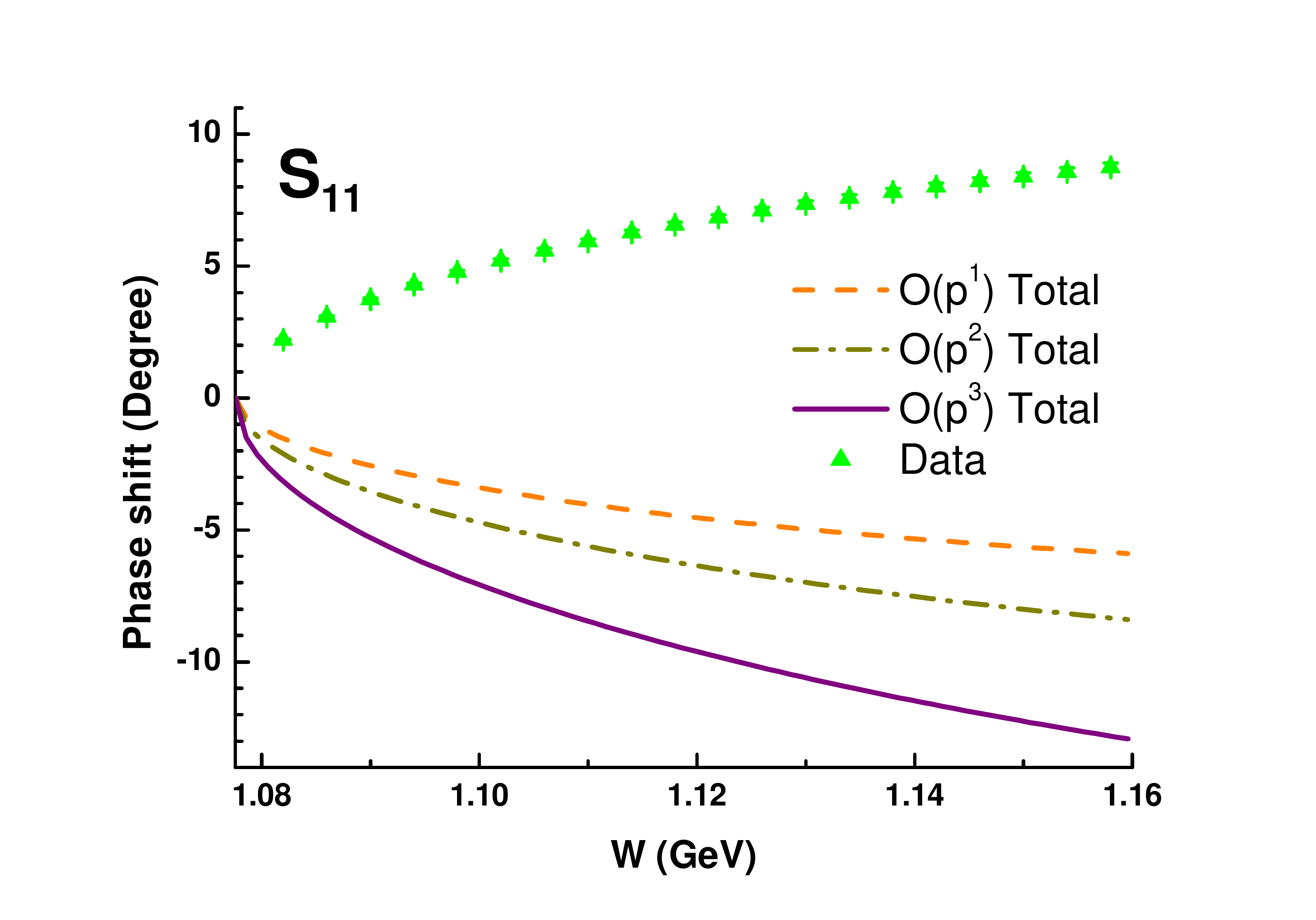}
\caption{PKU-representation analysis of the $S_{11}$ channel. Left: kown contributions vs the data. Right: known contributions with {\it l.h.c.}s estimated at different chiral orders ($s_c=-0.08$ GeV$^2$). }
\label{fig:S11ana}
\end{figure}

The known-pole and cut contributions in $S_{11}$ channel are shown in the left panel of Fig.~\ref{fig:S11ana}. It can be observed that the $S_{11}$ channel suffers from a large discrepancy between the RS data and the sum of all the known contributions. Unfortunately, this disagreement cannot be rescued by fine-tuning the involved parameters in the numerical computation. Instead, a hidden pole contribution, as a new dynamical structure, must be included so as to compensate such discrepancy, as suggested in the $\mathcal{O}(p^2)$ study carried out in Ref.~\cite{Wang:2017agd}. In the right panel of Fig.~\ref{fig:S11ana}, the known contributions are displayed order by order. Compared to the $\mathcal{O}(p^2)$ case, our thorough $\mathcal{O}(p^3)$ calculation now enlarges the discrepancy\footnote{The figure indicates that the convergence property of the current method is not so satisfactory, i.e. $\mathcal{O}(p^3)$ contributions do not bring obviously smaller changes to the phase shift than the $\mathcal{O}(p^2)$. This implies that the cut-off parameter value $s_c=-0.08$ GeV$^2$ is a bit larger than the validity region of ChPT. As already discussed in Sect.~\ref{known}, a larger cut-off parameter value is necessary. }, which strongly indicates the necessity of the hidden pole. Furthermore, the fit result is unique, at least for two poles. For example, if we add one resonance together with one virtual state, the virtual state will move automatically to the pseudo-threshold and the remaining resonance results in the same location as the one resonance scene. In addition, if only two virtual states are taken into consideration, they will automatically collide on the real axis and turn into one resonance. This property of the stability of pole positions is an imperceptible advantage of the present method.

\begin{table}[htbp]
\begin{center}
 \begin{tabular}  {| c | c | c | c ||}
  \hline
  $s_{c}$ (GeV$^2$)  & Pole position (MeV) & $\chi^2/\text{d.o.f}$ \\
  \hline
  $-0.08$ & $814(3)-i\,141(8)$ & $1.46$ \\
  \hline
  $-1.00$ &$882(2)-i\,190(4)$ & $1.31$ \\
  \hline
  $-9.00$ &$960(2)-i\,192(2)$ & $1.14$ \\
  \hline
  $-25.0$ &$976(2)-i\,187(1)$ & $1.14$ \\
  \hline
 \end{tabular}\\
 \caption{The $S_{11}$ hidden pole fit with different choices of $s_{c}$. }\label{tab:S11p3sh}
\end{center}
\end{table}

\begin{figure}[htbp]
\centering
\includegraphics[width=0.48\textwidth]{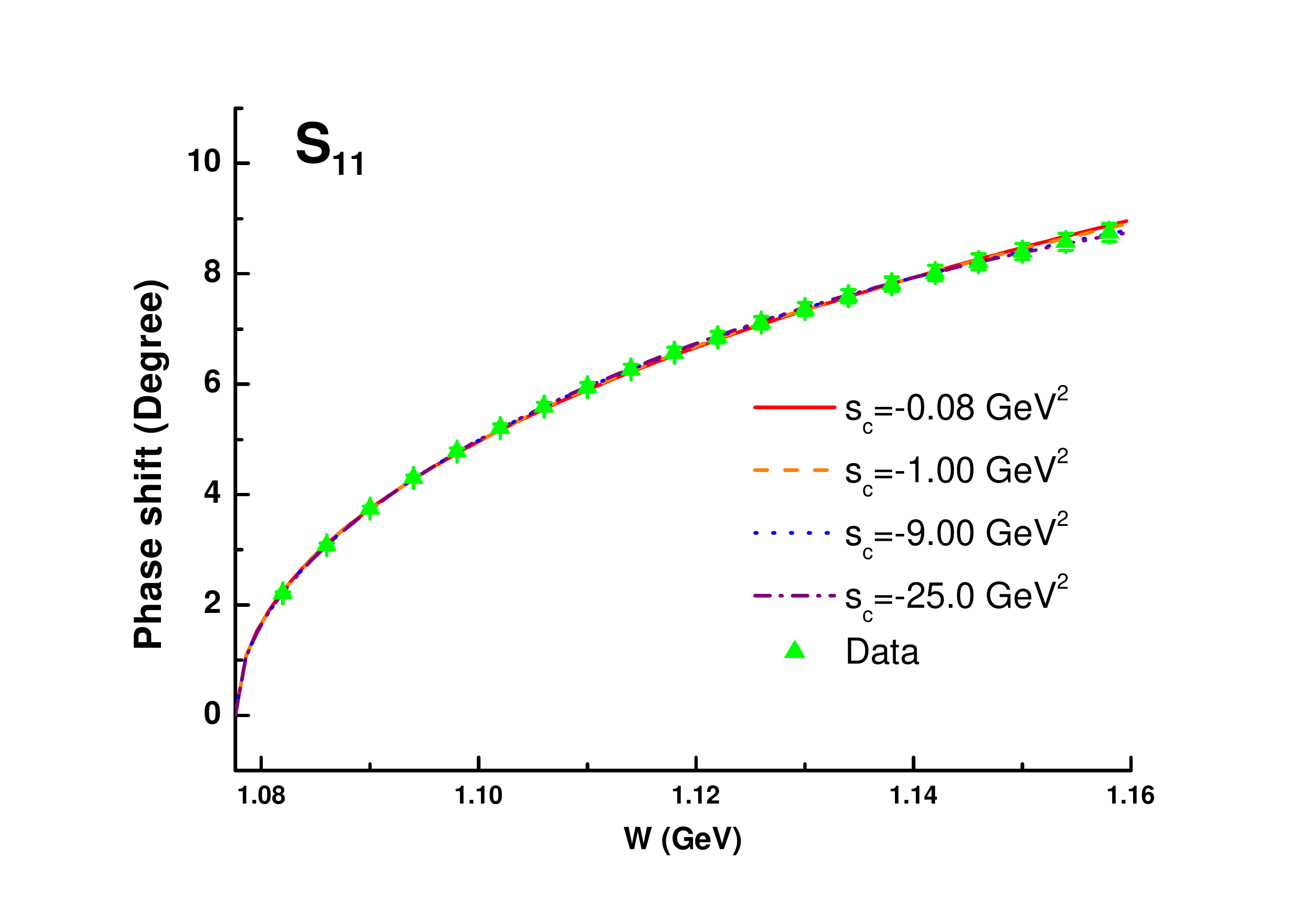}
\caption{Fit with an extra hidden pole in the $S_{11}$ channel. }
\label{fig:S11p3sh}
\end{figure}

To proceed, the position of the hidden pole can be determined by a fit to the RS phase shift data. The fit results in a crazy resonance below threshold, which is quite stable against the variation of the cut-off parameters. The hidden pole positions with various cut-off parameters are listed in Table.~\ref{tab:S11p3sh}\footnote{The chi-squares here are calculated using the bootstrap method, and so as the other five channels. }. The fit results are plotted in Fig.~\ref{fig:S11p3sh} and the RS data are well described due to the inclusion of the hidden pole\footnote{{The $S_{11}$ hidden pole has a residue close to the $\mathcal{O}(p^2)$ case. }}.

The pole position of this crazy resonance is given by
\begin{equation}\label{S11hp}
z=895(81)(2)-{i}\,164(23)(4)\ \mbox{MeV .}
\end{equation}
The numbers in the first brackets are systematical errors responsible for the variation of the cut-off parameter $s_c$ (see Table~\ref{tab:S11p3sh}), while the ones in the second brackets are the averaged statistical uncertainties from fitting\footnote{In the fits for different channels the statistical errors are much more smaller than the systematical errors and hence are negligible. Such small errors may come form the
way we use to handle the data and may not be so trustworthy. Anyhow, the systematical errors from the evaluation of left-hand cuts are of course more important physically. }. By adding the two uncertainties in quadrature, our final reported result is
\begin{equation}\label{S11hp}
z=895(81)-{i}\,164(23)\ \mbox{MeV .}
\end{equation}
It is compatible with the determination reported by the $\mathcal{O}(p^2)$ analyses~\cite{Wang:2017agd}, i.e. $861(53)-130(75)i$ MeV\footnote{{Note that we have also tried to fit the data of a larger energy region, i.e. below $\sqrt{s}=1.45$ GeV. Within expectation, the hidden pole remains stable. }}.

\subsubsection{$P_{11}$ channel}

\begin{figure}[htbp]
\centering
\includegraphics[width=0.48\textwidth]{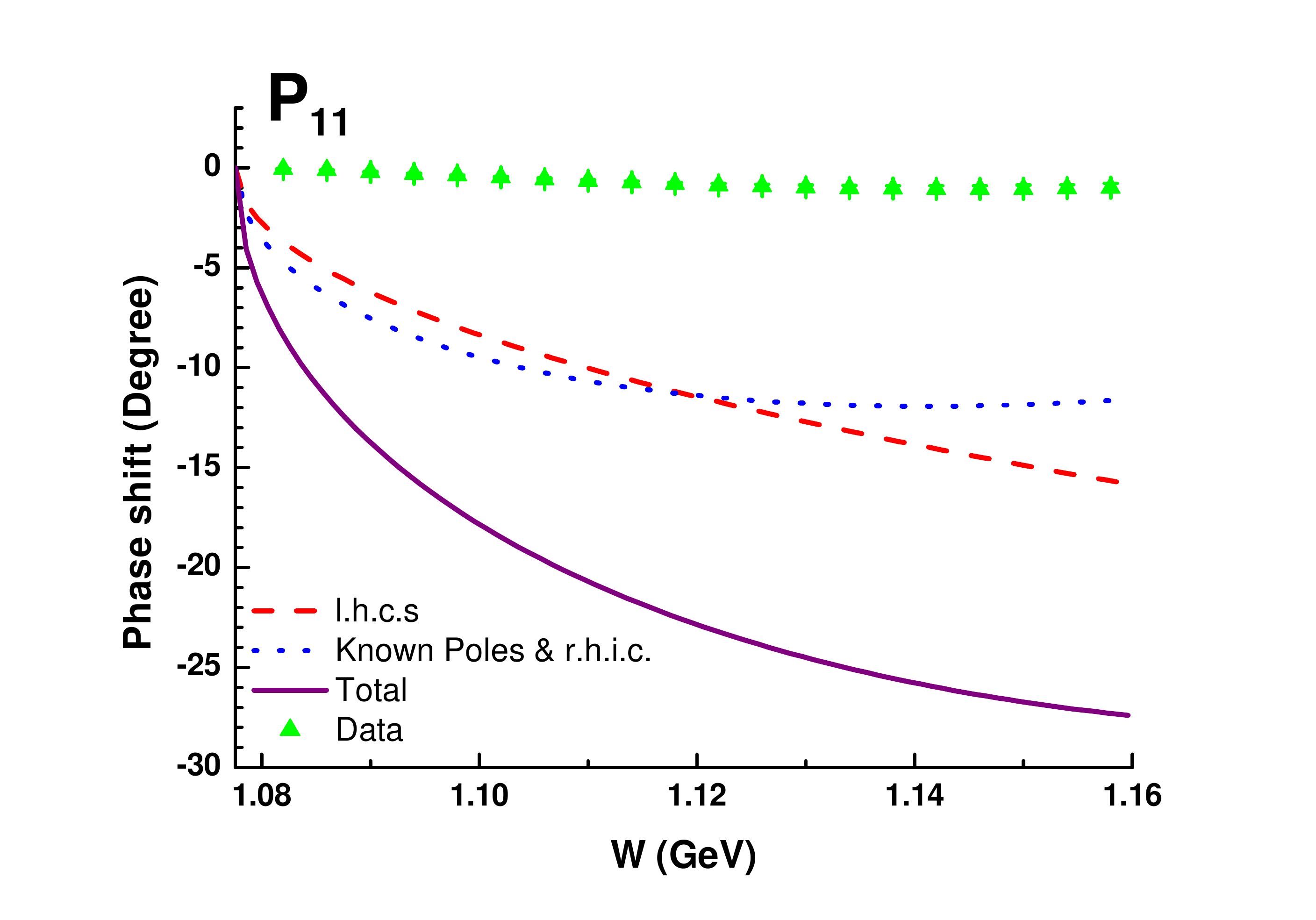}
\includegraphics[width=0.48\textwidth]{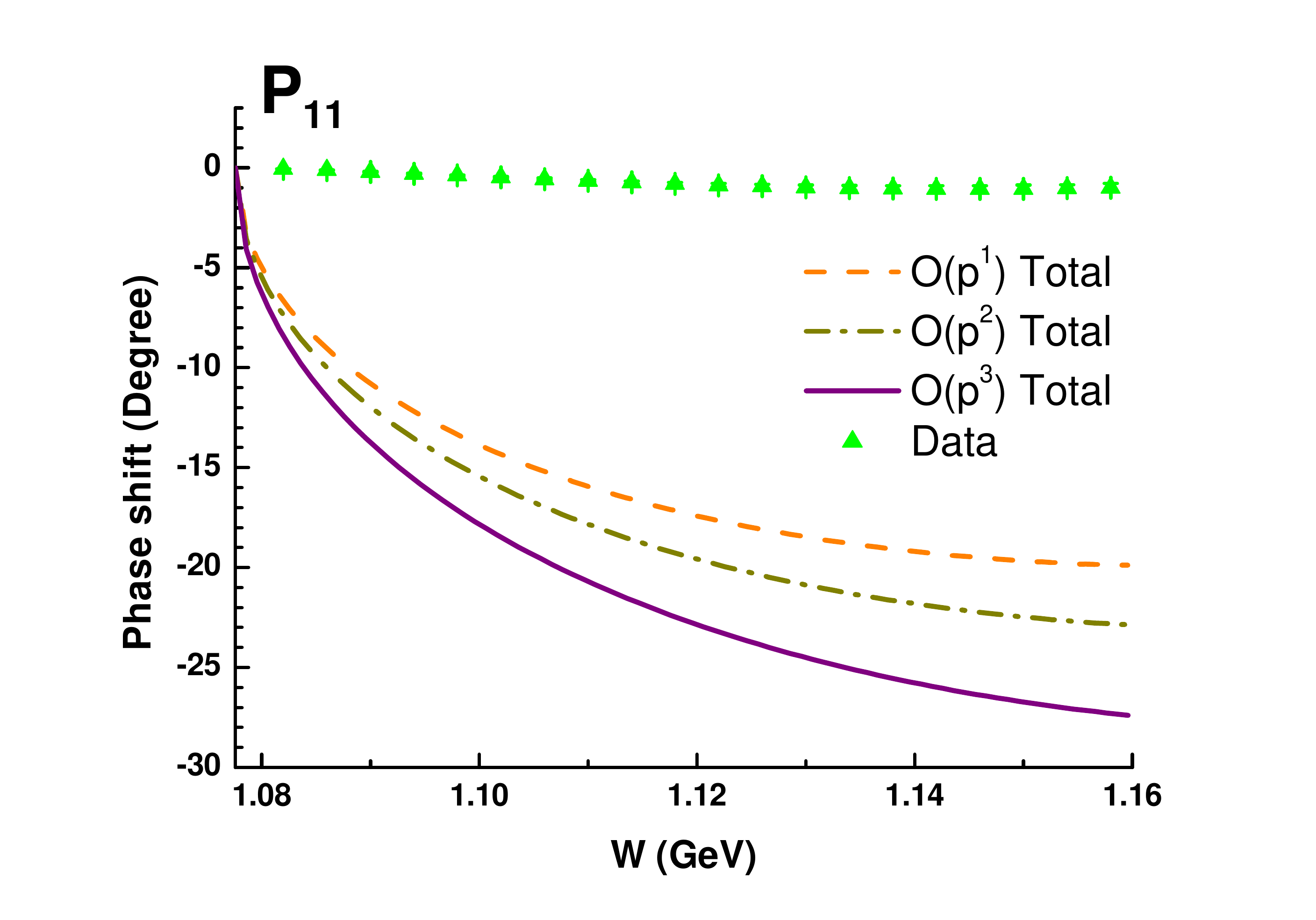}
\caption{PKU-representation analysis of the $P_{11}$ channel. Left: kown contributions vs the data. Right: known contributions with {\it l.h.c.}s estimated at different chiral orders ($s_c=-0.08$ GeV$^2$). }
\label{fig:P11ana}
\end{figure}

Likewise in the $P_{11}$ channel, the RS data is far from being described by the known contributions as shown in the left panel of Fig.~\ref{fig:P11ana}. Furthermore, it can be seen from the right panel of Fig.~\ref{fig:P11ana} that the discrepancy in $P_{11}$ channel becomes larger when the chiral order increases. Hence, the same prescription proposed in Ref.~\cite{Wang:2017agd} should be employed to resolve the inconsistence problem in the $P_{11}$ channel.

\begin{table}[htbp]
\begin{center}
 \begin{tabular}  { | c | c | c  |}
  \hline
  $s_{c}$ (GeV$^2$)  & Pole position (MeV) &  $\chi^2/\text{d.o.f}$\\
  \hline
  $-1.00$ &$983$ & $1.63$ \\
  \hline
  $-9.00$ &$962(1)$ & $1.37$ \\
  \hline
  $-25.0$ &$948(1)$ & $2.08$ \\
  \hline
 \end{tabular}\\
 \caption{The $P_{11}$ hidden pole fit with different choices of $s_{c}$. }\label{tab:P11p3sh}
\end{center}
\end{table}

\begin{figure}[htbp]
\centering
\includegraphics[width=0.48\textwidth]{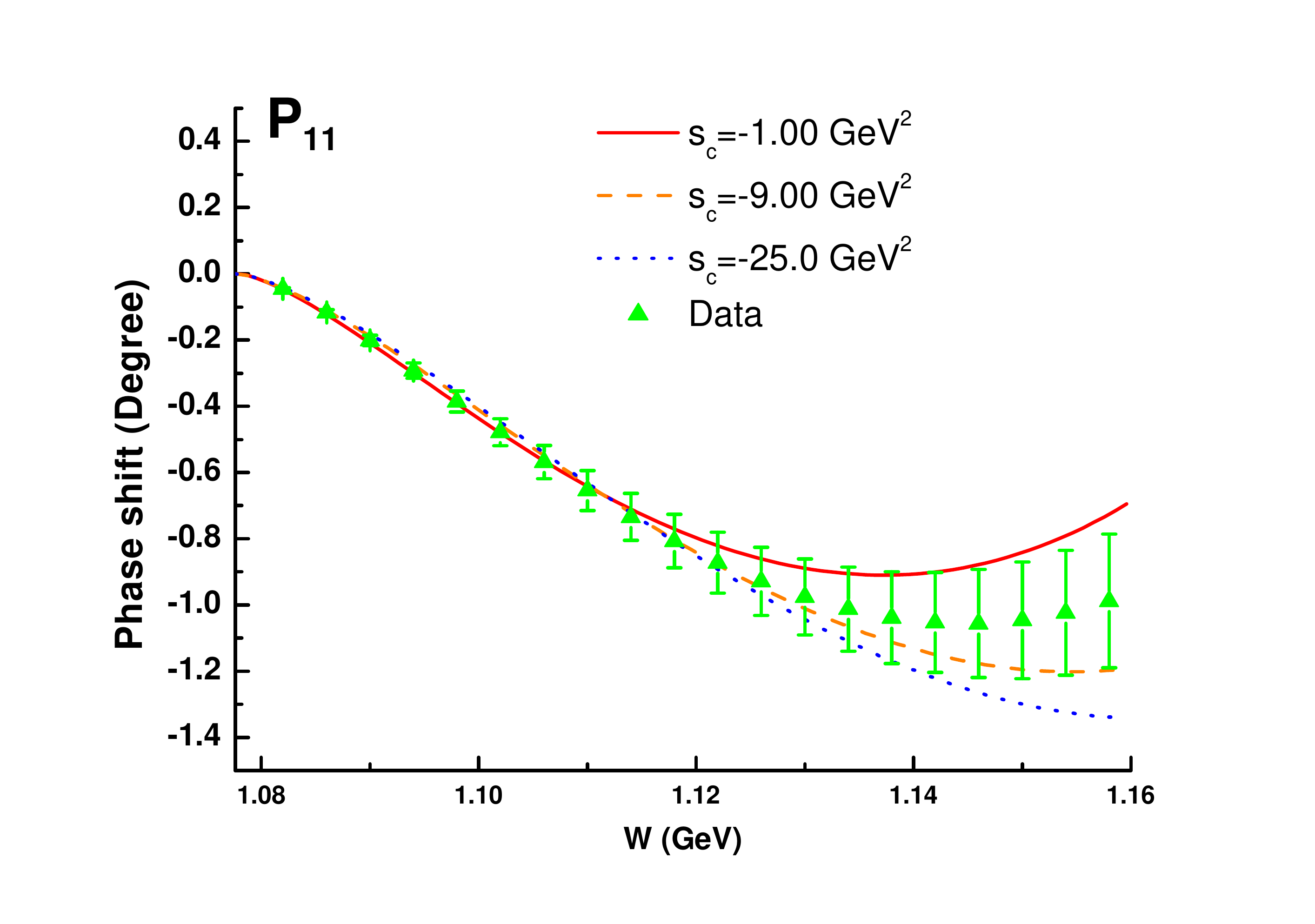}
\caption{The fit with two extra virtual states in $P_{11}$ channel. }
\label{fig:P11p3sh}
\end{figure}

Following Ref.~\cite{Wang:2017agd}, two virtual states are added to the $S$ matrix, and one stays in the near-threshold region as the major hidden contribution, while the other is nearly absorbed by the pseudo-threshold $(M_N-m_\pi)^2$.\footnote{Even if we put a resonance (usually appearing in pair according to analyticity), it automatically converts into two virtual states. This amazing fact supports crucially  the stability of our whole program. As already discussed in Ref.~\cite{Wang:2017agd}, one can also pin down the location of the virtual state by perturbation calculation or subthreshold expansion, and the two results coincide with each other, implying that such a way of estimating \textit{l.h.c.}s is fully justified. } The locations of the near-threshold virtual state with different cut-off parameters are listed in Table.~\ref{tab:P11p3sh}, and the corresponding phase shifts are shown in Fig.~\ref{fig:P11p3sh}. Note that a $P$-wave threshold constraint, i.e. $\delta(k)=\mathcal{O}(k^3)$ for $3$-momentum $k$, has been considered in the fits.

It should be pointed out that no satisfied fit can be achieved with $s_c=-0.08$ GeV$^2$. Nevertheless, if $|s_c|$ becomes a bit larger, e.g. $s_c=-1.00$ GeV$^2$, the fit quality is significantly improved. However, this is not true for the $\mathcal{O}(p^2)$ fit in Ref.~\cite{Wang:2017agd}. Therein a rather large $|s_c|$ like $s_c=-9.00$ GeV$^2$ has to be adopted. This improvement should be owing to the full consideration of the various {\it l.h.c.}s, such as the circular cut. As a consequence, the low-energy fit is now no longer sensitive to the physics in high energy region. Finally, based on the results in Table.~\ref{tab:P11p3sh}, we report the location for the extra near-threshold virtual pole:
\begin{equation}\label{P11hp}
v=966(18)(1)\ \mbox{MeV }=966(18)\ \mbox{MeV. }
\end{equation}
{Note that similar to what is discussed in Ref.~\cite{Wang:2017agd}, the $\mathcal{O}(p^3)$ perturbative $S$ matrix gives a first-sheet zero at $\sqrt{s}=973$ MeV, which is compatible with the fit result Eq.~\eqref{P11hp}. }

In principle, one may also perform fit with the cut-off parameter $s_c$ released, meanwhile, $\theta_c$ is determined by Eq.~\ref{thetac}. However,
due to the existence of many local-minimums, such fit does not converge and the resultant statistical errors are untrustworthy any more.

{The physical mechanism of the $P_{11}$ virtual state is already discussed in Ref.~\cite{Wang:2017agd}: it is a partner of the nucleon pole, which is different from the $S_{11}$ hidden pole, i.e. (likely) from potential mechanism. }
\subsubsection{$S_{31}$ channel}
\begin{figure}[htbp]
\centering
\includegraphics[width=0.48\textwidth]{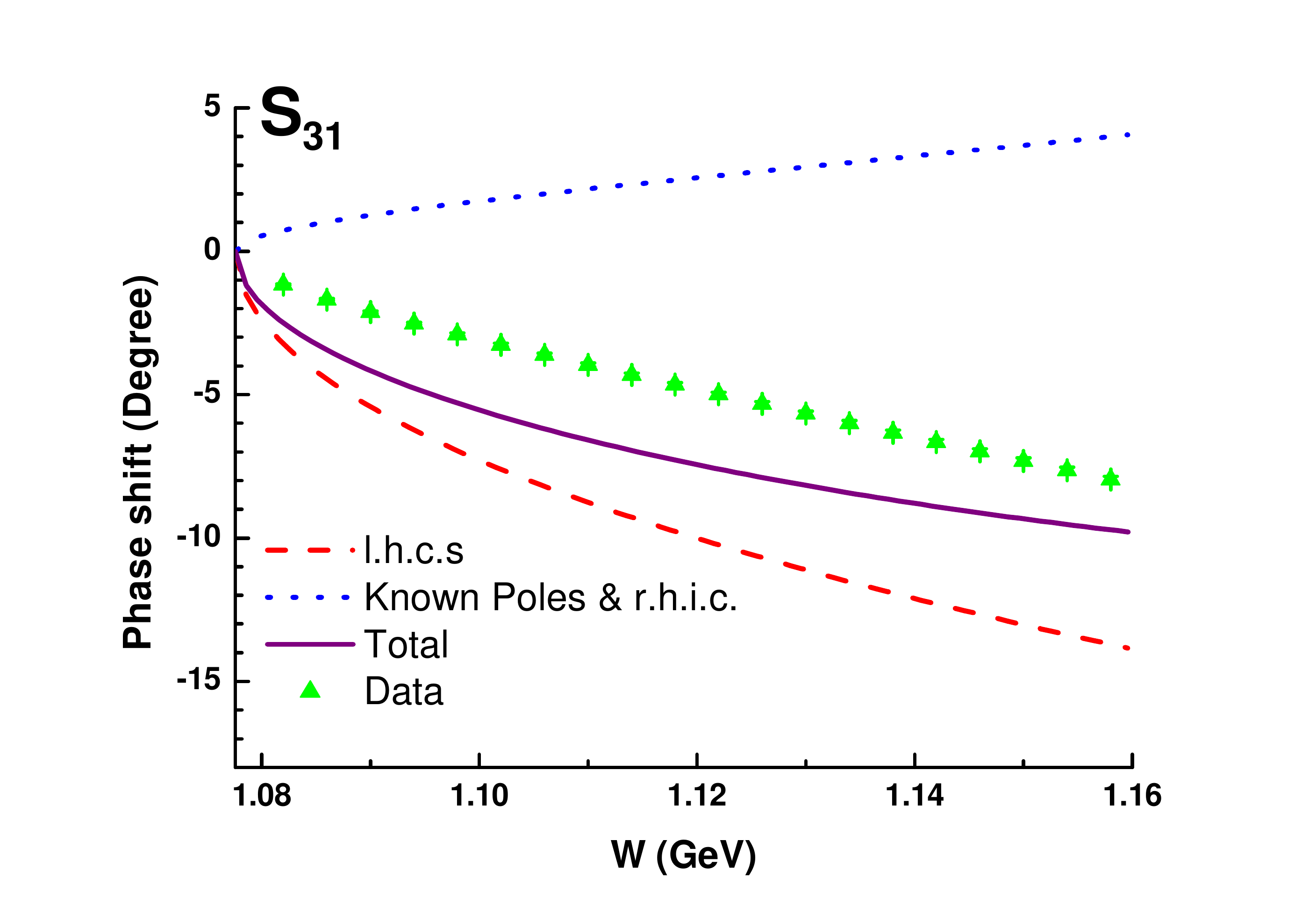}
\includegraphics[width=0.48\textwidth]{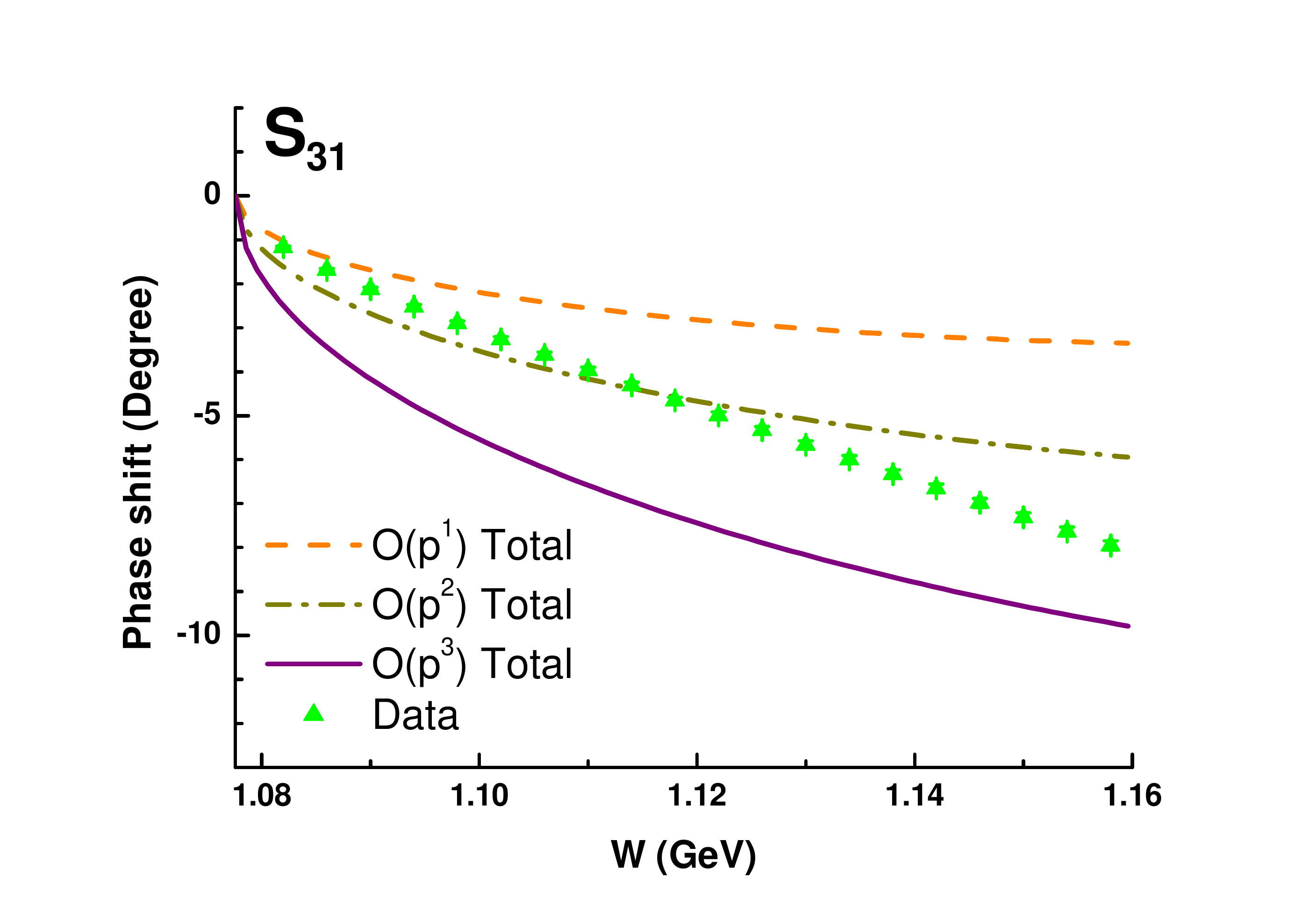}
\caption{PKU-representation analysis of the $S_{31}$ channel. Left: kown contributions vs the data. Right: known contributions with \textit{l.h.c.}s estimated in different chiral orders ($s_c=-0.08$ GeV$^2$). }
\label{fig:S31ana}
\end{figure}

Case in $S_{31}$ channel is quite interesting. As the left panel of Fig.~\ref{fig:S31ana} shows, firstly one finds some disagreements between the known contributions and the data, but those disagreements are minor: at least the known terms give the correct sign of the phase shift, and it seems that fine-tuning of the cut-off parameter $s_c$ can make the curve match the first two data points near threshold, which is totally different to the cases in $S_{11}$ and $P_{11}$ channels.
Unexpectedly, situation becomes awful once $s_c$ is released as a fitting parameter. The RS data cannot be fitted well due to the mismatching of line shape -- the data curve is too ``straight'' to be described by the current formula. Besides, this line-shape mismatching can not be cured by means of truncating the chiral series, since $\mathcal{O}(p^i)\ (i=1,2,3)$ calculation all give similar line shapes, see the right panel of Fig.~\ref{fig:S31ana}.

This observation may indicate that the evaluation of \textit{l.h.c.}s is not accurate enough, which has various possible origins: some additional minor cuts like the left-hand cuts from $t$- or $u$-channel \textit{resonance} exchange, improper shadow pole positions of the known resonances due to the narrow width approximation (see Eq.~\eqref{eq:shpole}), uncertainties from LECs, or some quite distant hidden contributions. It is impossible to take all the factors into consideration and do an all-embracing analyses.

\begin{figure}[htbp]
\centering
\includegraphics[width=0.48\textwidth]{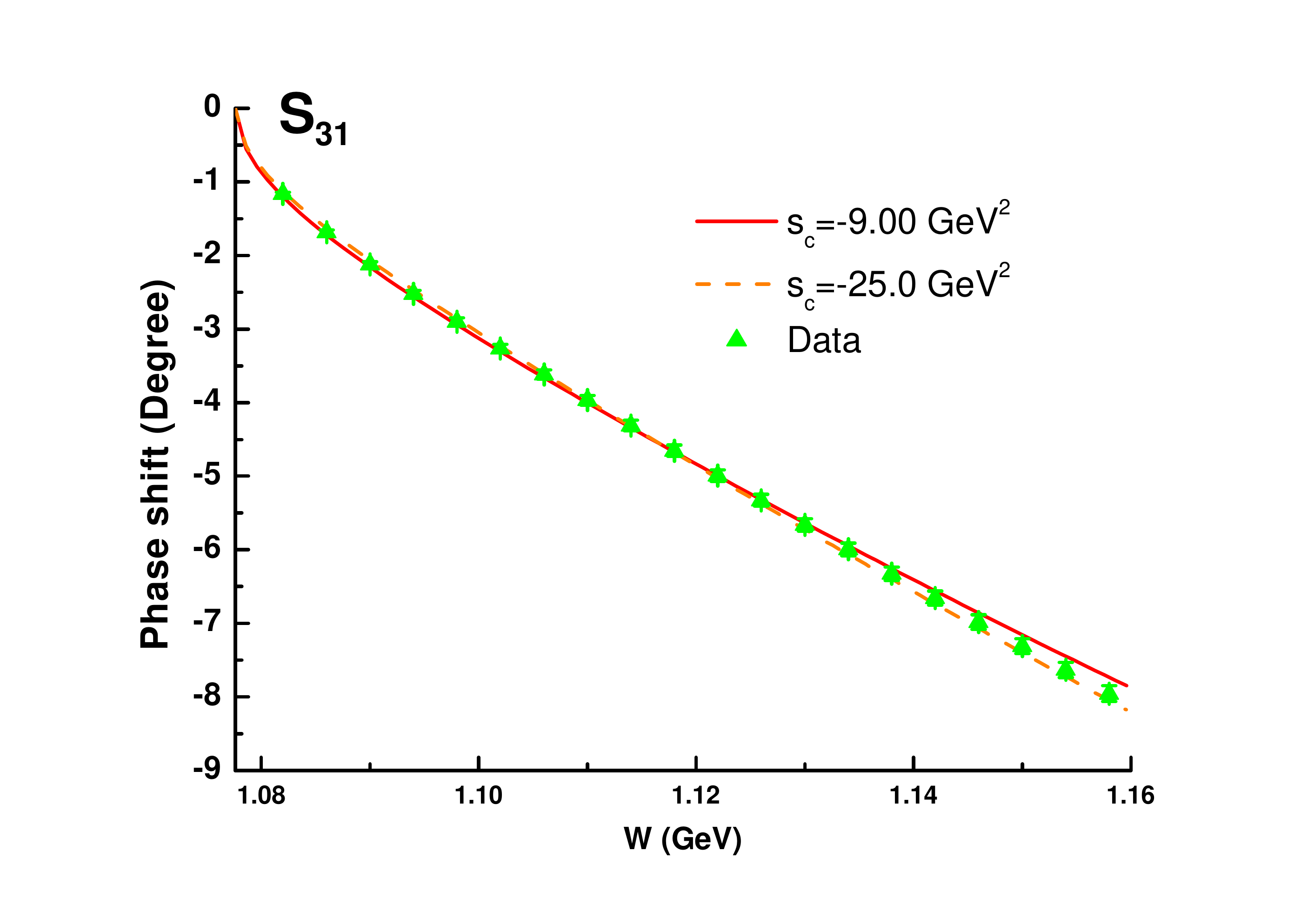}
\caption{The fit with one extra virtual state as background in $S_{31}$ channel. }
\label{fig:S31p3sh}
\end{figure}

However if one insists on getting a good fit to the data, the easiest way is to employ a pole term as an extra background, similar to what have been done in the $P_{11}$ channel. Here one virtual state is put into the $S$ matrix, and the results are shown in Table.~\ref{tab:S31fit}. A good fit needs a larger $|s_c|$ value, e.g. when $|s_c|>9$ GeV$^2$ one gets $\chi^2/\text{d.o.f}<3$. At the same time, the line shape becomes better when $|s_c|$ is large enough, see Fig.~\ref{fig:S31p3sh}. It should be emphasized here the $\mathcal{O}(p^2)$ fit fails to give a good chi-square when $|s_c|\leq25$ GeV$^2$.
\begin{table}[htbp]
\begin{center}
 \begin{tabular}  {| c | c | c | }
  \hline
  $s_{c}$ (GeV$^2$)  & Pole position (MeV) & $\chi^2/\text{d.o.f}$ \\
  \hline
  $-9.00$ &$875$ & $2.13$ \\
  \hline
  $-25.0$ &$887$ & $1.86$ \\
  \hline
 \end{tabular}\\
 \caption{The $S_{31}$ fit with an extra virtual pole as background. }\label{tab:S31fit}
\end{center}
\end{table}

It is interesting to find that the location of the virtual state serving as background is not so far away from the threshold (although it lies on the left of the $u$-channel nucleon cut), and seems stable, which implies that the extra corrections of \textit{l.h.c.}s in this channel is rather important.

\subsubsection{$P_{31}$ channel}

\begin{figure}[htbp]
\centering
\includegraphics[width=0.48\textwidth]{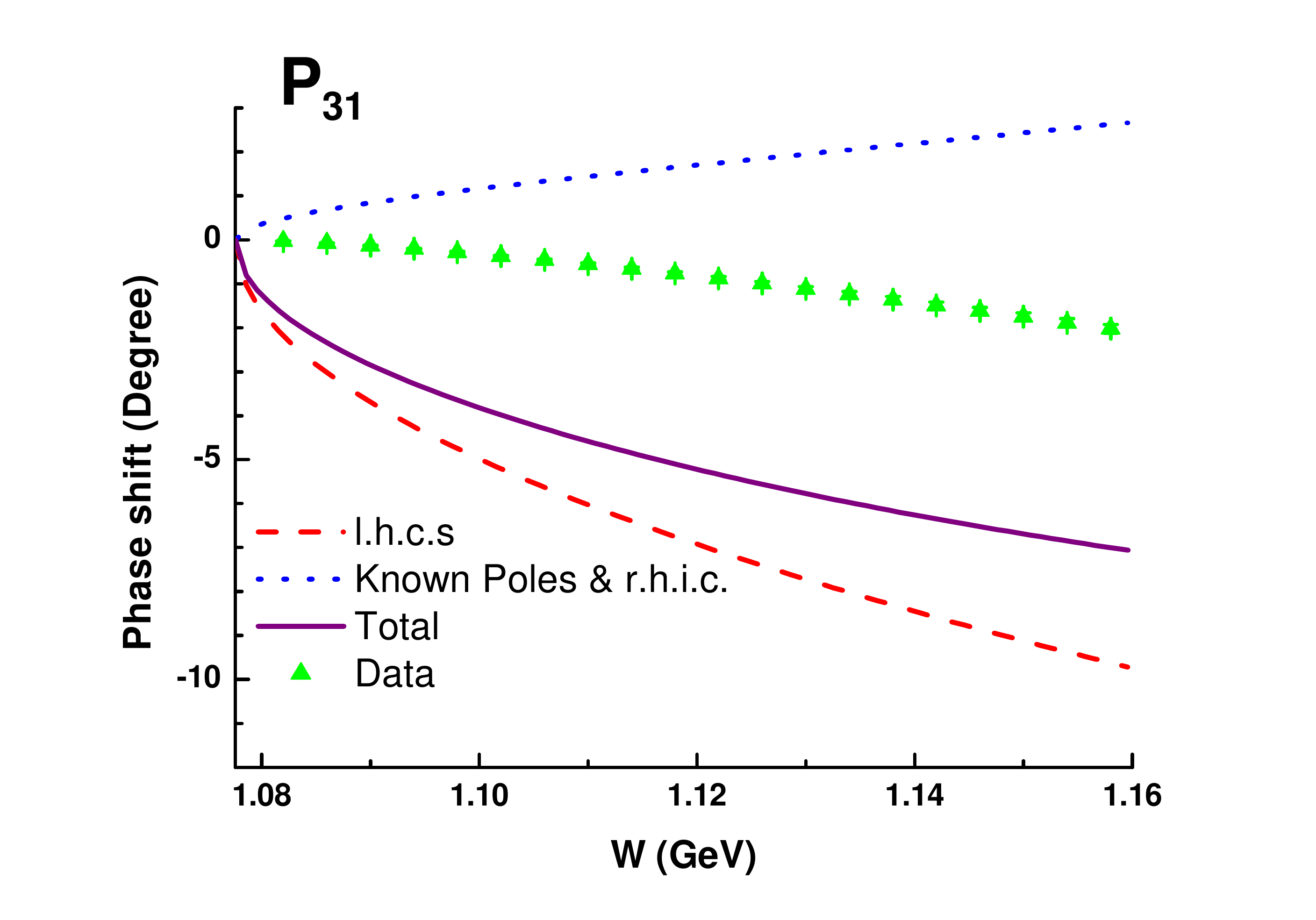}
\includegraphics[width=0.48\textwidth]{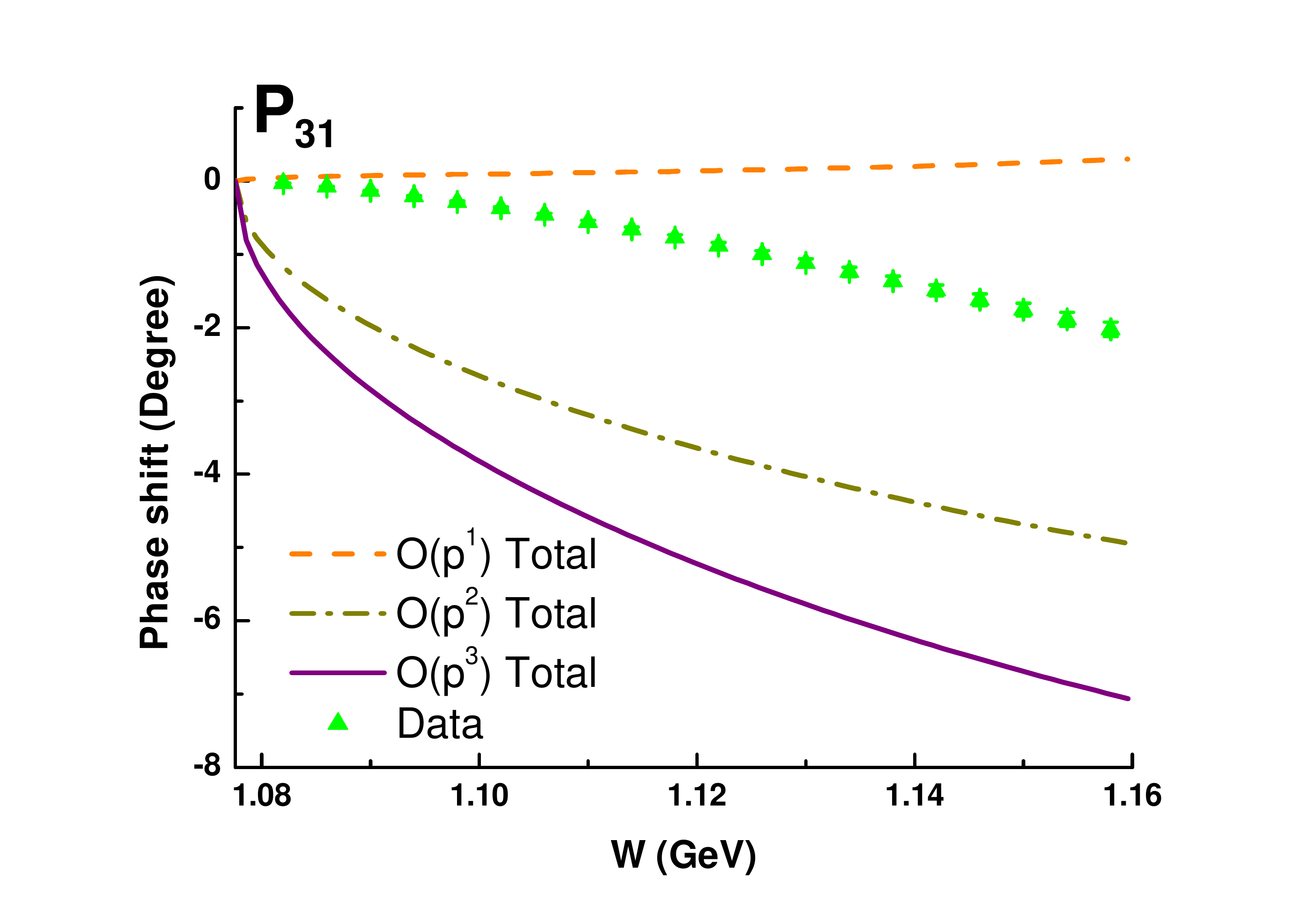}
\caption{PKU-representation analysis of the $P_{31}$ channel. Left: kown contributions vs the data. Right: known contributions with \textit{l.h.c.}s estimated in different chiral orders ($s_c=-0.08$ GeV$^2$). }
\label{fig:P31ana}
\end{figure}

Similar to the $S_{31}$ channel, there exist some minor discrepancy between the known contributions and the data in $P_{31}$ channel, see the left panel of Fig.~\ref{fig:P31ana}. The discrepancy remains at each chiral order, as can be seen from the right panel of Fig.~\ref{fig:P31ana}.

\begin{figure}[htbp]
\centering
\includegraphics[width=0.48\textwidth]{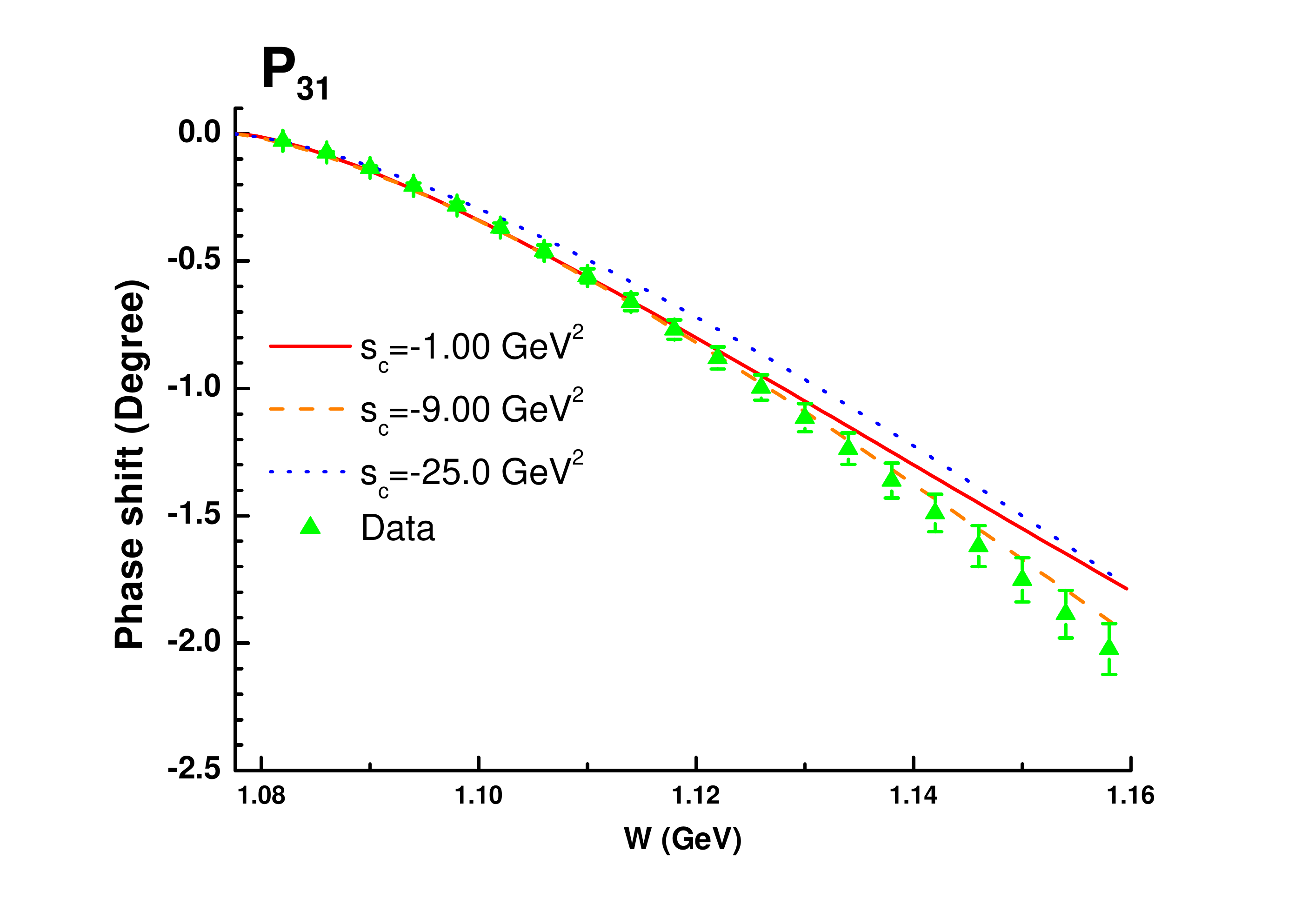}
\caption{The fit with extra poles as background in $P_{31}$ channel. }
\label{fig:P31p3sh}
\end{figure}

The $s_c$ parameter can be tuned to be $0.14$ GeV$^2$ to satisfy the $P$-wave threshold constraint, i.e. $\delta(k)=\mathcal{O}(k^3)$, but the tuning of $s_c$ does not work since it gives a disastrous fit except at threshold. This drives us to add extra poles again to simulate the missing corrections of the \textit{l.h.c.}s. The fit with extra poles are shown in Table.~\ref{tab:P31fit}, and plotted in Fig.~\ref{fig:P31p3sh}.

Only when $s_c$ is around $-9$ GeV$^2$ a good fit can be achieved. Furthermore, if one uses the \textit{l.h.c.}s from $\mathcal{O}(p^2)$ level and perform the same fit, $|s_c|$ should be rather large e.g. $|s_c|>16$ GeV$^2$, to get a result with the quality as good as $s_c=-1$ GeV$^2$ here.

\begin{table}[htbp]
\begin{center}
 \begin{tabular}  {| c | c | c |}
  \hline
  $s_{c}$ (GeV$^2$)  & Pole position (MeV) &  $\chi^2/\text{d.o.f}$ \\
  \hline
  $-1.00$ & $832,\ 799$ & $4.62$ \\
  \hline
  $-9.00$ &$784(1)-i\,164(2)$ & $1.80$ \\
  \hline
  $-25.0$ &$807(1)-i\,219(2)$ & $4.35$ \\
  \hline
 \end{tabular}\\
 \caption{The $P_{31}$ fit with two virtual poles or an extra resonance as background.  The fit in the framework of PKU representation can distinguish the type and the number of the poles.}\label{tab:P31fit}
\end{center}
\end{table}

\subsubsection{$P_{13}$ channel}

\begin{figure}[htbp]
\centering
\includegraphics[width=0.48\textwidth]{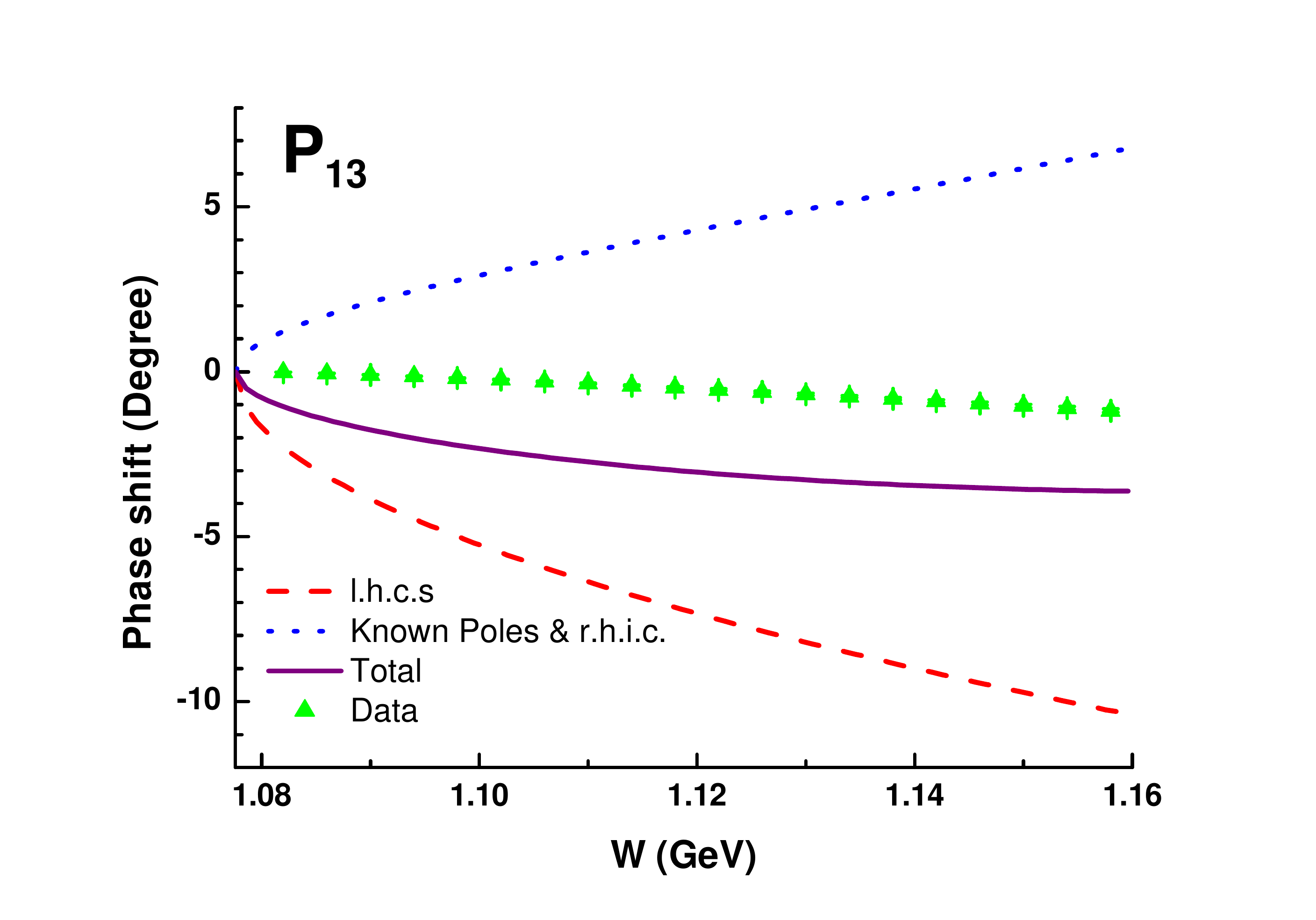}
\includegraphics[width=0.48\textwidth]{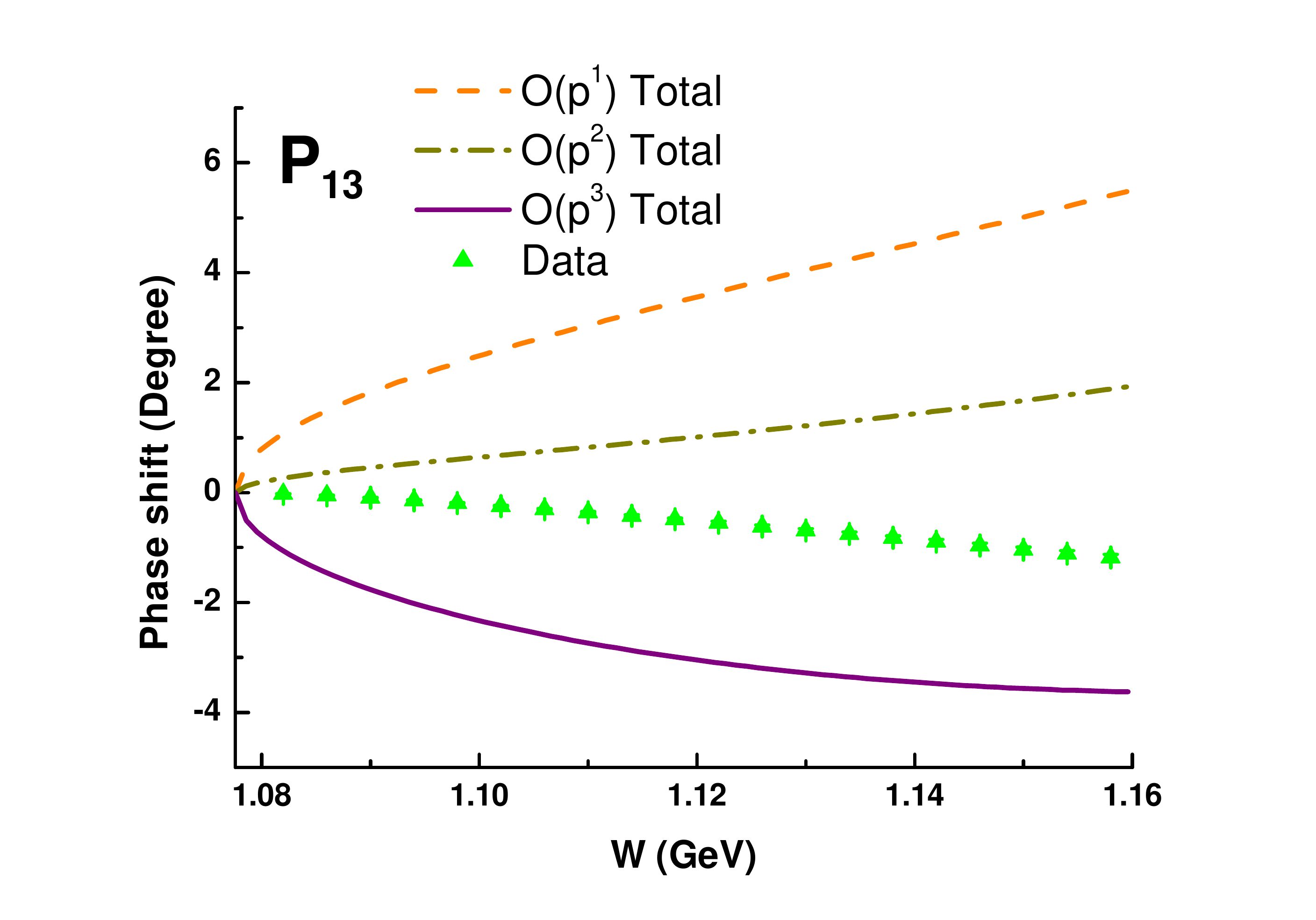}
\caption{PKU analysis of $P_{13}$ channel. Left: kown contributions vs the data. Right: known contributions with \textit{l.h.c.}s estimated in different chiral orders ($s_c=-0.08$ GeV$^2$). }
\label{fig:P13ana}
\end{figure}

The situation in $P_{13}$ channel is similar to that in $P_{31}$ channel: the left panel of Fig.~\ref{fig:P13ana} is the comparison between known terms and the data, and the right panel is the known contributions at different chiral orders.

\begin{figure}[htbp]
\centering
\includegraphics[width=0.48\textwidth]{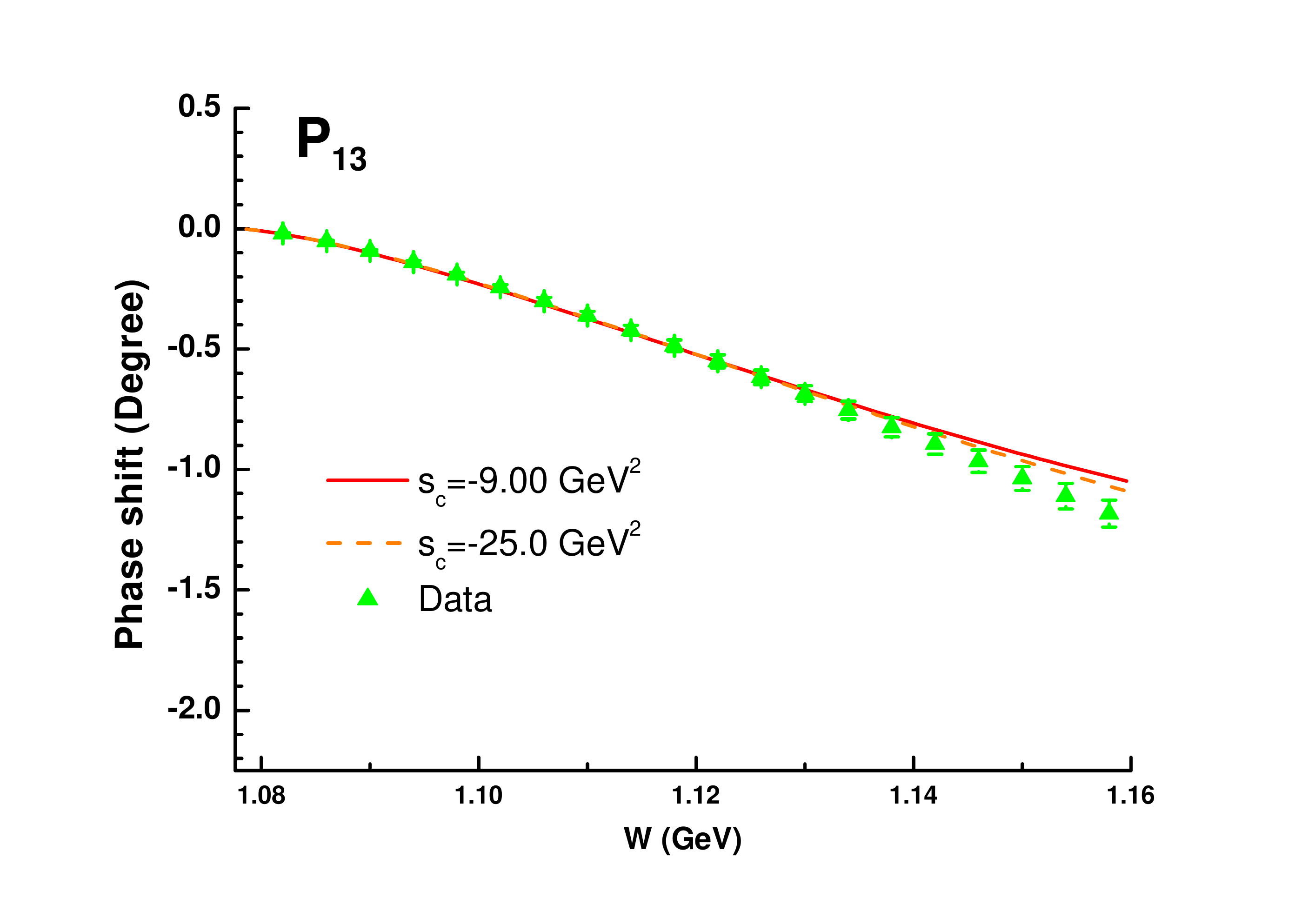}
\caption{The fit with one extra resonance as background in $P_{13}$ channel. }
\label{fig:P13p3sh}
\end{figure}

The fit results with an extra resonance as background are shown in Table.~\ref{tab:P13fit} and Fig.~\ref{fig:P13p3sh}, and the fit quality is worse than the $P_{31}$ channel. However, $\mathcal{O}(p^2)$ estimations can never give a result with $\chi^2/\text{d.o.f}<10$ for any $s_c$ value.

\begin{table}[htbp]
\begin{center}
 \begin{tabular}  {| c | c | c |}
  \hline
  $s_{c}$ (GeV$^2$)  & Pole position (MeV) & $\chi^2/\text{d.o.f}$ \\
  \hline
  $-9.00$ &$755(1)-i\,116(1)$ & $3.64$ \\
  \hline
  $-25.0$ &$756(1)-i\,167(2)$ & $2.48$ \\
  \hline
 \end{tabular}\\
 \caption{The $P_{13}$ fit with an extra resonance as background. }\label{tab:P13fit}
\end{center}
\end{table}

\subsubsection{$P_{33}$ channel}
\begin{figure}[htbp]
\centering
\includegraphics[width=0.48\textwidth]{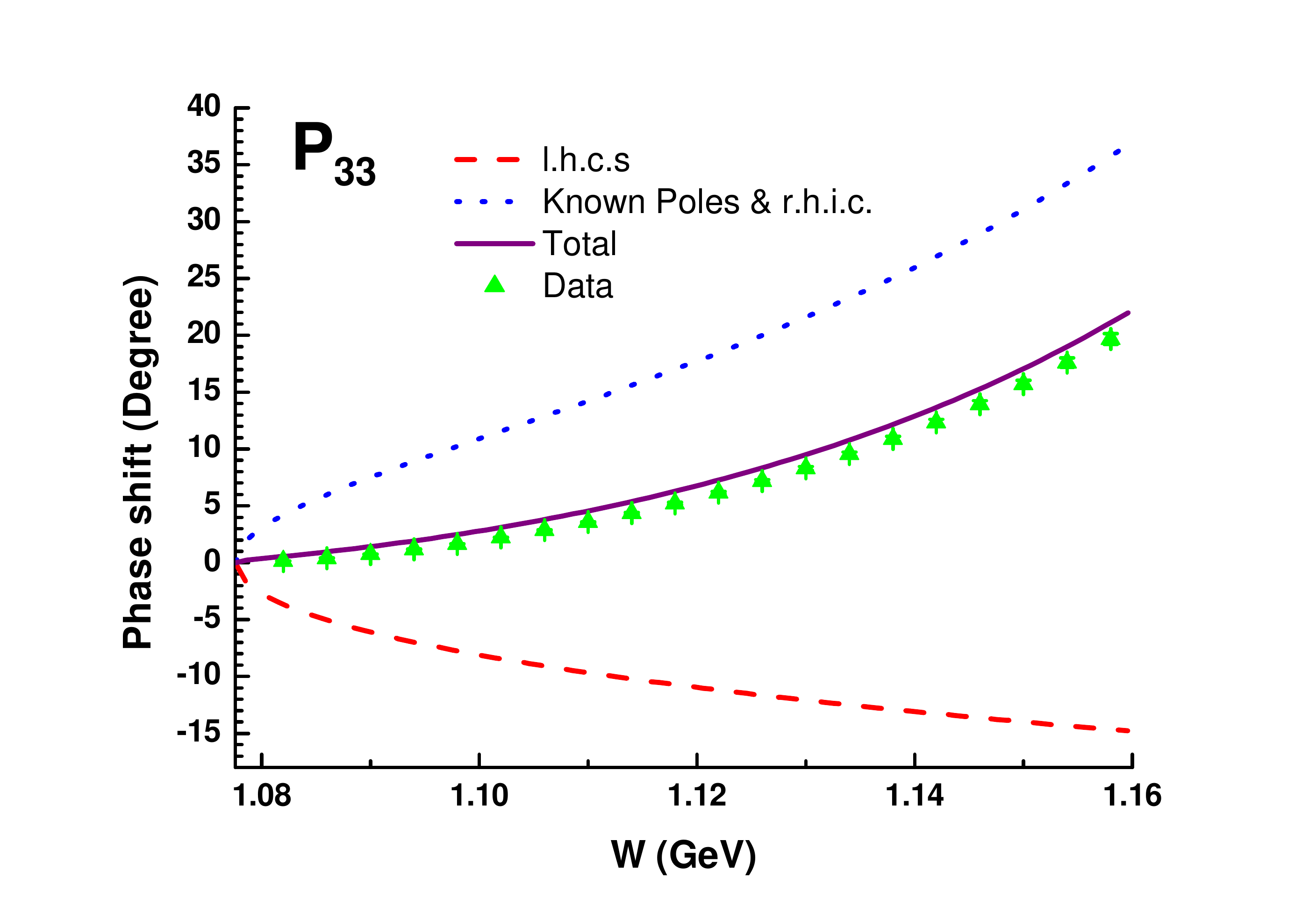}
\includegraphics[width=0.48\textwidth]{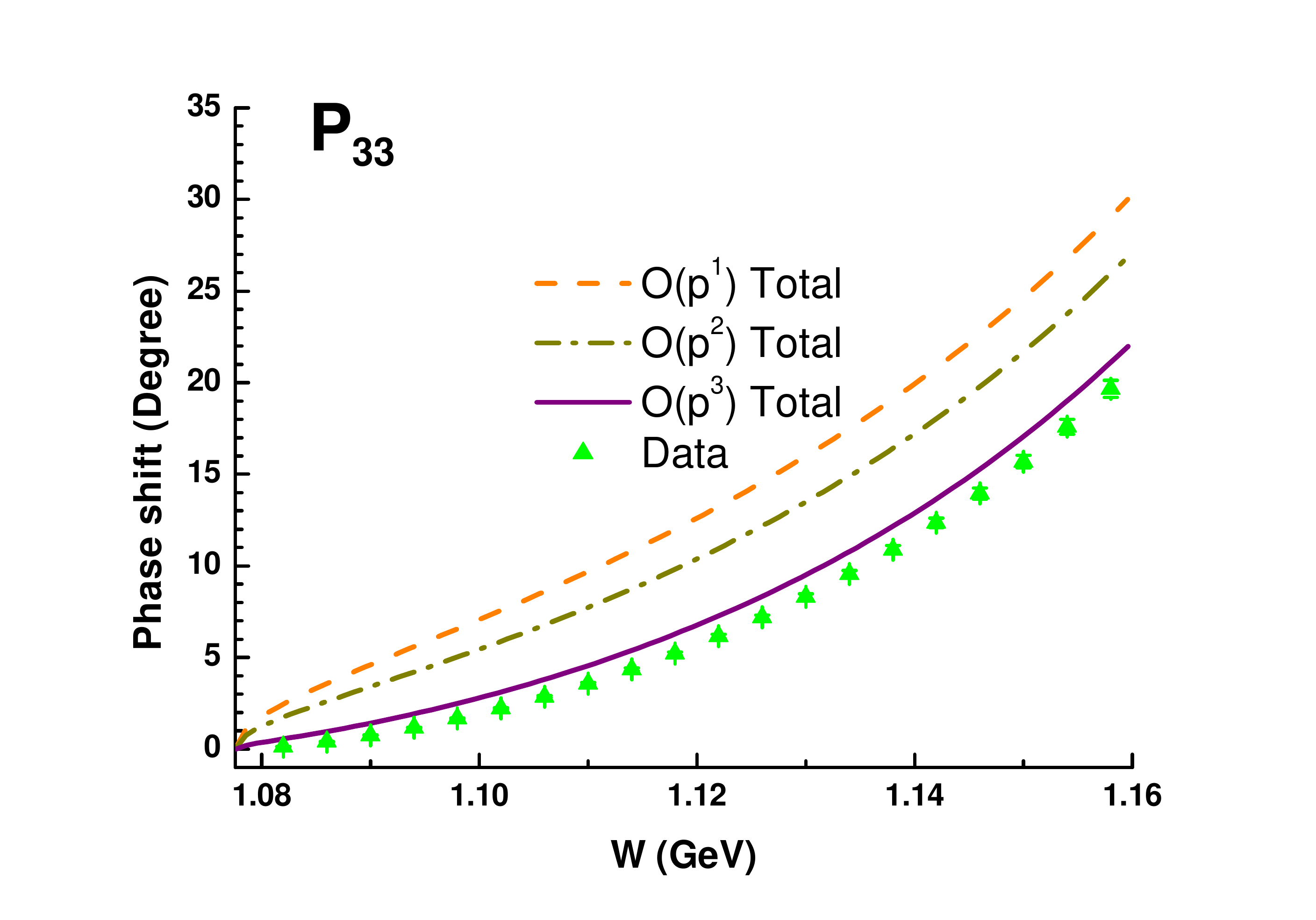}
\caption{PKU-representation analysis of $P_{33}$ channel. Left: kown contributions vs the data. Right: known contributions with \textit{l.h.c.}s estimated in different chiral orders ($s_c=-0.08$ GeV$^2$). }
\label{fig:P33ana}
\end{figure}

The discrepancy between known contributions and data in $P_{33}$ channel is the smallest among the six channels. The comparison between the known terms and the data is shown in the left panel of Fig.~\ref{fig:P33ana}. The higher the chiral order of the BChPT amplitudes used in the estimation of {\it l.h.c.}s is, the better the description of the phase shift becomes, see the right panel of Fig.~\ref{fig:P33ana}.

\begin{figure}[htbp]
\centering
\includegraphics[width=0.48\textwidth]{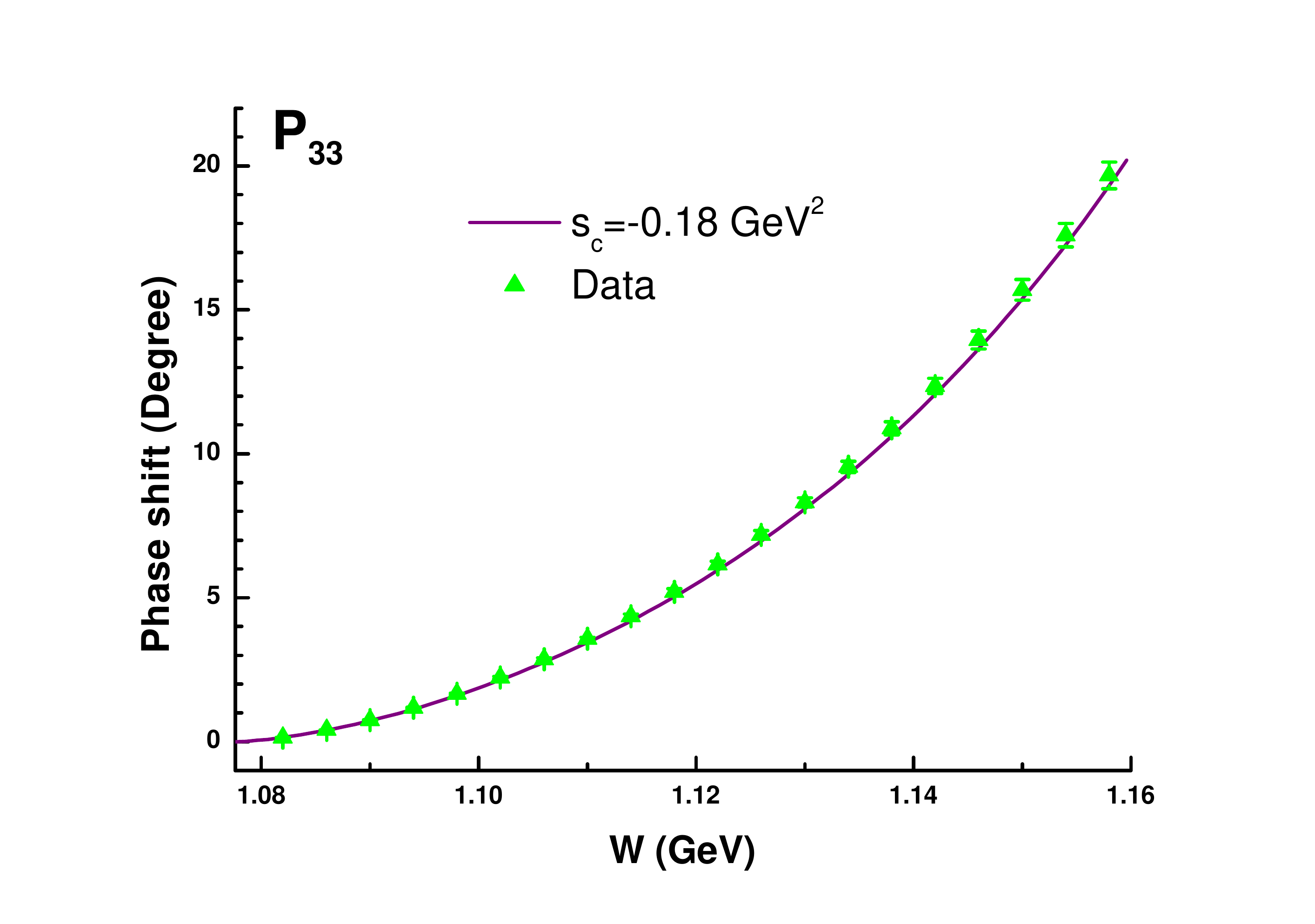}
\caption{The known contributions with $s_c=-0.18$ GeV$^2$ and the data in $P_{33}$ channel. }
\label{fig:P33p3sh}
\end{figure}

In this channel the $P$-wave threshold constraint can be satisfied by tuning $s_c$ to be $-0.18$ GeV$^2$. This is a parameter-free calculation. It is amazing that the data are automatically described with a quality $\chi^2/\text{d.o.f}=3.39$, see Fig.~\ref{fig:P33p3sh}. One can also use a different strategy by varying the pole position of $\Delta(1232)$: its location is only shifted slightly (from $1.211-0.050i$ GeV to $1.208-0.047i$ GeV), and the $\chi^2/d.o.f$ is reduced to $1.78$.

\subsection{Further remarks on the $S_{31}$, $P_{31}$ and $P_{13}$ channels}
From the above discussions, it is found that, to obtain good description of the data in $S_{31}$, $P_{31}$ and $P_{13}$ channels, one still needs some extra terms, which are for the moment simulated by single pole terms. Nevertheless, one should be cautious to regard them as physically existing poles. The following criterion might be helpful to distinguish a real pole from a fake one.
\begin{itemize}
    \item {\it Large discrepancies between the known contributions and the data}. As already mentioned in previous $\mathcal{O}(p^2)$ analyses, the discrepancies in $S_{11}$ and $P_{11}$ channels are crucial since even if the left-hand cut terms are switched off, the discrepancies are still significant, which forces us to add new poles. On the contrary, the discrepancies in other channels are minor, because at least one can fine-tune the cut-off parameter to make the scattering lengths from the known terms conform with the data. To be conservative, we prefer the missing contributions in those channels to be corrections of the background term {\it a priori}.
    \item {\it Stability of the pole positions}. It is obvious that the $S_{11}$ hidden pole stays stable and fits the data well for all the cut-off values we chose. On the contrary , the poles in $S_{31}$, $P_{31}$ and $P_{13}$ channels only fit the data when the cut-off parameters are large, and some of them are not so stable (e.g. in $P_{31}$ channel the poles change from virtual states to resonances).
    \item {\it Alternative physical mechanism in support of the poles}. As pointed out in previous $\mathcal{O}(p^2)$ analyses, the $P_{11}$ pole can be regarded as the shadow pole of the nucleon, and the $S_{11}$ pole may be of potential dynamical nature. However we fail, at least at the moment, to find a possible physical interpretation of the poles in $S_{31}$, $P_{31}$ and $P_{13}$ channels. Actually we have tried to use square-well potential as a toy model to fit the $S_{31}$ data but it failed: the potential would become repulsive, leaving no nearby poles, meanwhile the line-shape cannot be well fitted.
\end{itemize}

\begin{figure}[htbp]
\centering
\includegraphics[width=0.6\textwidth]{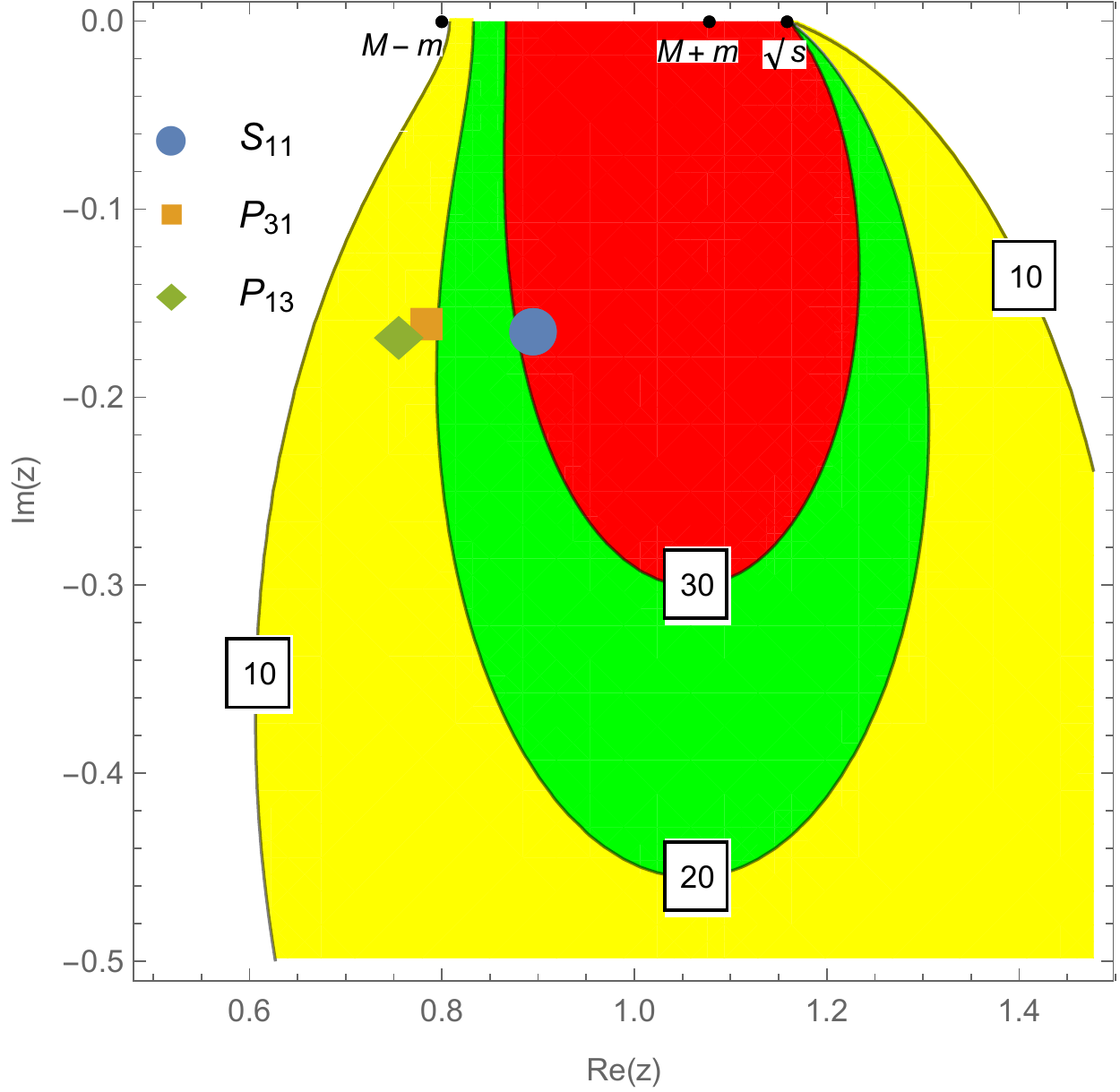}
\caption{The relationship between the location of a resonance pole $z$, and its contribution to the phase shift at $\sqrt{s}=1.16$ GeV. The regions of $\delta<10^\circ$, $10^\circ<\delta<20^\circ$, $20^\circ<\delta<30^\circ$ and $\delta>30^\circ$ are marked in white, yellow, green and red, in order. The $P_{31}$ and $P_{13}$ extra poles are taken from the best fits, and $S_{11}$ pole is the averaged value. }
\label{fig:resdel}
\end{figure}

The poles in $S_{31}$, $P_{31}$ and $P_{13}$ channels do not meet all the three criterion raised above, hence we can not claim those extra poles are truly existing in reality. It is also worth noticing that the extra contributions in $S_{31}$, $P_{31}$ and $P_{13}$ channels are indeed much smaller than the $S_{11}$ and $P_{11}$ hidden poles. For $S_{11}$, $P_{31}$ and $P_{13}$ channels, the comparison of the phase shifts contributing from the extra resonances is shown in Fig.~\ref{fig:resdel}. For $P_{11}$ and $S_{31}$ channels where virtual states are added,  the extra $P_{11}$ state contributes to the phase shift about two times as large as the $S_{31}$ one.
\subsection{Threshold parameters\label{subsec:thrpara}}
All the $S$- and $P$-wave phase shifts have been analyzed by means of the PKU representation with proper inputs taking from $\mathcal{O}(p^3)$ BChPT. For some channels, especially $P_{31}$ and $P_{13}$ waves, distant resonances have to be incorporated artificially as background to simulate any other underlying contributions. Nonetheless, our resultant predictions are in good agreement with the data in the low-energy region close to threshold, which therefore can reliably be applied to the extraction of threshold parameters, namely the coefficients of effective range expansion (ERE). As usual, the ERE is defined by
\begin{equation}\label{ERA}
k^{2L+1}\cot\delta=-\frac{1}{a_L}+\frac{r_L}{2}k^2+\mathcal{O}(k^4)\ \mbox{, }
\end{equation}
where $L$ is the angular momentum, $k$ is modulus of the 3-momentum in the center of mass frame, and $\delta$ is the phase shift. The coefficient $a_L$ in Eq.~\ref{ERA} is usually called scattering length (or volume for $P$- waves), while $r_L$ is termed as effective range. For $L=0$ they are both in dimension of length, characterizing the effective size of the target and the range of the interacting potential as well known in potential scattering theory, respectively. For a better extraction of threshold parameters, the fit results with $\chi^2/d.o.f.>5$ are excluded.

For the scattering lengths (volumes) $a_L$ and the parameters $r_L$, Table.~\ref{tab:ERAal} and Table.~\ref{tab:ERArl} compiles our numerical results (the $a_L$ parameters are in comparison with the ones from Roy-Steiner analyses of Ref.~\cite{Hoferichter:2015hva}), respectively. Except the $P_{33}$ channel, the ERE coefficients are presented as ranges, responsible for the systematical uncertainties originating from the various fits under consideration.
\begin{table}[htbp]
\begin{center}
 \begin{tabular}  {| c | c | c ||}
  \hline
  Channel  & $a_L$ from PKU analyses & $a_L$ from Roy-Steiner analyses \\
  \hline
  $S_{11}$ & $(-169.0\sim -165.4)$ & $-169.9(19.4)$\\
  \hline
  $S_{31}$ & $(79.7\sim 86.0)$ & $86.3(10.4)$\\
  \hline
  $P_{11}$ & $(51.0\sim 67.3)$ & $70.7(4.3)$\\
  \hline
  $P_{31}$ & $(45.1\sim 57.0)$ & $41.0(3.1)$\\
  \hline
  $P_{13}$ & $(30.1\sim 31.1)$ & $29.4(3.9)$\\
  \hline
  $P_{33}$ & $-198.8$ & $-211.5(2.8)$\\
  \hline
 \end{tabular}\\
 \caption{$S$- and $P$-wave scattering lengths (in the unit of $10^{-3}m_{\pi}^{-1}$) and volumes (in the unit of $10^{-3}m_{\pi}^{-3}$). }\label{tab:ERAal}
\end{center}
\end{table}

\begin{table}[htbp]
\begin{center}
 \begin{tabular}  {| c | c ||}
  \hline
  Channel  & $r_L$ from PKU analyses \\
  \hline
  $S_{11}$ & $(1.0\sim 2.5)$\\
  \hline
  $S_{31}$ & $(15.2\sim 22.3)$\\
  \hline
  $P_{11}$ & $(39.0\sim 206.6)$\\
  \hline
  $P_{31}$ & $(-227.0\sim -40.3)$\\
  \hline
  $P_{13}$ & $(-31.2\sim -29.9)$\\
  \hline
  $P_{33}$ & $4.8$\\
  \hline
 \end{tabular}\\
 \caption{$S$- and $P$-wave effective ranges (in the unit of $m_{\pi}^{-1}$ for $S$- waves and $m_{\pi}$ for $P$- waves). }\label{tab:ERArl}
\end{center}
\end{table}

Furthermore, the non-relativistic limit of PKU representation (i.e., Ning Hu representation~\cite{Hu:1948zz}) is
\begin{equation}\label{Hure}
S(k)=\prod_{n}\frac{p_n+k}{p_n-k}\times e^{-2iRk}\ \mbox{, }
\end{equation}
where $k$ is the 3-momentum in the center of mass frame, $p_n$ denotes the pole position on the complex $k$ plane, and the exponential term is the background term with $R$ being a constant parameter. According to Ref.~\cite{Regge:1958ft}, $R$ may be positive definite and approximately equal to the potential range. Here one can also investigate the $R$ parameters using the results from PKU representation. The values of $R$ parameters in each $S$- and $P$- wave channels (calculated under the best fit of the cut-off parameter $s_c$) are exhibited in Table.~\ref{tab:HuR}.
\begin{table}[htbp]
\begin{center}
 \begin{tabular}  {| c | c | c | c | c | c | c ||}
  \hline
  Channel  & $S_{11}$ & $S_{31}$ & $P_{11}$ & $P_{31}$ & $P_{13}$ & $P_{33}$\\
  \hline
  $R$ parameter & $0.56$ & $0.47$ & $0.52$ & $0.38$ & $0.38$ & $0.30$\\
  \hline
 \end{tabular}\\
 \caption{$S$- and $P$-wave $R$ parameters from Eq.~\eqref{Hure} (in the unit of $m_{\pi}^{-1}$). }\label{tab:HuR}
\end{center}
\end{table}
Note that in $\pi N$ scatterings the potential can be regarded as the effect of two-pion exchanges in $t$ channel, thus $R$ should be roughly $1/(2m_\pi)$, hence the results in Table.~\ref{tab:HuR} turn out to be reasonable. Similarly, application of the simplified version of PKU representation to nucleon-nucleon scatterings can be found in Ref.~\cite{Wang:2018gul}.

\section{Conclusions and outlook}\label{con}
In summary, this paper, as a follow-up of Ref.~\cite{Wang:2017agd}, applies PKU representation to the analyses of elastic $\pi N$ scattering processes in $S$- and $P$- waves, with the left-hand cut contributions calculated from the results obtained in covariant baryon chiral perturbation theory up to $\mathcal{O}(p^3)$ level. It is found generally that the fit quality has been greatly improved compared with the previous $\mathcal{O}(p^2)$ results. The existence of the hidden states in $S_{11}$ and $P_{11}$ channels are firmly established irrespective of the chiral order as well as other numerical details.

The other four channels are also investigated quantitatively: a good description of the data in $P_{33}$ wave can be achieved with a cut-off parameter determined by the threshold constraint, while in the other three waves some finer structures are needed that are temporarily simulated by extra poles. As byproducts, the parameters of effective range expansion for all the channels are calculated.

Although the analyses of low-energy pion-nucleon scatterings based on PKU representation with left-hand cuts simply from perturbative calculation are completed, the present work could still be improved in future. The major weakness of the present work, though not vital in drawing the current conclusions as we believe, comes from the uncertainty when calculating distant left-hand cuts. Hence through our work, we only cautiously draw physical conclusions extracted from discrepancies at qualitative level, and avoid any results drawn only depending on minor discrepancies between data and the cuts. It is worth stressing that our major conclusions are actually established based on the negative definiteness of the background contribution, which is supported by quantum mechanical scattering theory. The $S_{11}$ and $P_{11}$ poles turn out to be very stable and hence trustworthy. There have been successful examples for this approach: the $f_0(500)$ pole and the $K^*(800)$ pole positions in Refs.~\cite{Zhou:2006wm,Zhou:2004ms} are found in very good agreements with the results from Roy-like equation analyses(see Refs.~\cite{Caprini:2005zr,DescotesGenon:2006uk}). One may improve the calculation by invoking resonance plus Reggeon exchange model to estimate the distant left-hand cuts. Constrains like crossing symmetry may also be incorporated. Finally, the physical meaning of the newly found hidden poles are still open for discussions.

\section*{Acknowledgments}
We would like to thank one referee for the constructive suggestions and conments which are helpful in formulating the present version of this paper. One of the authors (YFW) acknowledges Helmholtz-Institut f\"{u}r Strahlen- und Kernphysik of Bonn University, for warm hospitality where part of this work is being done; and thanks Bastian Kubis and Ulf-G. Mei{\ss}ner for helpful discussions. This work is supported in part by National Nature Science Foundations of China (NSFC) under Contract Nos. 10925522, 11021092; and by the Spanish Ministerio de Econom\'ia y Competitividad (MINECO) and the European Regional Development Fund (ERDF), under contracts FIS2017-84038-C2-1-P, FIS2017-84038-C2-2-P, SEV-2014-0398.

\appendix{\center\bf\huge Appendices}
\section{Relativistic BChPT amplitudes up to $\mathcal{O}(p^3)$\label{sec:BChPTcalc}}
\subsection{Tree-level $A$ and $B$ functions}\label{app:fftree}

\begin{figure}[htbp]
\centering
\includegraphics[width=0.4\textwidth]{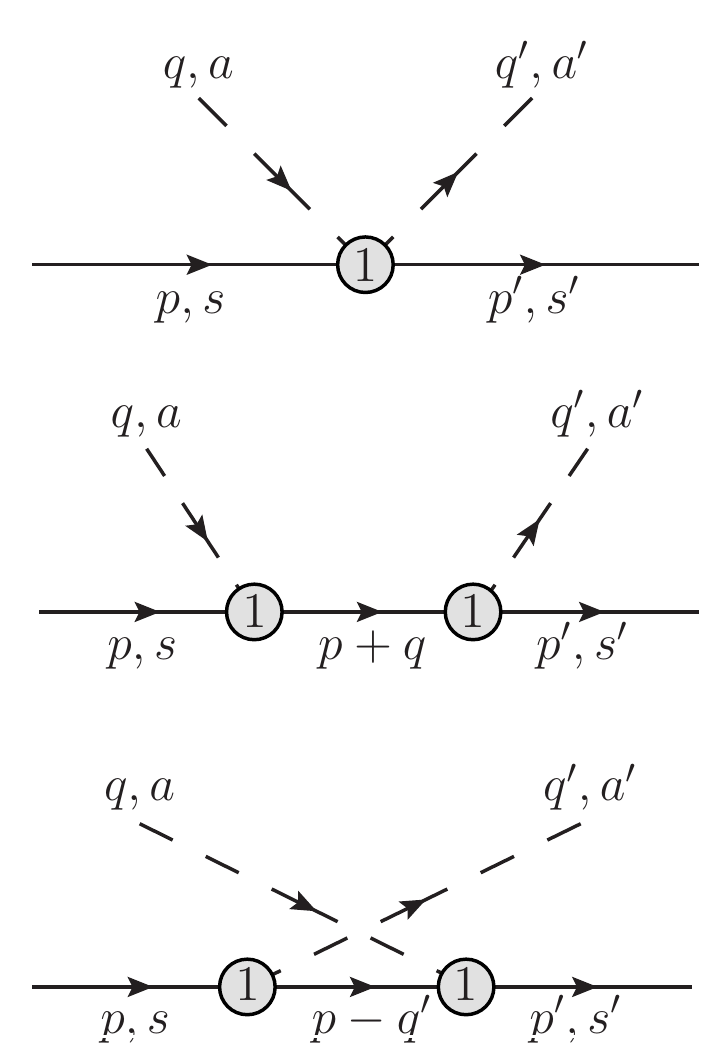}
\caption{The tree diagrams at $\mathcal{O}(p^1)$ level. }
\label{fig:treediap1}
\end{figure}
\begin{figure}[htbp]
\centering
\includegraphics[width=0.4\textwidth]{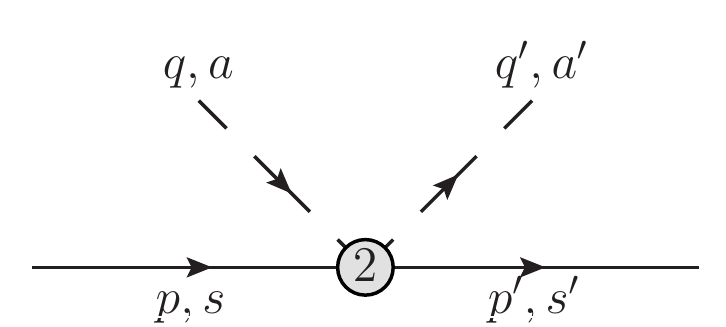}
\caption{The tree diagrams at $\mathcal{O}(p^2)$ level. }
\label{fig:treediap2}
\end{figure}
\begin{figure}[htbp]
\centering
\includegraphics[width=0.8\textwidth]{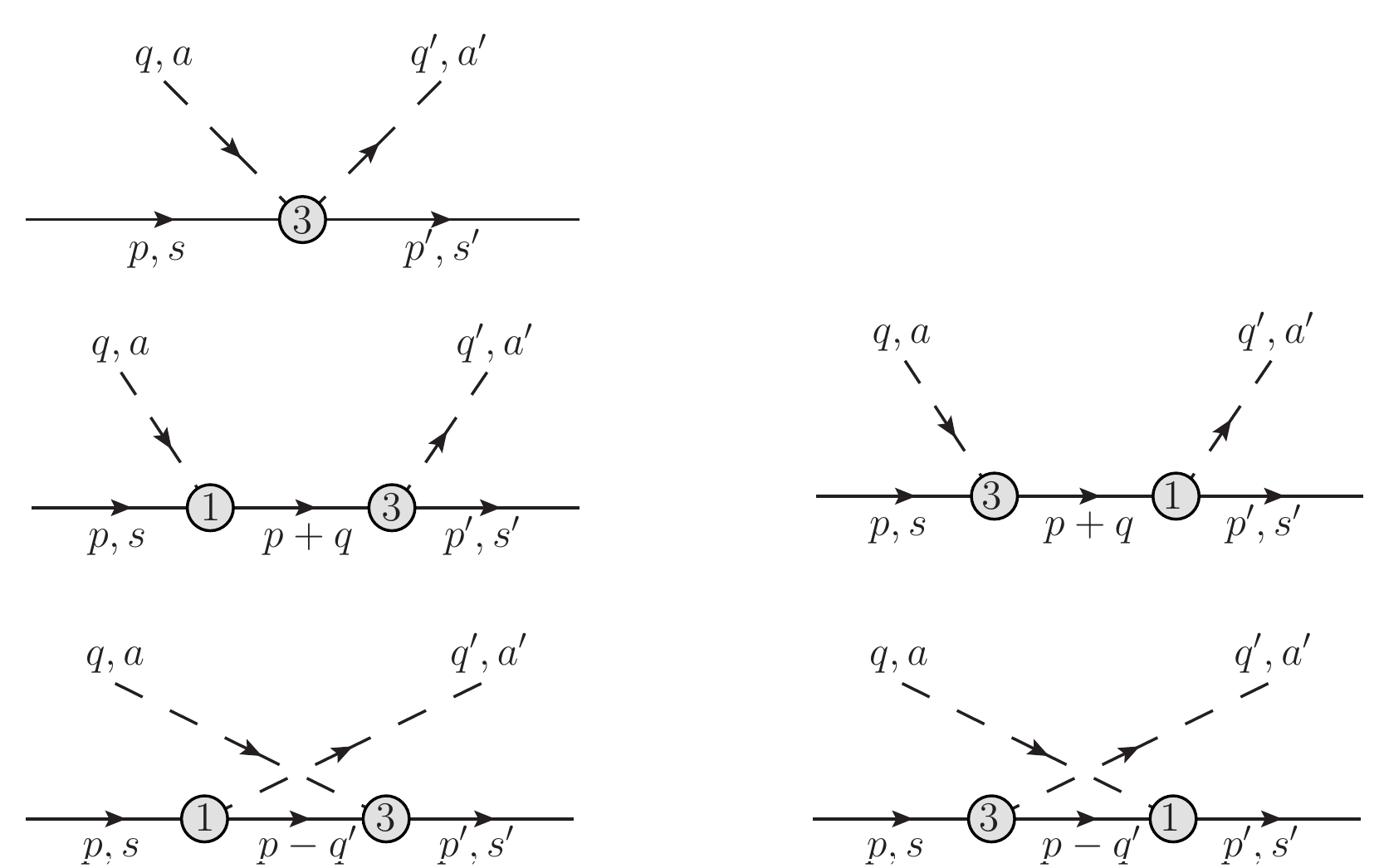}
\caption{The tree diagrams at $\mathcal{O}(p^3)$ level. }
\label{fig:treediap3}
\end{figure}

The tree diagrams of $\mathcal{O}(p^i)$ ($i=1,2,3$) are shown in Figs.~\ref{fig:treediap1}, \ref{fig:treediap2} and ~\ref{fig:treediap3}, in order, where the pions are represented by dashed lines and the nucleons by solid lines. The circled numbers $N$ stand for vertices of $\mathcal{O}(p^N)$. The $A$ and $B$ amplitudes corresponding to those diagrams are given as follows.

\begin{itemize}[leftmargin=0.5cm]
\item{At $\mathcal{O}(p^1)$:}
\begin{equation}\label{p1ABres}
\begin{split}
&A_1^{1/2}=\frac{g^2M}{F^2}\ \mbox{, }\\
&B_1^{1/2}=\frac{1-g^2}{2F^2}-\frac{3M^2g^2}{F^2(s-M^2)}-\frac{M^2g^2}{F^2}\frac{1}{u-M^2}\ \mbox{, }\\
&A_1^{3/2}=\frac{g^2M}{F^2}\ \mbox{, }\\
&B_1^{3/2}=-\frac{1-g^2}{2F^2}+\frac{2M^2g^2}{F^2(u-M^2)}\ \mbox{; }
\end{split}
\end{equation}
\item{At $\mathcal{O}(p^2)$:}
\begin{equation}
\begin{split}
&A_2^{1/2}=-\frac{4c_1 m^2}{F^2}+\frac{c_2(s-u)^2}{8M^2F^2}+\frac{c_3}{F^2}(2m^2-t)-\frac{c_4(s-u)}{F^2}\ \mbox{, }\\
&B_2^{1/2}=\frac{4Mc_4}{F^2}\ \mbox{, }\\
&A_2^{3/2}=-\frac{4c_1 m^2}{F^2}+\frac{c_2(s-u)^2}{8M^2F^2}+\frac{c_3}{F^2}(2m^2-t)+\frac{c_4(s-u)}{2F^2}\ \mbox{, }\\
&B_2^{3/2}=-\frac{2Mc_4}{F^2}\ \mbox{; }
\end{split}
\end{equation}
\item{At $\mathcal{O}(p^3)$ (Born diagram):}
\begin{equation}\label{p3Bterm}
\begin{split}
&A_{3B}^{1/2}=-\frac{Mg}{F^2}\times 4m^2(d_{18}-2d_{16})\ \mbox{, }\\
&B_{3B}^{1/2}=\frac{4m^2g(d_{18}-2d_{16})}{F^2}\times\frac{su+M^2(2u-3M^2)}{(s-M^2)(u-M^2)}\ \mbox{, }\\
&A_{3B}^{3/2}=-\frac{Mg}{F^2}\times 4m^2(d_{18}-2d_{16})\ \mbox{, }\\
&B_{3B}^{3/2}=\frac{2m^2g(d_{18}-2d_{16})}{F^2}\times\frac{u+3M^2}{u-M^2}\ \mbox{; }
\end{split}
\end{equation}
\item{At $\mathcal{O}(p^3)$ (contact diagram):}
\begin{equation}
\begin{split}
&A_{3C}^{1/2}=-\frac{(d_{14}-d_{15})(s-u)^2}{4MF^2}+\frac{(d_1+d_2)}{MF^2}(s-u)(2m^2-t)+\frac{d_3}{8M^3F^2}(s-u)^3+\frac{4m^2 d_5}{MF^2}(s-u)\ \mbox{, }\\
&B_{3C}^{1/2}=\frac{(d_{14}-d_{15})(s-u)}{F^2}\ \mbox{, }\\
&A_{3C}^{3/2}=-\frac{(d_{14}-d_{15})(s-u)^2}{4MF^2}-\frac{(d_1+d_2)}{2MF^2}(s-u)(2m^2-t)-\frac{d_3}{16M^3F^2}(s-u)^3-\frac{2m^2 d_5}{MF^2}(s-u)\ \mbox{, }\\
&B_{3C}^{3/2}=\frac{(d_{14}-d_{15})(s-u)}{F^2}\ \mbox{; }
\end{split}
\end{equation}
\end{itemize}

Note that Eq.~\eqref{p3Bterm} can be absorbed in Eq.~\eqref{p1ABres} by redefinition of the $g$ coupling. From the $A,\ B$ functions above one can readily get the analytical expressions of the amplitudes in each channel after partial wave projection, as done in Appendix.~B of Ref.~\cite{Wang:2017agd}.
\subsection{One-loop $A$ and $B$ functions}\label{app:ffloop}
\subsubsection{Diagrams}
\begin{figure}[htbp]
\centering
\includegraphics[width=0.8\textwidth]{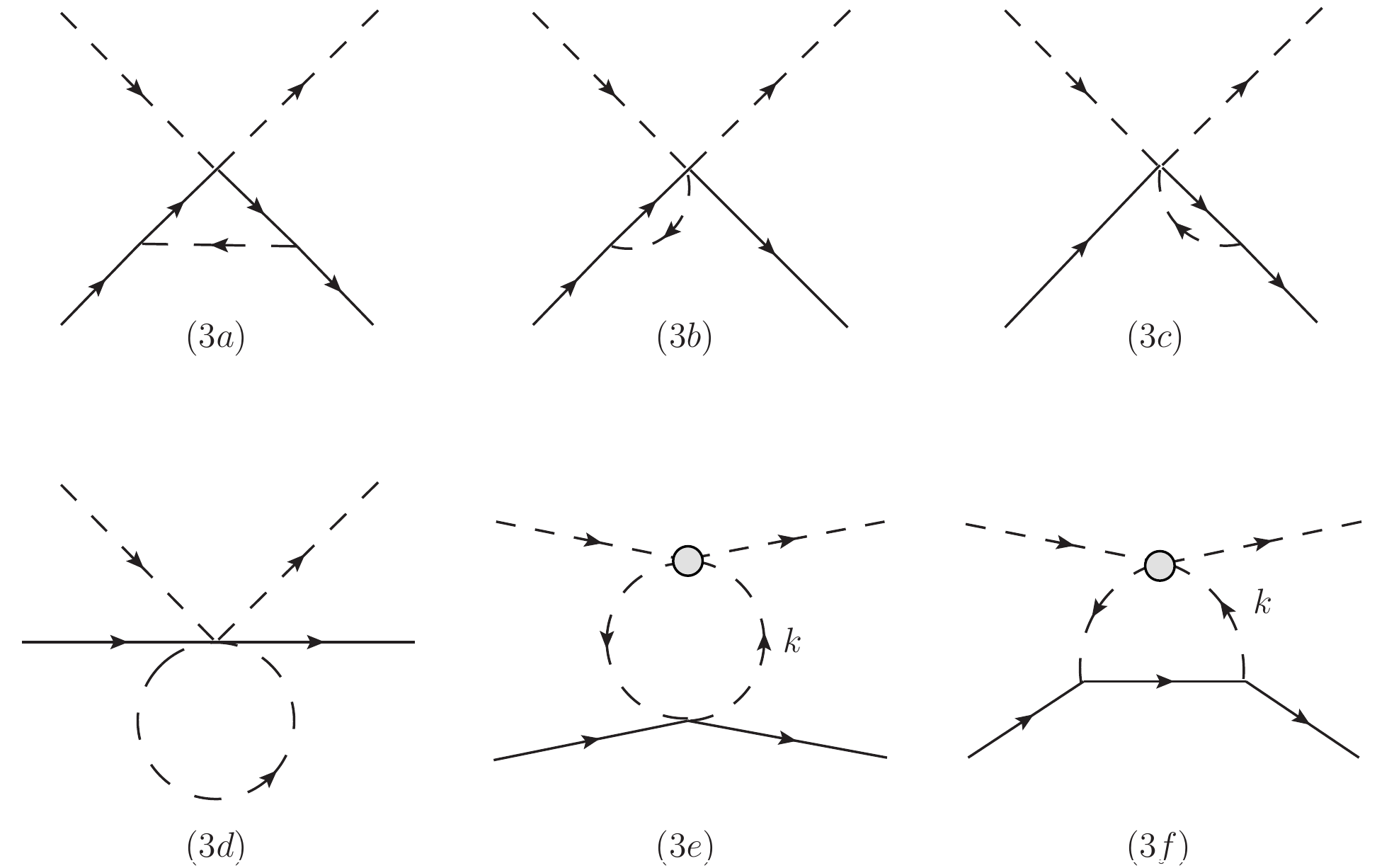}
\caption{One-loop diagrams at $\mathcal{O}(p^3)$ level of contact type (first group $G_1$). }
\label{fig:p3lC}
\end{figure}

\begin{figure}[htbp]
\centering
\includegraphics[width=0.8\textwidth]{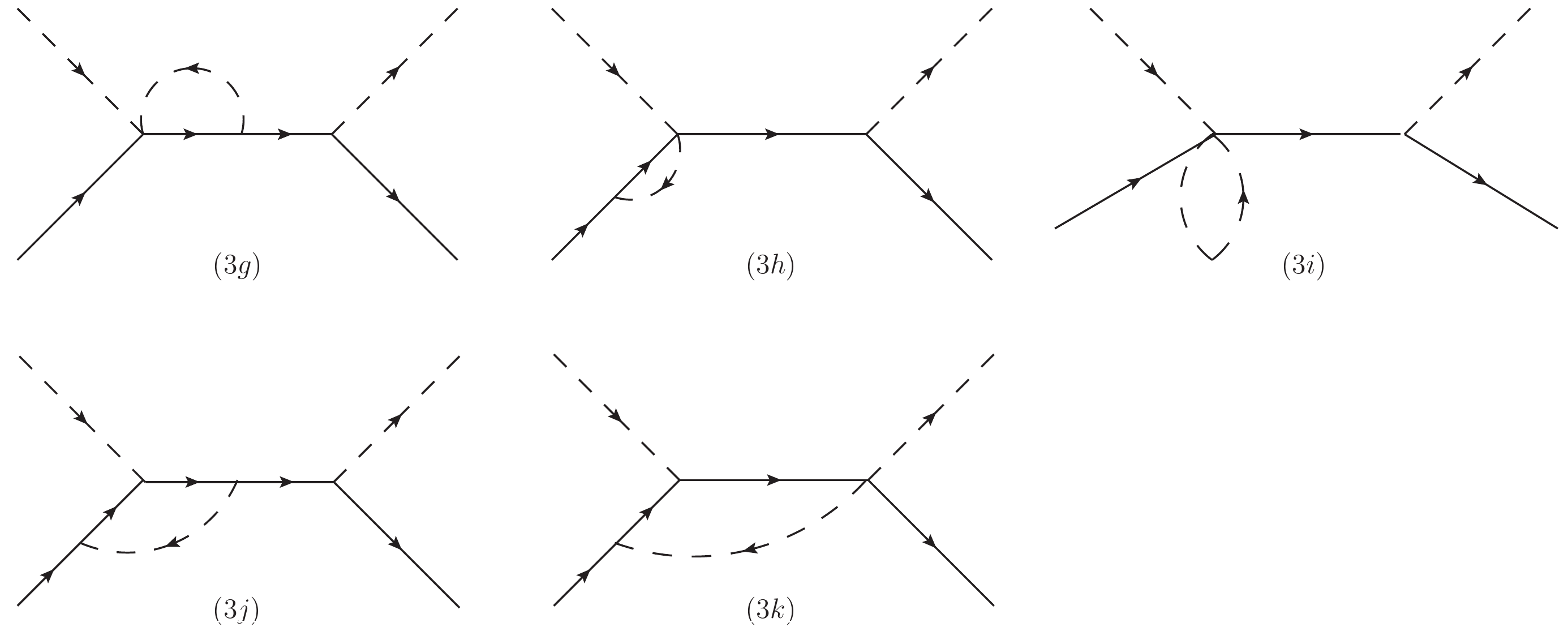}
\caption{One-loop diagrams at $\mathcal{O}(p^3)$ level of Born type I (second group $G_2$). }
\label{fig:p3lB1}
\end{figure}

\begin{figure}[htbp]
\centering
\includegraphics[width=0.6\textwidth]{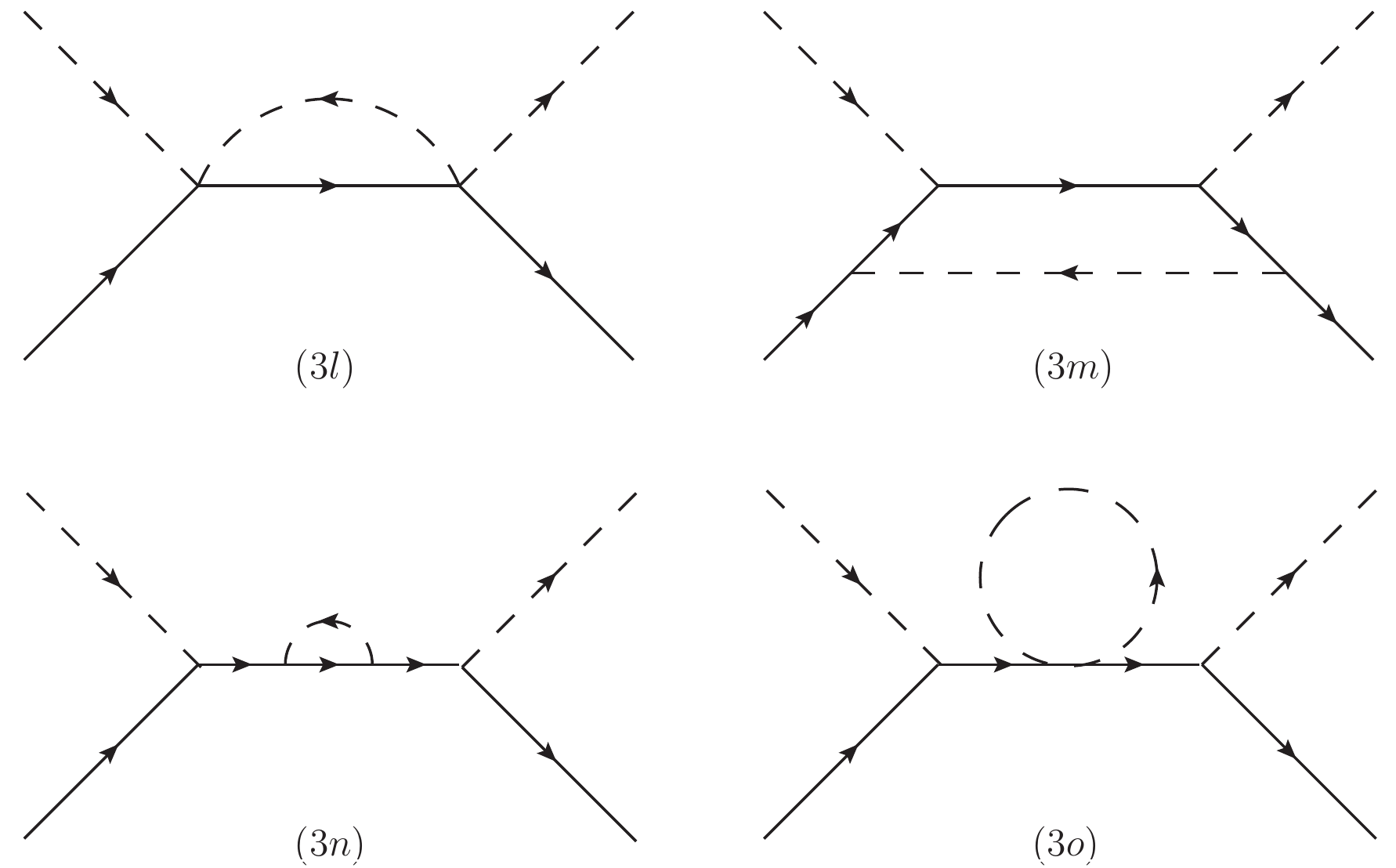}
\caption{One-loop diagrams at $\mathcal{O}(p^3)$ level of Born type II (third group $G_3$). }
\label{fig:p3lB2}
\end{figure}

According to the power counting rule of BChPT, for a loop diagram with $L$ independent loops, $I_{\pi}$ internal pion lines, $I_N$ internal nucleon lines, and $V_n$ vertices of $\mathcal{O}(p^n)$, its chiral order should be $\mathcal{O}(p^{d_{\chi}})$, where
\begin{equation}\label{power}
d_{\chi}=4L-2I_{\pi}-I_N+\sum_n nV_n\ \mbox{. }
\end{equation}
Consequently, the diagrams, which are classified into three groups and displayed in Figs.~\ref{fig:p3lC}, \ref{fig:p3lB1} and \ref{fig:p3lB2}, are under our consideration.

To be specific, the first group ($G_1$) contains the diagrams of contact type, which do not have companions related by crossing symmetry; the second ($G_2$) and third groups ($G_3$) are Born diagrams, which have corresponding $u$-channel crossing diagrams but are not shown explicitly; moreover, the time-reversal counterparts of the $G_2$ diagrams are not displayed either. The sum of those diagrams is
\begin{equation}\label{ABtotal}
\begin{split}
&A^+_{\mathcal{O}(p^3)\ \text{loop}}=\sum_{d_1\in G_1}A^+_{d_1}(s,t,u)+2\sum_{d_2\in G_2}[A^+_{d_2}(s,t,u)+A^+_{d_2}(u,t,s)]+\sum_{d_3\in G_3}[A^+_{d_3}(s,t,u)+A^+_{d_3}(u,t,s)]\ \mbox{, }\\
&B^+_{\mathcal{O}(p^3)\ \text{loop}}=\sum_{d_1\in G_1}B^+_{d_1}(s,t,u)+2\sum_{d_2\in G_2}[B^+_{d_2}(s,t,u)-B^+_{d_2}(u,t,s)]+\sum_{d_3\in G_3}[B^+_{d_3}(s,t,u)-B^+_{d_3}(u,t,s)]\ \mbox{, }\\
&A^-_{\mathcal{O}(p^3)\ \text{loop}}=\sum_{d_1\in G_1}A^-_{d_1}(s,t,u)+2\sum_{d_2\in G_2}[A^-_{d_2}(s,t,u)-A^-_{d_2}(u,t,s)]+\sum_{d_3\in G_3}[A^-_{d_3}(s,t,u)-A^-_{d_3}(u,t,s)]\ \mbox{, }\\
&B^-_{\mathcal{O}(p^3)\ \text{loop}}=\sum_{d_1\in G_1}B^-_{d_1}(s,t,u)+2\sum_{d_2\in G_2}[B^-_{d_2}(s,t,u)+B^-_{d_2}(u,t,s)]+\sum_{d_3\in G_3}[B^-_{d_3}(s,t,u)+B^-_{d_3}(u,t,s)]\ \mbox{. }
\end{split}
\end{equation}
\subsubsection{Definition of the loop functions}
In this paper we use Passarino-Veltman notations~\cite{Passarino:1978jh} to handle the loop functions. Definitions of the loop functions in dimension $D\to4$ are as follows ($\nu$ is the energy scale and $\epsilon=2-D/2$).
\begin{itemize}[leftmargin=0.5cm]
\item{One point function}\\
\begin{equation}
A_0(m^2)\equiv -16\pi^2 \text{i}\int\frac{d^Dk\nu^{2\epsilon}}{(2\pi)^D}\frac{1}{k^2-m^2}\ \mbox{. }
\end{equation}

\item{Two point functions}\\
\begin{equation}
\begin{split}
&B_0(p^2;m_1^2,m_2^2)\equiv -16\pi^2 \text{i}\int\frac{d^Dk\nu^{2\epsilon}}{(2\pi)^D}\frac{1}{(k^2-m_1^2)\big[(p+k)^2-m_2^2\big]}\ \mbox{, }\\
&B^\mu(p;m_1^2,m_2^2)\equiv -16\pi^2 \text{i}\int\frac{d^Dk\nu^{2\epsilon}}{(2\pi)^D}\frac{k^\mu}{(k^2-m_1^2)\big[(p+k)^2-m_2^2\big]}\\
&\equiv p^\mu B_1(p^2;m_1^2,m_2^2)\ \mbox{, }\\
&B^{\mu\nu}(p;m_1^2,m_2^2)\equiv -16\pi^2 \text{i}\int\frac{d^Dk\nu^{2\epsilon}}{(2\pi)^D}\frac{k^\mu k^\nu}{(k^2-m_1^2)\big[(p+k)^2-m_2^2\big]}\\
&\equiv g^{\mu\nu} B_{00}(p^2;m_1^2,m_2^2)+p^\mu p^\mu B_{11}(p^2;m_1^2,m_2^2)\ \mbox{, }
\end{split}
\end{equation}

\item{Three point functions}\\
\begin{equation}
\begin{split}
&C_0(p_1^2,p_2^2,p_3^2;m_1^2,m_2^2,m_3^2)\equiv -16\pi^2 \text{i}\int\frac{d^Dk\nu^{2\epsilon}}{(2\pi)^D}\frac{1}{(k^2-m_1^2)\big[(p_1+k)^2-m_2^2\big]\big[(k-p_3)^2-m_3^2\big]}\ \mbox{, }\\
&C^\mu(p_1,p_2;m_1^2,m_2^2,m_3^2)\equiv -16\pi^2 \text{i}\int\frac{d^Dk\nu^{2\epsilon}}{(2\pi)^D}\frac{k^\mu}{(k^2-m_1^2)\big[(p_1+k)^2-m_2^2\big]\big[(k-p_3)^2-m_3^2\big]}\\
&\equiv p_1^\mu C_1(p_1^2,p_2^2,p_3^2;m_1^2,m_2^2,m_3^2)-p_3^\mu C_2(p_1^2,p_2^2,p_3^2;m_1^2,m_2^2,m_3^2)\ \mbox{, }\\
&C^{\mu\nu}(p_1,p_2;m_1^2,m_2^2,m_3^2)\equiv -16\pi^2 \text{i}\int\frac{d^Dk\nu^{2\epsilon}}{(2\pi)^D}
\frac{k^\mu k^\nu}{(k^2-m_1^2)\big[(p_1+k)^2-m_2^2\big]\big[(k-p_3)^2-m_3^2\big]}\\
&\equiv g^{\mu\nu}C_{00}(p_1^2,p_2^2,p_3^2;m_1^2,m_2^2,m_3^2)+p_1^\mu p_1^\nu C_{11}(p_1^2,p_2^2,p_3^2;m_1^2,m_2^2,m_3^2)\\
&-(p_1^\mu p_3^\nu+p_3^\mu p_1^\nu)C_{12}(p_1^2,p_2^2,p_3^2;m_1^2,m_2^2,m_3^2)+p_3^\mu p_3^\nu C_{22}(p_1^2,p_2^2,p_3^2;m_1^2,m_2^2,m_3^2)\ \mbox{, }
\end{split}
\end{equation}
where $p_3=-(p_1+p_2)$.

\item{Four point functions}\\
\begin{equation}
\begin{split}
&D_0(p_1^2,p_2^2,p_3^2,p_4^2,p_{12}^2,p_{23}^2;m_1^2,m_2^2,m_3^2,m_4^2)\\
&\equiv -16\pi^2 \text{i}\int\frac{d^Dk\nu^{2\epsilon}}{(2\pi)^D}
\frac{1}{(k^2-m_1^2)\big[(k+p_1)^2-m_2^2\big]\big[(k+p_{12})^2-m_3^2\big]\big[(k-p_4)^2-m_4^2\big]}\ \mbox{, }\\
&D^\mu(p_1,p_2,p_3;m_1^2,m_2^2,m_3^2,m_4^2)\\
&\equiv -16\pi^2 \text{i}\int\frac{d^Dk\nu^{2\epsilon}}{(2\pi)^D}
\frac{k^\mu}{(k^2-m_1^2)\big[(k+p_1)^2-m_2^2\big]\big[(k+p_{12})^2-m_3^2\big]\big[(k-p_4)^2-m_4^2\big]}\\
&\equiv p_1^\mu D_1(p_1^2,p_2^2,p_3^2,p_4^2,p_{12}^2,p_{23}^2;m_1^2,m_2^2,m_3^2,m_4^2)+p_{12}^\mu D_2(p_1^2,p_2^2,p_3^2,p_4^2,p_{12}^2,p_{23}^2;m_1^2,m_2^2,m_3^2,m_4^2)\\
&-p_4^\mu D_3(p_1^2,p_2^2,p_3^2,p_4^2,p_{12}^2,p_{23}^2;m_1^2,m_2^2,m_3^2,m_4^2)\ \mbox{, }
\end{split}
\end{equation}
where $p_{12}=p_1+p_2$, $p_{23}=p_2+p_3$, $p_4=-(p_1+p_2+p_3)$.
\end{itemize}
\subsubsection{Results of the one-loop amplitudes}
\begin{table}[htbp]
\begin{center}
\begin{tabular}{| c | c |}
\hline
Full notations & Abbreviations\\
\hline
$A_0(m^2)$& $a_0$\\
\hline
$A_0(M^2)$& $A_0$\\
\hline
$B_{..}(M^2;m^2,M^2)$& $B_{..}^M$\\
\hline
$B_{..}(s\backslash u;m^2,M^2)$& $B_{..}^{s\backslash u}$\\
\hline
$B_{..}(m^2;M^2,M^2)$& $B_{..}^m$\\
\hline
$B_{..}(t;M^2,M^2)$& $B_{..}^T$\\
\hline
$B_{..}(t;m^2,m^2)$& $B_{..}^t$\\
\hline
$C_{..}(M^2,t,M^2;m^2,M^2,M^2)$& $C_{..}^t$\\
\hline
$C_{..}(M^2,M^2,t;m^2,M^2,m^2)$& $C_{..}^{t2m}$\\
\hline
$C_{..}(M^2,m^2,s\backslash u;m^2,M^2,M^2)$& $C_{..}^{s\backslash u}$\\
\hline
$C_{..}(t,m^2,m^2;M^2,M^2,M^2)$& $C_{..}^T$\\
\hline
$D_{..}(M^2,t,m^2,s,M^2,m^2;m^2,M^2,M^2,M^2)$& $D_{..}$\\
\hline
$D_{..}(M^2,t,m^2,u,M^2,m^2;m^2,M^2,M^2,M^2)$& $D_{..}^u$\\
\hline
\end{tabular}\\
\caption{Abbreviations of the loop functions. }\label{tab:loopfuncabbr}
\end{center}
\end{table}

For brevity, the abbreviations of the loop functions in Table.~\ref{tab:loopfuncabbr} will be used from now on. All the $A,\ B$ functions corresponding to the diagrams in Figs.~\ref{fig:p3lC}, \ref{fig:p3lB1} and \ref{fig:p3lB2} are shown diagram by diagram in the following.
\begin{itemize}[leftmargin=0.5cm]
\item{\textbf{(3a)}}\\
\begin{equation}
\begin{split}
&A_{3a}^+=0\ \mbox{, }\\
&B_{3a}^+=0\ \mbox{, }\\
&A_{3a}^-=\frac{g^2M^3}{32F^4\pi^2}(s-u)\Big(C_{11}^t+C_{12}^t\Big)\ \mbox{, }\\
&B_{3a}^-=\frac{g^2}{128F^4\pi^2}\Big[a_0-4M^2B_1^M+4M^2(2C_{00}^t-B_0^T-m^2C_0^t)\Big]\ \mbox{. }
\end{split}
\end{equation}
\item{\textbf{(3b)}}\\
\begin{equation}
\begin{split}
&A_{3b}^+=\frac{Mg^2}{48F^4\pi^2}(A_0+m^2B_0^M)\ \mbox{, }\\
&B_{3b}^+=0\ \mbox{, }\\
&A_{3b}^-=0\ \mbox{, }\\
&B_{3b}^-=-\frac{g^2}{64F^4\pi^2}(A_0+m^2B_0^M)\ \mbox{. }
\end{split}
\end{equation}
\item{\textbf{(3c)}}\\
\begin{equation}
\begin{split}
&A_{3c}^+=\frac{Mg^2}{48F^4\pi^2}(A_0+m^2B_0^M)\ \mbox{, }\\
&B_{3c}^+=0\ \mbox{, }\\
&A_{3c}^-=0\ \mbox{, }\\
&B_{3c}^-=-\frac{g^2}{64F^4\pi^2}(A_0+m^2B_0^M)\ \mbox{. }
\end{split}
\end{equation}
\item{\textbf{(3d)}}\\
\begin{equation}
\begin{split}
&A_{3d}^+=0\ \mbox{, }\\
&B_{3d}^+=0\ \mbox{, }\\
&A_{3d}^-=0\ \mbox{, }\\
&B_{3d}^-=\frac{5}{24F^4}\times\frac{1}{16\pi^2}a_0\ \mbox{. }
\end{split}
\end{equation}
\item{\textbf{(3e)}}\\
\begin{equation}
\begin{split}
&A_{3e}^+=0\ \mbox{, }\\
&B_{3e}^+=0\ \mbox{, }\\
&A_{3e}^-=0\ \mbox{, }\\
&B_{3e}^-=-\frac{1}{16F^4\pi^2}B_{00}^t\ \mbox{. }
\end{split}
\end{equation}
\item{\textbf{(3f)}}\\
\begin{equation}
\begin{split}
&A_{3f}^+=-\frac{Mg^2}{96F^4\pi^2}\big[4(A_0+m^2B_0^M)+3(m^2-2t)(B_0^t-2M^2C_1^{t2m})\big]\ \mbox{, }\\
&B_{3f}^+=0\ \mbox{, }\\
&A_{3f}^-=\frac{g^2M^3}{16F^4\pi^2}(s-u)C_{11}^{t2m}\ \mbox{, }\\
&B_{3f}^-=\frac{g^2}{16F^4\pi^2}(B_{00}^t+4M^2C_{00}^{t2m})\ \mbox{. }
\end{split}
\end{equation}
\item{\textbf{(3g)}}\\
\begin{equation}
\begin{split}
&A_{3g}^\pm=-\frac{Mg^2}{64F^4\pi^2}\big[(s-M^2)B_1^s+2(A_0+m^2B_0^s)\big]\ \mbox{, }\\
&B_{3g}^\pm=\frac{g^2}{64F^4\pi^2}\big[(s+M^2)B_1^s+\frac{s+3M^2}{s-M^2}(A_0+m^2B_0^s)\big]\ \mbox{. }
\end{split}
\end{equation}
\item{\textbf{(3h)}}\\
\begin{equation}
\begin{split}
&A_{3h}^\pm=-\frac{Mg^2}{32F^4\pi^2}(A_0+m^2B_0^M)\ \mbox{, }\\
&B_{3h}^\pm=\frac{g^2}{128F^4\pi^2}\frac{s+7M^2}{s-M^2}(A_0+m^2B_0^M)\ \mbox{. }
\end{split}
\end{equation}
\item{\textbf{(3i)}}\\
\begin{equation}
\begin{split}
&A_{3i}^\pm=\frac{Mg^2}{96F^4\pi^2}a_0\ \mbox{, }\\
&B_{3i}^\pm=-\frac{g^2}{192F^4\pi^2}\frac{s+3M^2}{s-M^2}a_0\ \mbox{. }
\end{split}
\end{equation}
\item{\textbf{(3j)}}\\
\begin{equation}
\begin{split}
&A_{3j}^\pm=\frac{Mg^4}{256F^4\pi^2(s-M^2)}\Big\{2s a_0-2M^2A_0+2(sm^2-M^4+sM^2)B_0^s\\
&-2m^2(s+M^2)B_0^M+4M^2(s-M^2)B_0^m-(s+M^2)^2B_1^s\\
&+2m^2(s-M^2)(s+3M^2)C_0^s\\
&-2(s+M^2)\big[m^4+(s-M^2)(2M^2-m^2)\big]C_1^s\\
&+2\big[m^4(s+M^2)+(s-M^2)(sm^2+3M^2m^2-4M^4)\big]C_2^s\Big\}\ \mbox{, }\\
&B_{3j}^\pm=-\frac{g^4}{256F^4\pi^2(s-M^2)}\Big\{\frac{1}{2}\Big[(s+5M^2)a_0+(s+M^2)A_0+\big[m^2(s+M^2)-(s+5M^2)(s-M^2)\big]B_0^s\Big]\\
&+8M^2(s+M^2)B_0^m+M^2(s+M^2)B_1^s+8m^2M^2(s+M^2)C_0^s\\
&+4M^2(s-M^2)(2s+2M^2-m^2)C_2^s\Big\}\ \mbox{. }
\end{split}
\end{equation}
\item{\textbf{(3k)}}\\
\begin{equation}
\begin{split}
&A_{3k}^+=-\frac{Mg^2}{64F^4\pi^2}(s-M^2)\big[B_1^s+2(s-M^2-m^2)C_2^s+2(B_0^m+m^2C_0^s)\big]\ \mbox{, }\\
&B_{3k}^+=-\frac{g^2}{128F^4\pi^2}\big[-2a_0+(A_0+m^2B_0^M)-8M^2(B_0^m+m^2C_0^s)+2(s-M^2)(B_0^s+B_1^s-4M^2C_2^s)\big]\ \mbox{, }\\
&A_{3k}^-=0\ \mbox{, }\\
&B_{3k}^-=0\ \mbox{. }
\end{split}
\end{equation}
\item{\textbf{(3l)}}\\
\begin{equation}
\begin{split}
&A_{3l}^+=-\frac{M}{32F^4\pi^2}(s-M^2)B_1^s\ \mbox{, }\\
&B_{3l}^+=\frac{1}{128F^4\pi^2}\Big\{a_0-4\big[A_0+m^2B_0^s+(s-M^2)B_1^s\big]\Big\}\ \mbox{, }\\
&A_{3l}^-=-\frac{M}{64F^4\pi^2}(s-M^2)B_1^s\ \mbox{, }\\
&B_{3l}^-=\frac{1}{256F^4\pi^2}\Big\{a_0-4\big[A_0+m^2B_0^s+(s-M^2)B_1^s\big]\Big\}\ \mbox{. }
\end{split}
\end{equation}
\item{\textbf{(3m)}}\\
\begin{equation}
\begin{split}
&A_{3m}^+=\frac{3Mg^4}{256\pi^2F^4}\big\{4(s-M^2)B_0^m+(3s+M^2)B_1^s-4M^2B_1^M+8M^2tC_2^t\\
&+4m^2(s-M^2)C_0^s+8M^2(s-M^2-m^2)C_1^s+4(s+M^2)(s-M^2-m^2)C_2^s+8M^2(s-M^2)C_0^T\\
&+4M^2\big[(s-M^2-m^2)C_{11}^t+(s-u)C_{12}^t+(M^2+m^2-u)C_{22}^t\big]\\
&+8m^2M^2(s-M^2)D_0-8M^2(s-M^2)tD_2+8M^2(s-M^2)(s-M^2-m^2)D_3\big\}\ \mbox{, }\\
&B_{3m}^+=\frac{3g^4}{256\pi^2F^4}\big\{2M^2+(s-M^2)B_0^s-4M^2(B_0^m+B_1^M)+(s+M^2)B_1^s\\
&-8M^4C_1^t+4M^2(t-2M^2)C_2^t-4m^2M^2C_0^s+4M^2(s-M^2-m^2)(C_1^s+C_2^s)\\
&-4M^2\big[2C_{00}^t+M^2C_{11}^t+(2M^2-t)C_{12}^t+M^2C_{22}^t\big]\\
&+4M^2\big[(s+t-m^2-5M^2)C_0^T+tC_1^T+(t-2m^2)C_2^T\big]\\
&-16m^2M^4D_0+4M^2\big[m^4+(s-M^2)(2M^2-m^2)\big]D_1\\
&+4M^2\big[m^4-m^2(s+t-M^2)+(s-M^2)(2M^2-t)\big]D_2\\
&+4M^2\big[-m^4+(s-M^2)(s-3M^2-2m^2)\big]D_3\big\}\ \mbox{, }\\
&A_{3m}^-=-A_{3m}^+/3\ \mbox{, }\\
&B_{3m}^-=-B_{3m}^+/3\ \mbox{. }
\end{split}
\end{equation}
\item{\textbf{(3n)}}\\
\begin{equation}\label{3nnaive}
\begin{split}
&A_{3n}^\pm=\frac{3Mg^4}{256F^4\pi^2}\big[(3s+M^2)B_1^s+\frac{4(s+M^2)}{s-M^2}(A_0+m^2B_0^s)\big]\ \mbox{, }\\
&B_{3n}^\pm=-\frac{3g^4}{256F^4\pi^2}\frac{1}{(s-M^2)^2}\big[(s-M^2)(s^2+6M^2s+M^4)B_1^s+(s^2+10sM^2+5M^4)(A_0+m^2B_0^s)\big]\ \mbox{. }
\end{split}
\end{equation}
\item{\textbf{(3o)}}\\
\begin{equation}
A_{3o}^\pm=B_{3o}^\pm=0\ \mbox{. }
\end{equation}

\end{itemize}

\subsection{Renormalization}\label{app:renorm}
\subsubsection{UV cancellation}\label{app:renormdiv}
The UV divergent pieces in loop functions are dealt with in  dimensional regularization and expressed in terms of $R_\epsilon$ as defined in Eq.~\eqref{Repdef}, see Appendix.~\ref{app:exploop}. The $\overline{\text{MS}}-1$ subtraction scheme is employed in this paper, i.e. the counter terms should completely cancel the divergent pieces proportional to $R_\epsilon$. However, the diagram (3n) in Fig.~\ref{fig:p3lB2} needs to be treated specifically, since $\overline{\text{MS}}-1$ scheme itself can not erase the second order pole at $s=M^2$ in Eq.~\eqref{3nnaive}, which is a disaster of analyticity properties especially when investigating the $u$-channel nucleon cut. To remedy such a problem, one should regard the loop in the internal nucleon propagator in diagram (3n) as the renormalized self-energy $\Sigma^r(\slashed{p})$, which satisfies the on-shell renormalization conditions
\begin{equation}\label{onshell}
\begin{split}
&\Sigma^r(\slashed{p})\mid_{\slashed{p}=M}=0\ \mbox{, }\\
&\frac{d}{d\slashed{p}}\Sigma^r(\slashed{p})\mid_{\slashed{p}=M}=0\ \mbox{. }
\end{split}
\end{equation}
Such conditions can be fulfilled by choosing suitable values of the nucleon wave function renormalization constant $Z_N$ and the mass counter term $\delta M$.

To be specific, the bare parameters can be split as follows
\begin{equation}\label{counter1}
\begin{split}
&c_i^0\equiv c_i^r+\frac{\gamma_{c_i}M}{16\pi^2F^2}R_\epsilon\ \mbox{, }\\
&d_j^0\equiv d_j^r+\frac{\gamma_{d_i}}{16\pi^2F^2}R_\epsilon\ \mbox{, }\\
&g^0\equiv g^r+\frac{\gamma_{g}M^2F^2}{16\pi^2F^4}R_\epsilon\ \mbox{, }\\
&M^0\equiv M^r+\delta M\ \mbox{, }
\end{split}
\end{equation}
where the LECs with superscript ``0'' denote the bare quantities and  the ones with ``$r$'' stand for renormalized quantities. The corresponding coefficients $\gamma$ are
\begin{equation}\label{gamma1}
\begin{split}
&\gamma_{g}=-2g+g^3\ \mbox{, }\\
&\gamma_{c_1}=-\frac{3}{8}g^2\ \mbox{, }\\
&\gamma_{c_2}=\frac{1}{2}-g^2+\frac{1}{2}g^4\ \mbox{, }\\
&\gamma_{c_3}=\frac{1}{4}-\frac{3}{2}g^2+\frac{1}{4}g^4\ \mbox{, }\\
&\gamma_{c_4}=-\frac{1}{4}-\frac{1}{2}g^2+\frac{3}{4}g^4\ \mbox{, }\\
&\gamma_{d_1}+\gamma_{d_2}=\frac{1}{48}-\frac{1}{12}g^2+\frac{1}{16}g^4\ \mbox{, }\\
&\gamma_{d_3}=0\ \mbox{, }\\
&\gamma_{d_5}=\frac{1-g^2}{48}\ \mbox{; }\\
\end{split}
\end{equation}
and in addition,
\begin{equation}\label{ZdelM}
\begin{split}
&Z_N=1+f(M^2)+2M^2[f'(M^2)+g'(M^2)]\ \mbox{, }\\
&\delta M=-\frac{3g^2m^2}{64F^2\pi^2}(A_0+m^2 B_0^f)\ \mbox{, }
\end{split}
\end{equation}
where
\[
\begin{split}
&f(s)=\frac{3g^2}{64F^2\pi^2}[(s-M^2)B_1^s+A_0+m^2 B_0^s]\ \mbox{, }\\
&g(s)=\frac{3g^2}{64F^2\pi^2}(A_0+m^2 B_0^s)\ \mbox{, }\\
&B_0^f=B_0^M-1+R_\epsilon\ \mbox{. }
\end{split}
\]
See Table.~\ref{tab:loopfuncabbr} for the abbreviations of the involved loop functions. Note that except for $Z_N$ and $\delta M$, the other values in Eq.~\eqref{gamma1} are the same as in Ref.~\cite{Chen:2012nx,Siemens:2016hdi}. Lastly, the renormalized result of diagram (3n) reads
\begin{equation}\label{3nphysical}
\begin{split}
&\tilde{A}_{3n}^\pm=\frac{Mg^2}{4F^2}\left[2F_{3n}(s)+\frac{s+3M^2}{s-M^2}G_{3n}(s)\right]\ \mbox{, }\\
&\tilde{B}_{3n}^\pm=-\frac{g^2}{4F^2}\left[\frac{s+3M^2}{s-M^2}F_{3n}(s)+\frac{4M^2(s+M^2)}{(s-M^2)^2}G_{3n}(s)\right]\ \mbox{, }
\end{split}
\end{equation}
where
\begin{equation}
\begin{split}
&F_{3n}(s)=\frac{3g^2}{64F^2\pi^2}\left[(s-M^2)B_1^s+m^2(B_0^s-B_0^M)-2M^2\Big(B_1^M+2m^2\frac{\partial B_0^s}{\partial s}\mid_{s=M^2}\Big)\right]\ \mbox{, }\\
&G_{3n}(s)=\frac{3g^2}{64F^2\pi^2}\left[(s-M^2)B_1^s+2m^2(B_0^s-B_0^M)\right]\ \mbox{. }
\end{split}
\end{equation}
\subsubsection{PCB-term cancellation}\label{app:EOMS}
According to the spirit of EOMS scheme, the PCB terms hidden in the loop function come from the infrared regular parts and behave as polynomials of the external momenta. The infrared regular parts of related loop functions can be found in Ref.~\cite{Chen:2012nx}. The UV-renormalized LECs can further absorb the PCB terms in the way similarly to Eq.~\eqref{counter1}, i.e.,
\begin{equation}\label{counter2}
\begin{split}
&c_i^r\equiv c_i+\frac{\tilde{\gamma}_{c_i}M}{16\pi^2F^2}\equiv c_i+\delta c_i\ \mbox{, }\\
&g^r\equiv g+\frac{\tilde{\gamma}_{g}M^2}{16\pi^2F^2}\equiv g+\delta g\ \mbox{, }
\end{split}
\end{equation}
where the quantities with no superscripts refer to the physical ones. To cancel the power counting breaking terms, the $\tilde{\gamma}$ functions  are set to (see Ref.~\cite{Chen:2012nx})
\begin{equation}\label{gamma2}
\begin{split}
&\tilde{\gamma}_{g}=g^3\ \mbox{, }\\
&\tilde{\gamma}_{c_1}=\frac{3}{8}g^2\ \mbox{, }\\
&\tilde{\gamma}_{c_2}=-1-\frac{1}{2}g^4\ \mbox{, }\\
&\tilde{\gamma}_{c_3}=\frac{9}{4}g^4\ \mbox{, }\\
&\tilde{\gamma}_{c_4}=-\frac{1}{4}g^2(g^2+5)\ \mbox{. }
\end{split}
\end{equation}
The counterterms for the cancellations of the PCB terms can be obtained by the replacements: $c_i\to\delta c_i$ and $g^2\to 2g \delta g$, in the expressions of Fig.~\ref{fig:treediap1} (the last two diagrams) and Fig.~\ref{fig:treediap2}.

\subsection{Analytical expressions of loop integrals}\label{app:exploop}
In our numerical computations, the tensor integrals like $C_{ij}$ and $D_j$ in the amplitudes are first reduced to scalar functions with subscript ``0''. The relevant formulae are listed below.
\begin{itemize}[leftmargin=0.5cm]
\item{\textbf{Two point functions}}\\
\begin{align}
&B_1^u=\frac{1}{2u}\big[a_0-A_0-(u-M^2+m^2)B_0^u\big]\ \mbox{, }\\
&B_{00}^t=\frac{1}{18}(6m^2-t)+\frac{1}{6}a_0+\frac{1}{12}(4m^2-t)B_0^t\ \mbox{. }
\end{align}
\item{\textbf{Three point functions}}\\
\begin{enumerate}[leftmargin=0.5cm]
\item[{\bf 1.}]{\textbf{About $C_{\cdots}^t$}}\\
\begin{align}
&C_1^t=C_2^t=\frac{1}{t-4M^2}(B_0^T-B_0^M+m^2C_0^t)\ \mbox{, }\\
&C_{00}^t=\frac{1}{4}+\frac{1}{4(t-4M^2)}\big[(t-4M^2+2m^2)B_0^T-2m^2B_0^M+2m^2(t-4M^2+m^2)C_0^t\big]\ \mbox{, }\\
&C_{11}^t=C_{22}^t=\frac{M^2}{t(t-4M^2)}-\frac{t-2M^2}{2M^2t(t-4M^2)}(a_0-A_0)\nonumber\\
&+\frac{1}{2M^2(t-4M^2)^2t}\Big\{m^2(t^2-8M^2t+4M^4)B_0^M+M^2(4m^2M^2-8M^4+2m^2t+6M^2t-t^2)B_0^T\nonumber\\
&+2M^2m^2(2M^2t+m^2t+2m^2M^2-8M^4)C_0^t\Big\}\ \mbox{, }\\
&C_{12}^t=\frac{t-2M^2-2a_0+2A_0}{2t(t-4M^2)}-\frac{1}{t(t-4M^2)^2}\Big\{m^2(t+2M^2)B_0^M\nonumber\\
&+(M^2t-2m^2t+2m^2M^2-4M^4)B_0^T+m^2(6M^2t-2m^2t+2m^2M^2-8M^4-t^2)C_0^t\Big\}\ \mbox{. }
\end{align}
\item[{\bf 2.}]{\textbf{About $C_{\cdots}^{t2m}$}}\\
\begin{align}
&C_{1}^{t2m}=\frac{2(B_0^M-B_0^t)-(t-2m^2)C_0^{t2m}}{t-4M^2}\ \mbox{, }\\
&C_{00}^{t2m}=\frac{1}{4}+\frac{1}{4(t-4M^2)}\big[(t-2m^2)B_0^t-2(2M^2-m^2)B_0^M+2(M^2t-4M^2m^2+m^4)C_0^{t2m}\big]\ \mbox{, }\\
&C_{11}^{t2m}=\frac{1}{t-4M^2}+\frac{a_0-A_0}{M^2(t-4M^2)}+\frac{1}{M^2(t-4M^2)^2}\Big\{(10m^2M^2-4M^4-m^2t-2M^2t)B_0^M\nonumber\\
&+3M^2(t-2M^2)B_0^t+M^2(t^2+2M^2t-4m^2t-8m^2M^2+6m^4)C_0^{t2m}\Big\}\ \mbox{. }
\end{align}
\item[{\bf 3.}]{\textbf{About $C_{\cdots}^{s}$}}\\
\begin{align}
\hspace{-1.0cm}
&C_{1}^{s}=\frac{(s-M^2+m^2)B_0^m+(s+M^2-m^2)B_0^M-2sB_0^s-\big[s^2-2m^2s-(M^2-m^2)^2\big]C_0^s}
{\big[s-(M+m)^2\big]\big[s-(M-m)^2\big]}\ \mbox{, }\\
&C_{2}^{s}=\frac{1}{\big[s-(M+m)^2\big]\big[s-(M-m)^2\big]}\bigg\{-(s-M^2-m^2)B_0^m-2M^2B_0^M\nonumber\\
&\hspace{1cm}
+(s+M^2-m^2)B_0^s+\big[s(2M^2-m^2)-(M^2-m^2)(2M^2+m^2)\big]C_0^s\bigg\}
\ \mbox{. }
\end{align}
\item[{\bf 4.}]{\textbf{About $C_{\cdots}^T$}}\\
\begin{align}
&C_{1}^T=\frac{B_0^T-B_0^m+m^2C_0^T}{t-4m^2}\ \mbox{, }\\
&C_{2}^T=\frac{2(B_0^m-B_0^T)-(t-2m^2)C_0^T}{t-4m^2}\ \mbox{. }
\end{align}
\end{enumerate}
\item{\textbf{Four point functions}}\\
\begin{align}
&D_1=D_2=\frac{2sC_0^s-(s+M^2-m^2)C_0^t-(s-M^2+m^2)C_0^T+\big[s^2-2m^2s-(M^2-m^2)^2\big]D_0}{2\big[su-(M^2-m^2)^2\big]}\ \mbox{, }\\
&D_3=\frac{-2(s+M^2-m^2)C_0^s-(t-4M^2)C_0^t+(s-u)C_0^T}{2\big[su-(M^2-m^2)^2\big]}\nonumber\\
&+\frac{\big[st-(M^2-m^2)t-2s(2M^2-m^2)+2(M^2-m^2)(2M^2+m^2)\big]D_0}{2\big[su-(M^2-m^2)^2\big]}\ \mbox{. }
\end{align}
\end{itemize}

Now only the expressions of those scalar loop functions are needed. One point and two point functions are simple when the dimension $D$ is set to $4$:
\begin{equation}\label{Aexp}
A_0(m^2)=m^2\big(-R_\epsilon+\ln\frac{\nu^2}{m^2}\big)\ \mbox{, }
\end{equation}
where
\begin{equation}\label{Repdef}
R_\epsilon=-\frac{1}{\epsilon}+\gamma_E-\ln(4\pi)-1\ ,
\end{equation}
with $\gamma_E$ the Euler constant; and
\begin{equation}\label{Bexp}
\begin{split}
&B_0(p^2;m^2,M^2)=1-R_\epsilon+\ln\left(\frac{\nu^2}{M^2}\right)+\frac{1}{2p^2}\Bigg\{[p^2(1+\rho)-R_m]\ln\left[\frac{R_m+p^2(1-\rho)}{R_m-p^2(1+\rho)}\right]\\
&+[p^2(1-\rho)-R_m]\ln\left[\frac{R_m+p^2(1+\rho)}{R_m-p^2(1-\rho)}\right]\Bigg\}\ \mbox{, }
\end{split}
\end{equation}
where
\[
\rho=\frac{\sqrt{p^2-(M+m)^2}\sqrt{p^2-(M-m)^2}}{p^2},\ R_m=M^2-m^2\ \mbox{. }
\]
In Eqs.~\ref{Aexp} and \ref{Bexp} we set $\nu=M$. However, three and four point functions do not have such compact analytical expressions. The most commonly used representation of those functions in textbooks is the Feynman representation, which expresses the loop functions by integrals with respect to Feynman parameters. However in our calculation it is found that the Mandelstam spectral representation is more preferable numerically. In the Mandelstam spectral representation the loop functions are calculated by means of dispersion integrals:
\begin{equation}\label{loopfuncnew}
\begin{split}
&C_0^t(t)=\frac{1}{\pi}\int_{4M^2}^{+\infty}\frac{\text{Im}[C_0^t(t')]}{t'-t-\text{i}0^+}dt'\ \mbox{, }\\
&C_0^T(t)=\frac{1}{\pi}\int_{4M^2}^{+\infty}\frac{\text{Im}[C_0^T(t')]}{t'-t-\text{i}0^+}dt'\ \mbox{, }\\
&C_0^{t2m}(t)=\frac{1}{\pi}\int_{4m^2}^{+\infty}\frac{\text{Im}[C_0^{t2m}(t')]}{t'-t-\text{i}0^+}dt'\ \mbox{, }\\
&C_0^s(s)=\frac{1}{\pi}\int_{(M+m)^2}^{+\infty}\frac{\text{Im}[C_0^s(s')]}{s'-s-\text{i}0^+}ds'\ \mbox{, }\\
&D_0(s,t)=\frac{1}{\pi^2}\int_{(M+m)^2}^{+\infty}ds'\int_{(M+m)^2}^{+\infty}dt'\frac{\sigma_{D_0}(s',t')}{(s'-s-\text{i}0^+)(t'-t-\text{i}0^+)}\ \mbox{. }
\end{split}
\end{equation}
Those spectral functions can be obtained by Cutkosky rule in physical region\footnote{Generally speaking Cutkosky rule on physical cut may not give the complete spectral functions due to the existence of some anomalous singularities like the anomalous triangle singularity. Fortunately, in two-body scattering processes with stable particles such singularities do not appear on the physical sheet, so Cutkosky rule is valid. } and are listed below. For three point functions,
\begin{equation}\label{loopfuncspec3}
\begin{split}
&\text{Im}[C_0^t]=\frac{\pi}{t-4M^2}\rho_M(t)\ln\left(\frac{m^2}{t-4M^2+m^2}\right)\ \mbox{, }\\
&\text{Im}[C_0^T]=\frac{\pi}{\sqrt{t(t-4m^2)}}\ln\left(\frac{2m^2-t+\sqrt{t-4M^2}\sqrt{t-4m^2}}{2m^2-t-\sqrt{t-4M^2}\sqrt{t-4m^2}}\right)\ \mbox{, }\\
&\text{Im}[C_0^{t2m}]=\frac{\pi}{\sqrt{t(t-4M^2)}}\ln\left(\frac{2m^2-t+\sqrt{t-4M^2}\sqrt{t-4m^2}}{2m^2-t-\sqrt{t-4M^2}\sqrt{t-4m^2}}\right)\ \mbox{, }\\
&\text{Im}[C_0^{s}]=\frac{\pi}{s\rho(s)}\ln\left[\frac{M^2(s-\frac{(M^2-m^2)^2}{M^2})}{s(s-M^2-2m^2)}\right]\ \mbox{, }\\
\end{split}
\end{equation}
and for the four point function,
\begin{equation}\label{loopfuncspec4}
\begin{split}
&\sigma_{D_0}(s,t)=4\pi^2s\Theta[\Delta(s,t)]\times\sqrt{\frac{\lambda(s,m^2,M^2)}{\Delta(s,t)}}\ \mbox{, }\\
&\Delta(s,t)=(A_d^2-B_d^2z_s^2)^2-(A_d^2-B_d^2)^2\ \mbox{, }\\
&A_d=s^2-2m^2s-(M^2-m^2)^2\ \mbox{, }\\
&B_d=\lambda(s,m^2,M^2)\ \mbox{, }\\
&z_s=\frac{s}{\lambda(s,m^2,M^2)}\left[s+2t-2(M^2+m^2)+\frac{(M^2-m^2)^2}{s}\right]\ \mbox{. }
\end{split}
\end{equation}

\section{Uncertainties of left-hand cuts}\label{app:uncer}
\subsection{Uncertainties from cut-off parameters}

\begin{figure}[htbp]
\center
\subfigure[]{
\label{p3cutoff:subfig:S11}
\scalebox{1.2}[1.2]{\includegraphics[width=0.4\textwidth]{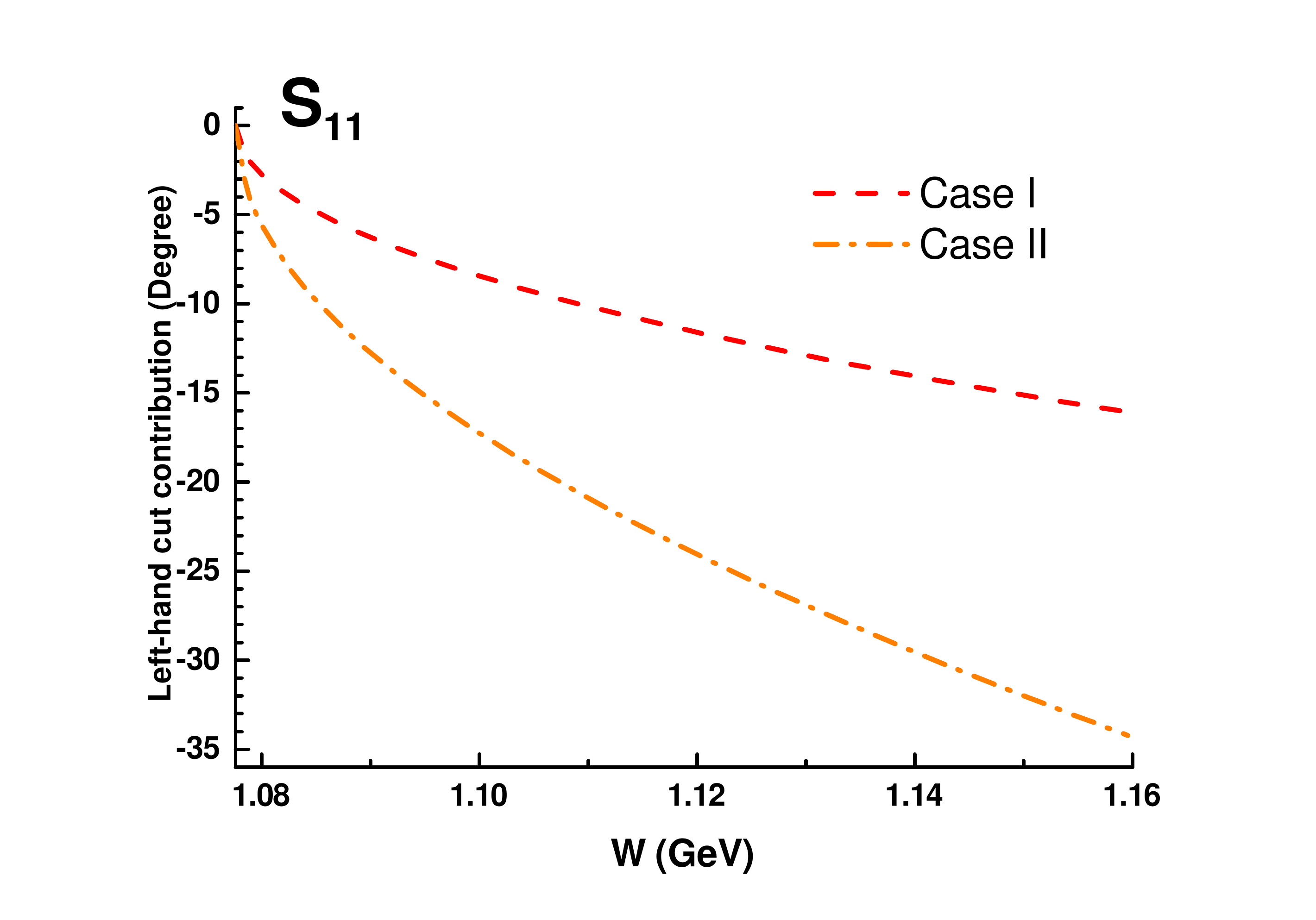}}}
\subfigure[]{
\label{p3cutoff:subfig:S31}
\scalebox{1.2}[1.2]{\includegraphics[width=0.4\textwidth]{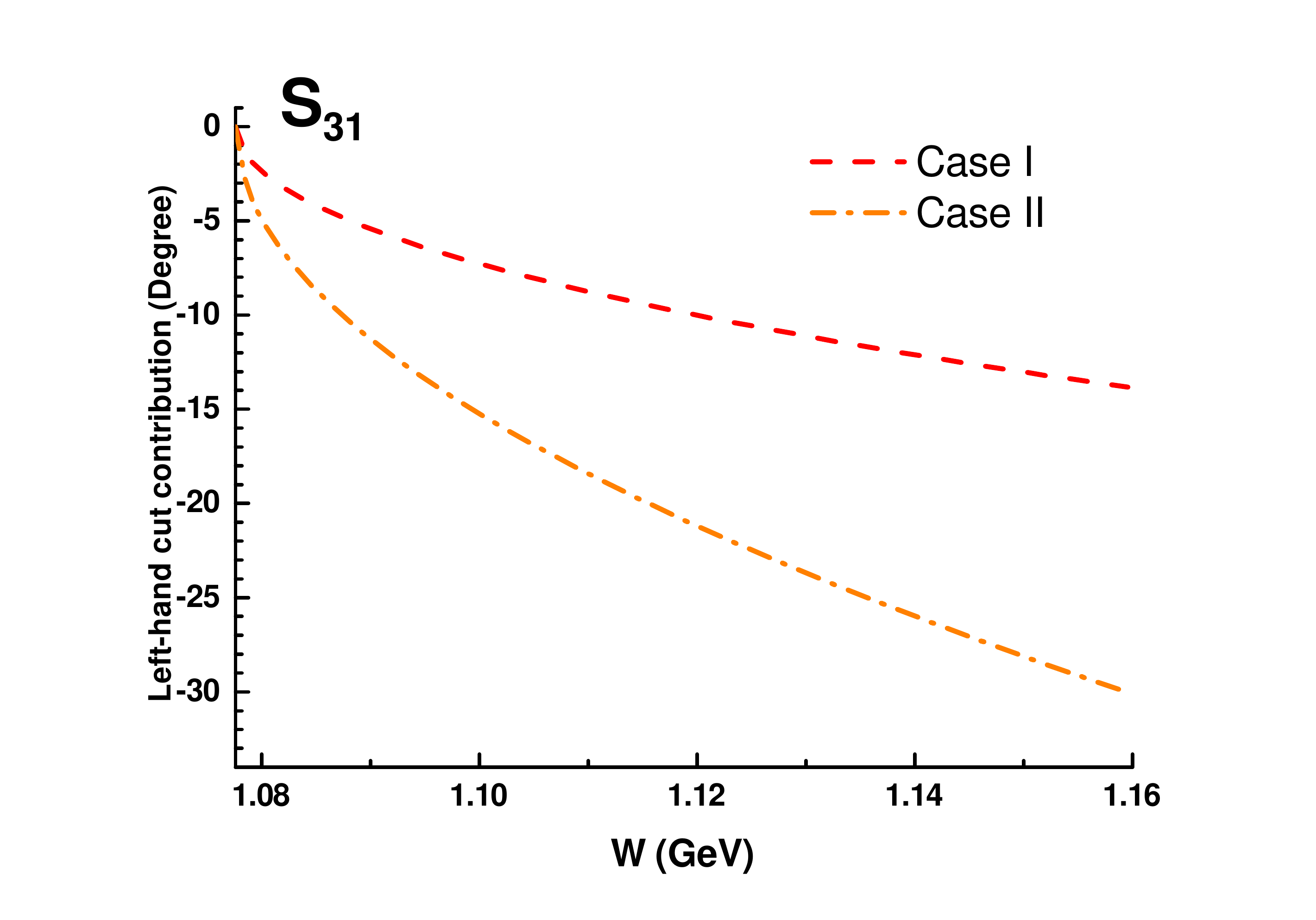}}}
\subfigure[]{
\label{p3cutoff:subfig:P11}
\scalebox{1.2}[1.2]{\includegraphics[width=0.4\textwidth]{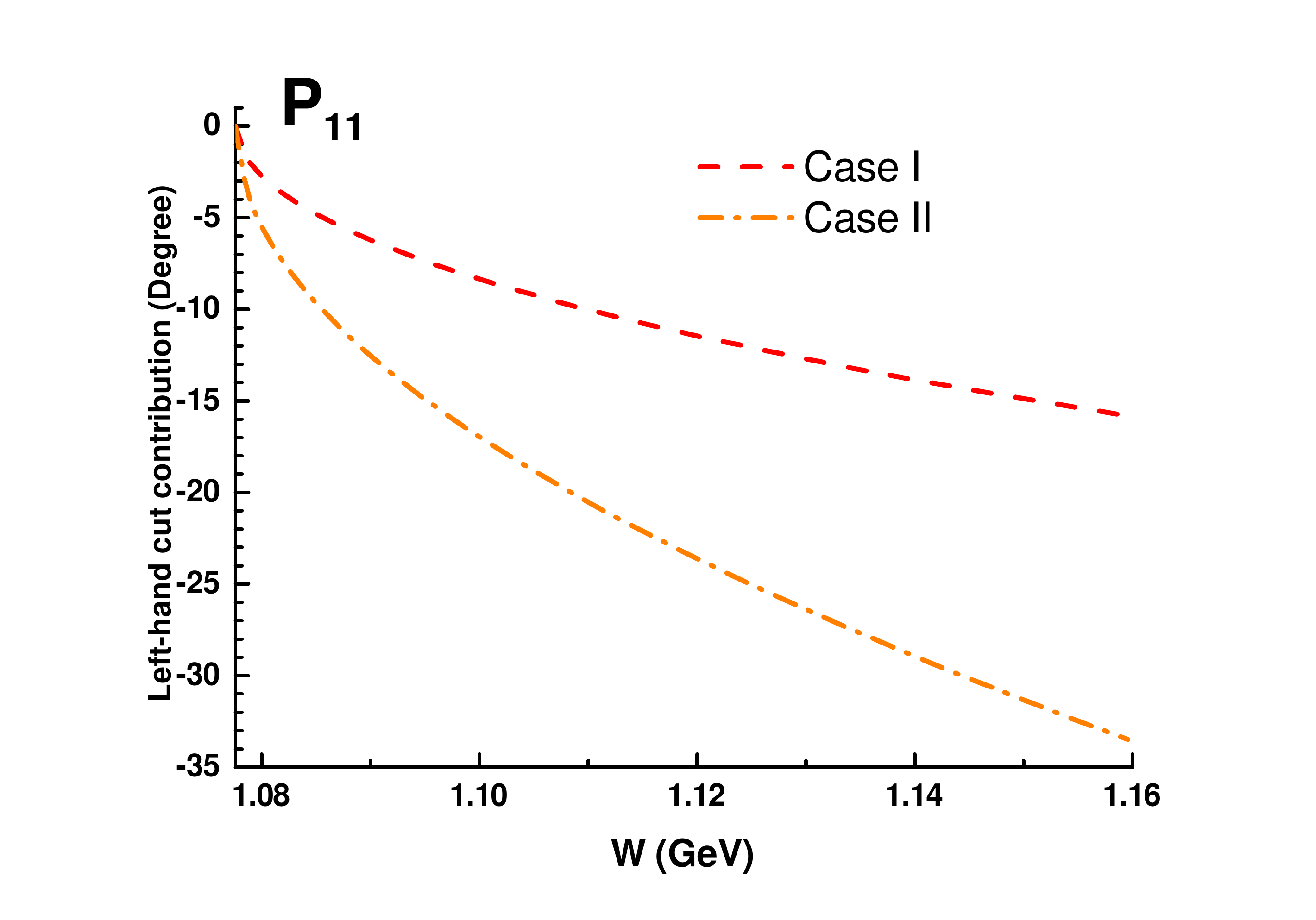}}}
\subfigure[]{
\label{p3cutoff:subfig:P31}
\scalebox{1.2}[1.2]{\includegraphics[width=0.4\textwidth]{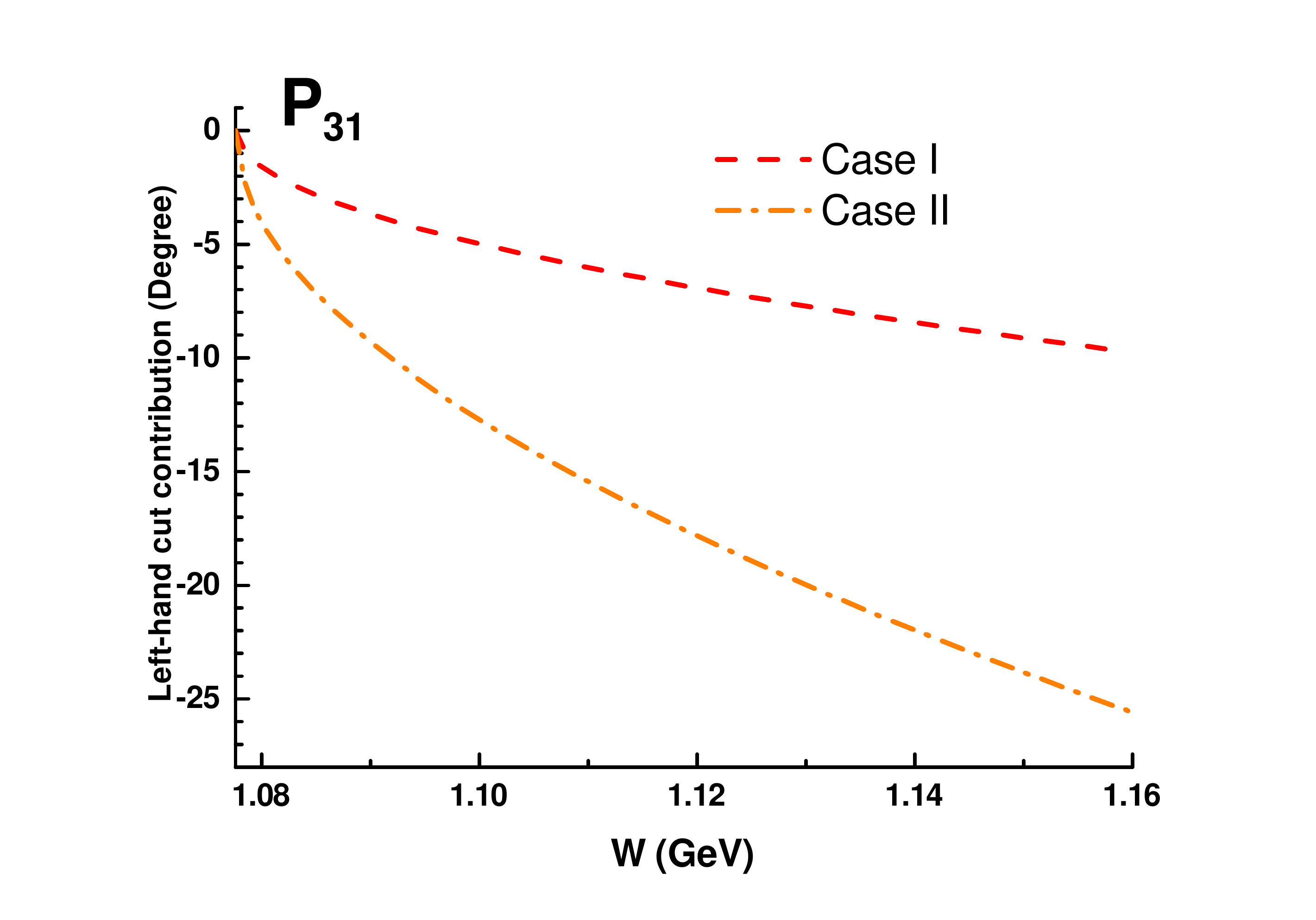}}}
\subfigure[]{
\label{p3cutoff:subfig:P13}
\scalebox{1.2}[1.2]{\includegraphics[width=0.4\textwidth]{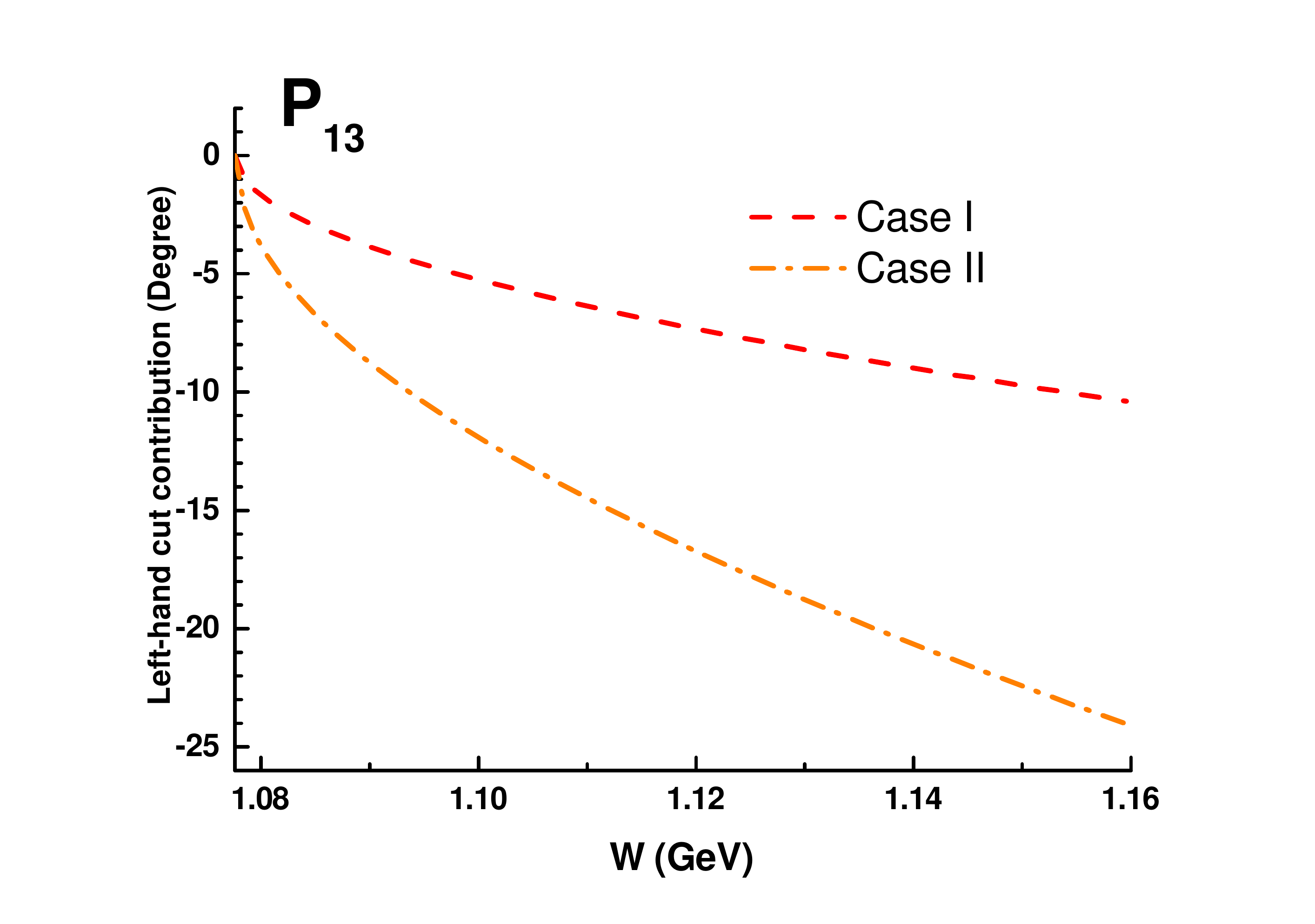}}}
\subfigure[]{
\label{p3cutoff:subfig:P33}
\scalebox{1.2}[1.2]{\includegraphics[width=0.4\textwidth]{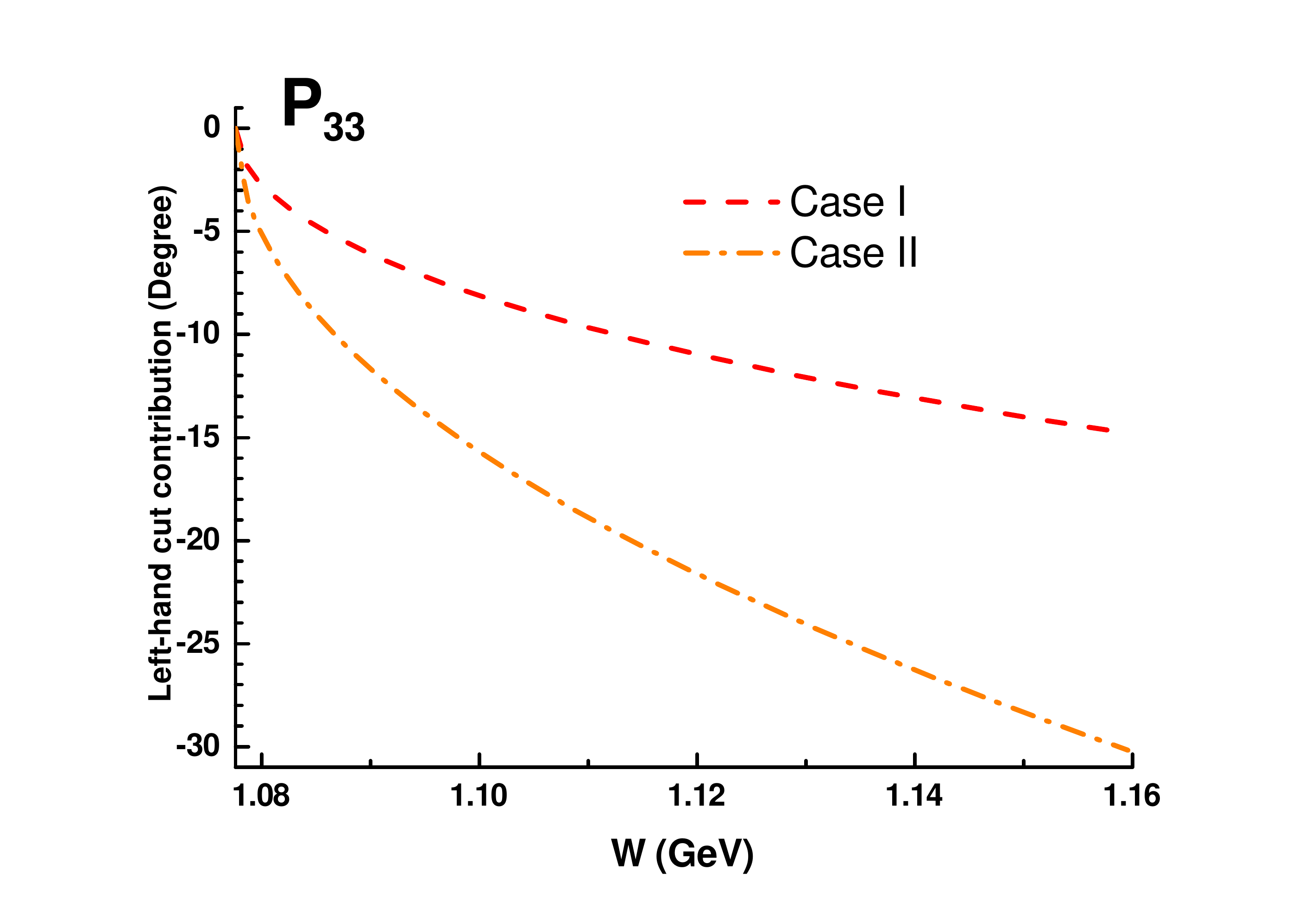}}}
\caption{The \textit{l.h.c.s} contribution with different values of cut-off parameters. }\label{fig:p3cutoff}
\end{figure}

In this subsection, the influence of the variations of the cut-off parameters is discussed.  The set $s_c=-0.08$ GeV$^2$ and $\theta_c=1.18$ radian is used in the text (denoted by Case I in Fig.~\ref{fig:p3cutoff}). Here we choose $s_c=-25$ GeV$^2$ and $\theta_c=\pi$ radian, and the consequent results are shown as Case II in Fig.~\ref{fig:p3cutoff}. The results show quite similar behavior as in $\mathcal{O}(p^2)$ calculation: lager $s_c$ (in absolute value) yields larger negative contributions of \textit{l.h.c.}s; different cut-off parameters do not alter the picture qualitatively.

\subsection{Left-hand cuts from unitarized amplitudes}

\begin{figure}[htbp]
\center
\subfigure[]{
\label{p3uni:subfig:S11}
\scalebox{1.2}[1.2]{\includegraphics[width=0.4\textwidth]{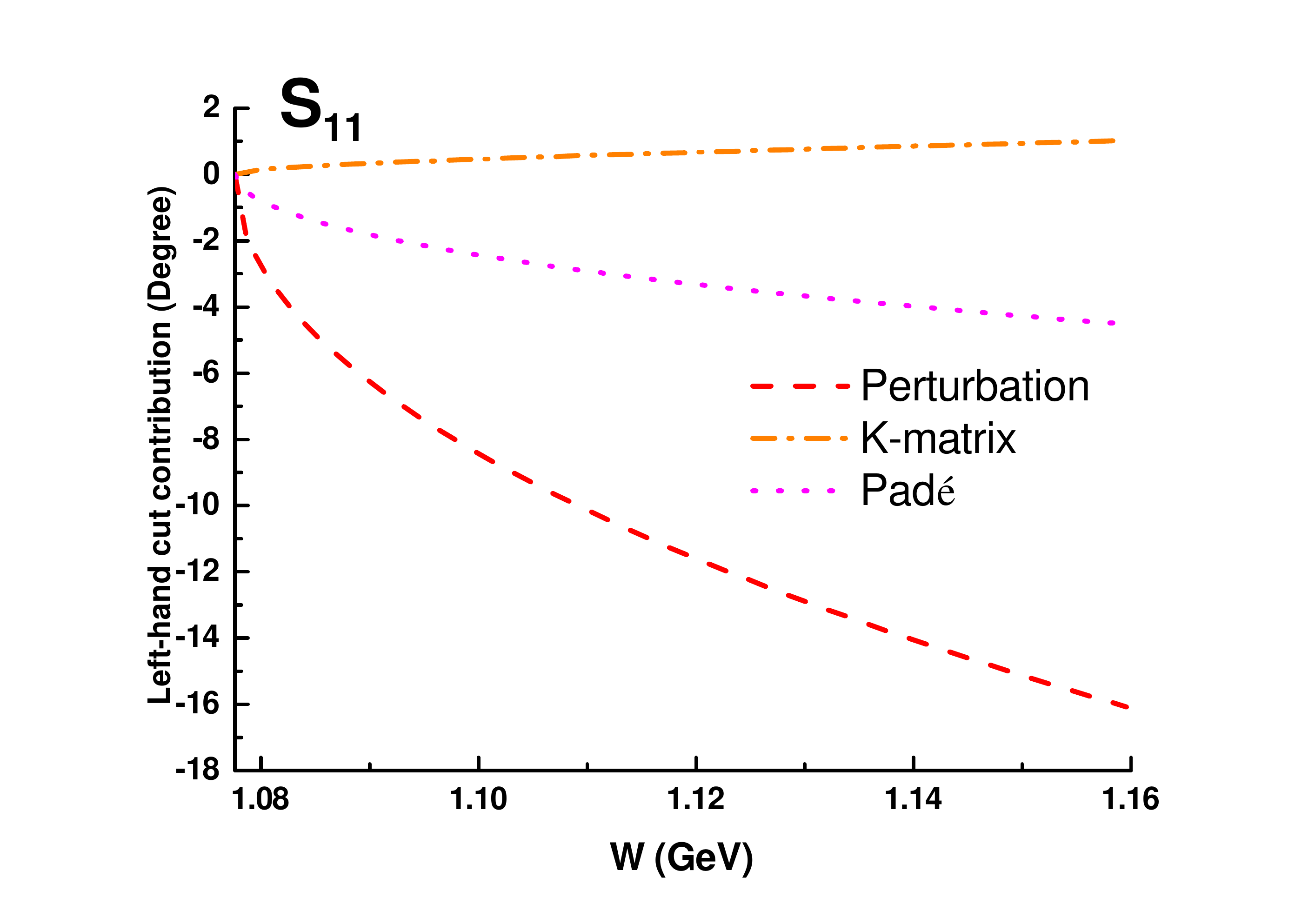}}}
\subfigure[]{
\label{p3uni:subfig:S31}
\scalebox{1.2}[1.2]{\includegraphics[width=0.4\textwidth]{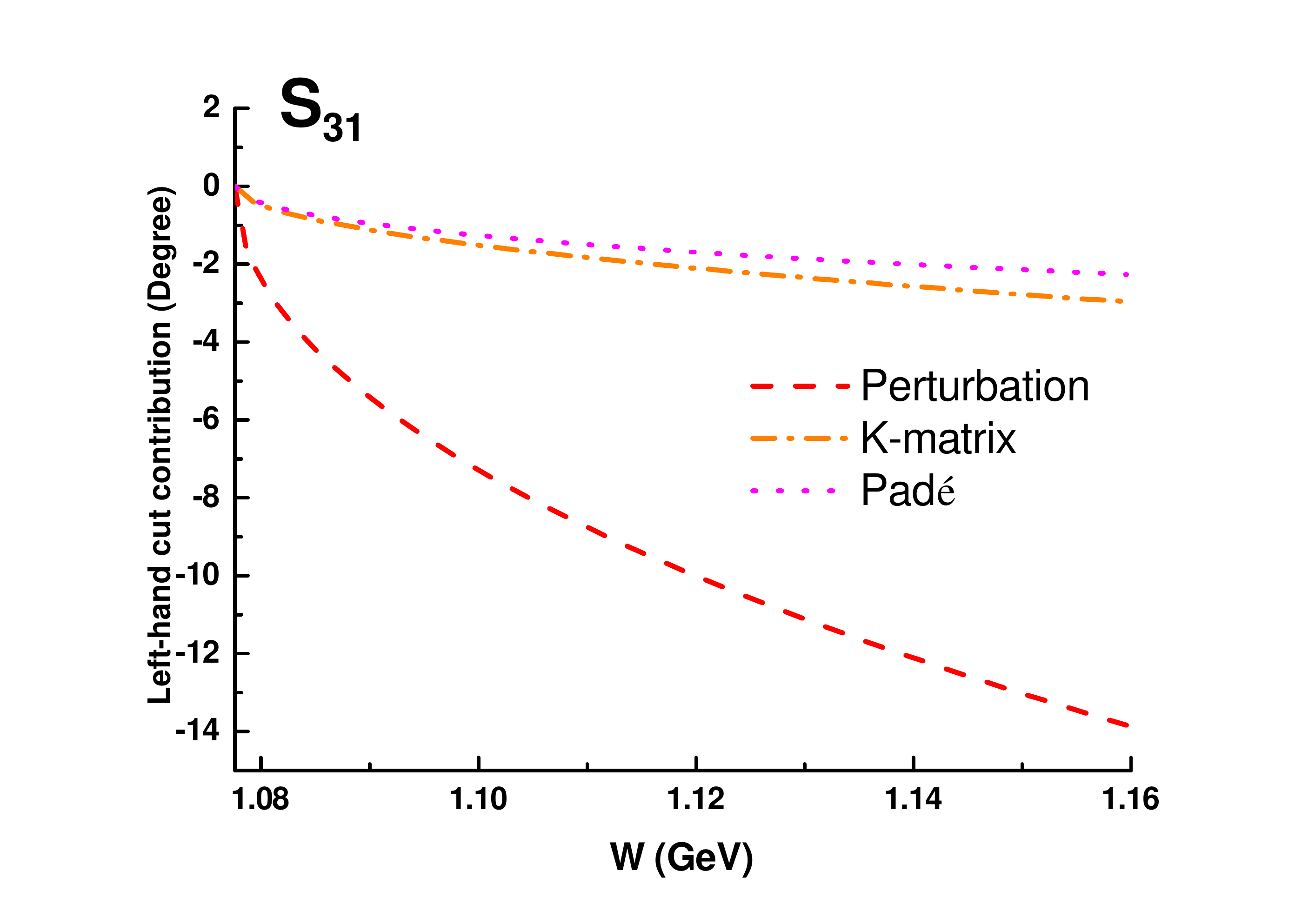}}}
\subfigure[]{
\label{p3uni:subfig:P11}
\scalebox{1.2}[1.2]{\includegraphics[width=0.4\textwidth]{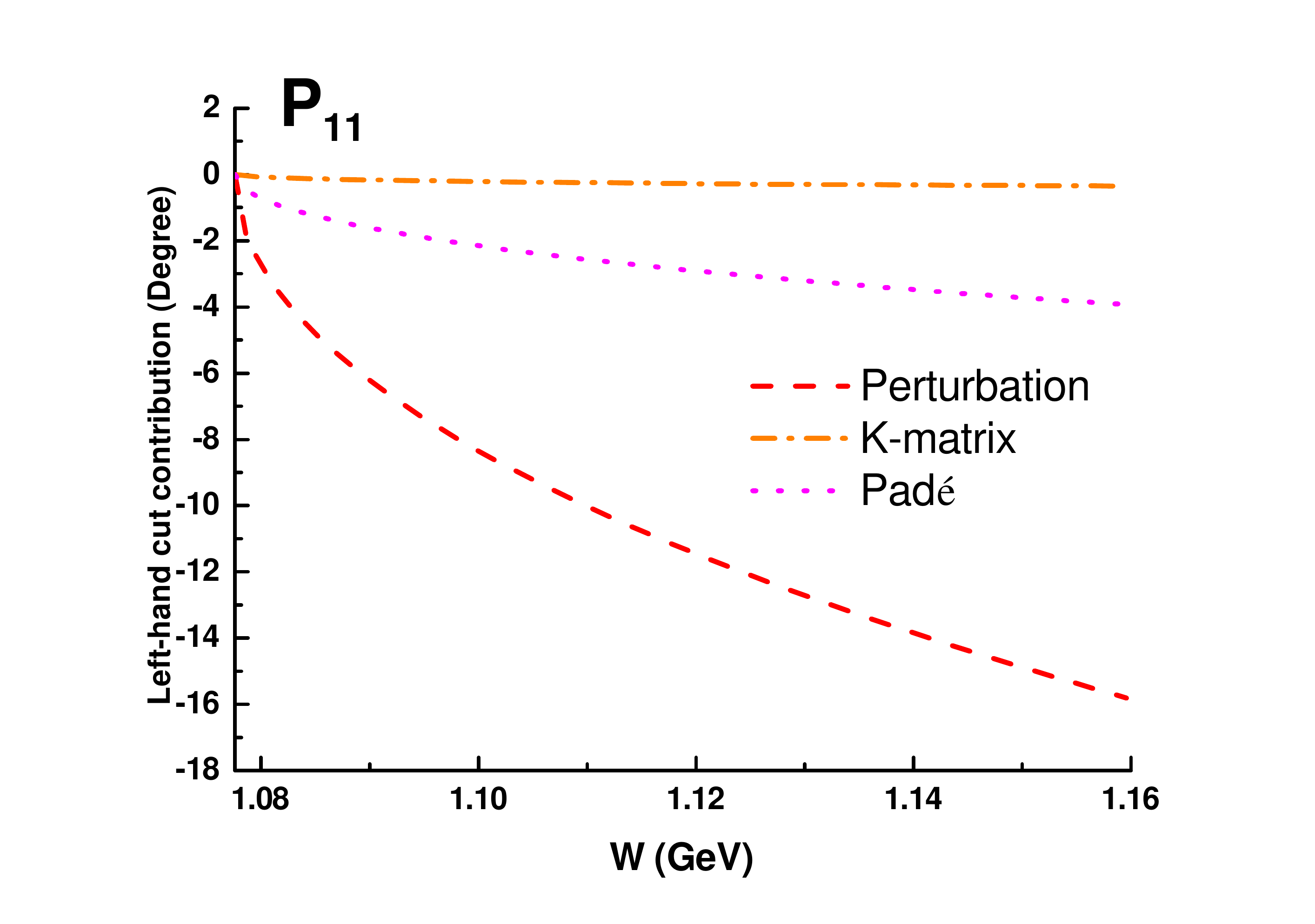}}}
\subfigure[]{
\label{p3uni:subfig:P31}
\scalebox{1.2}[1.2]{\includegraphics[width=0.4\textwidth]{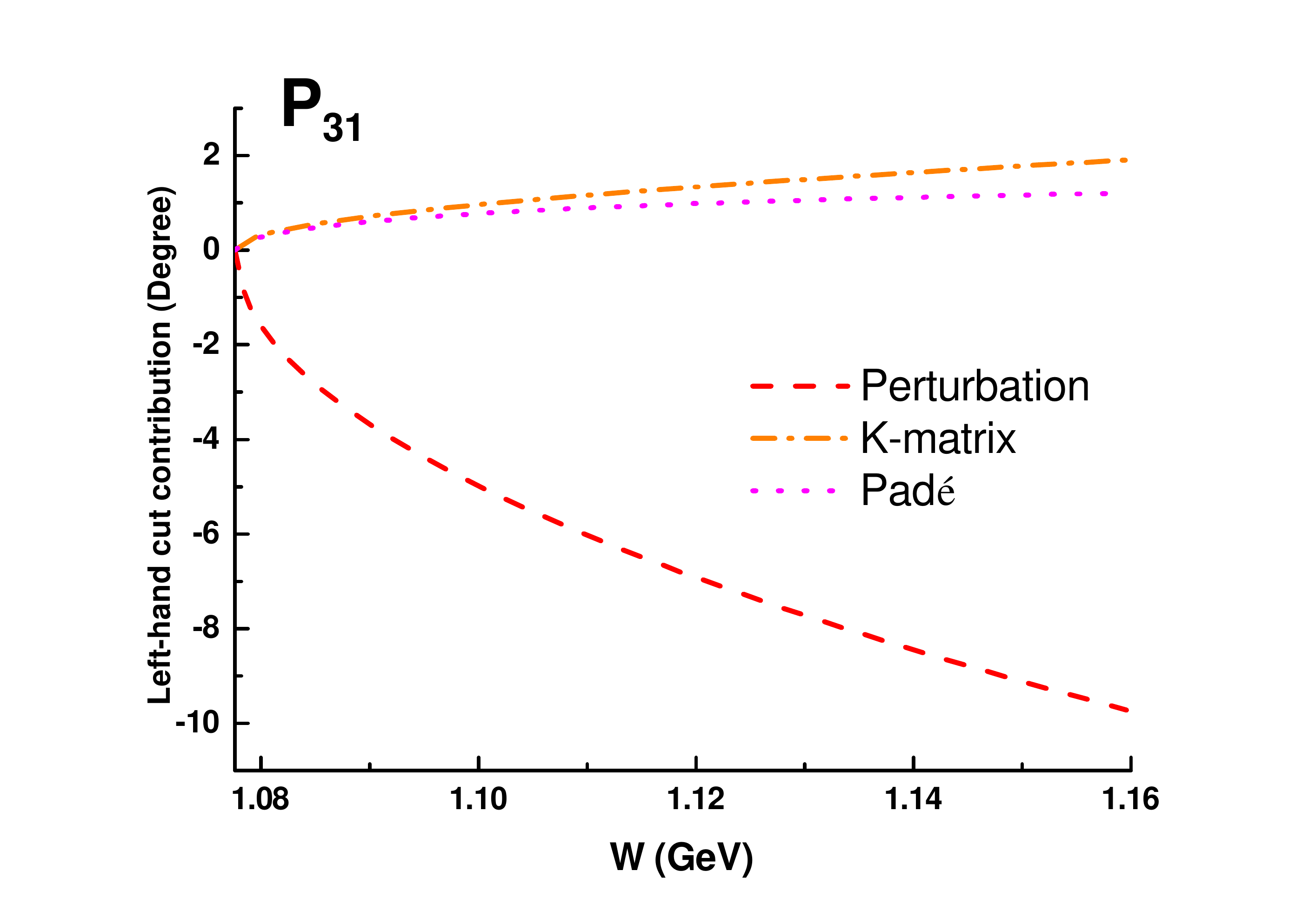}}}
\subfigure[]{
\label{p3uni:subfig:P13}
\scalebox{1.2}[1.2]{\includegraphics[width=0.4\textwidth]{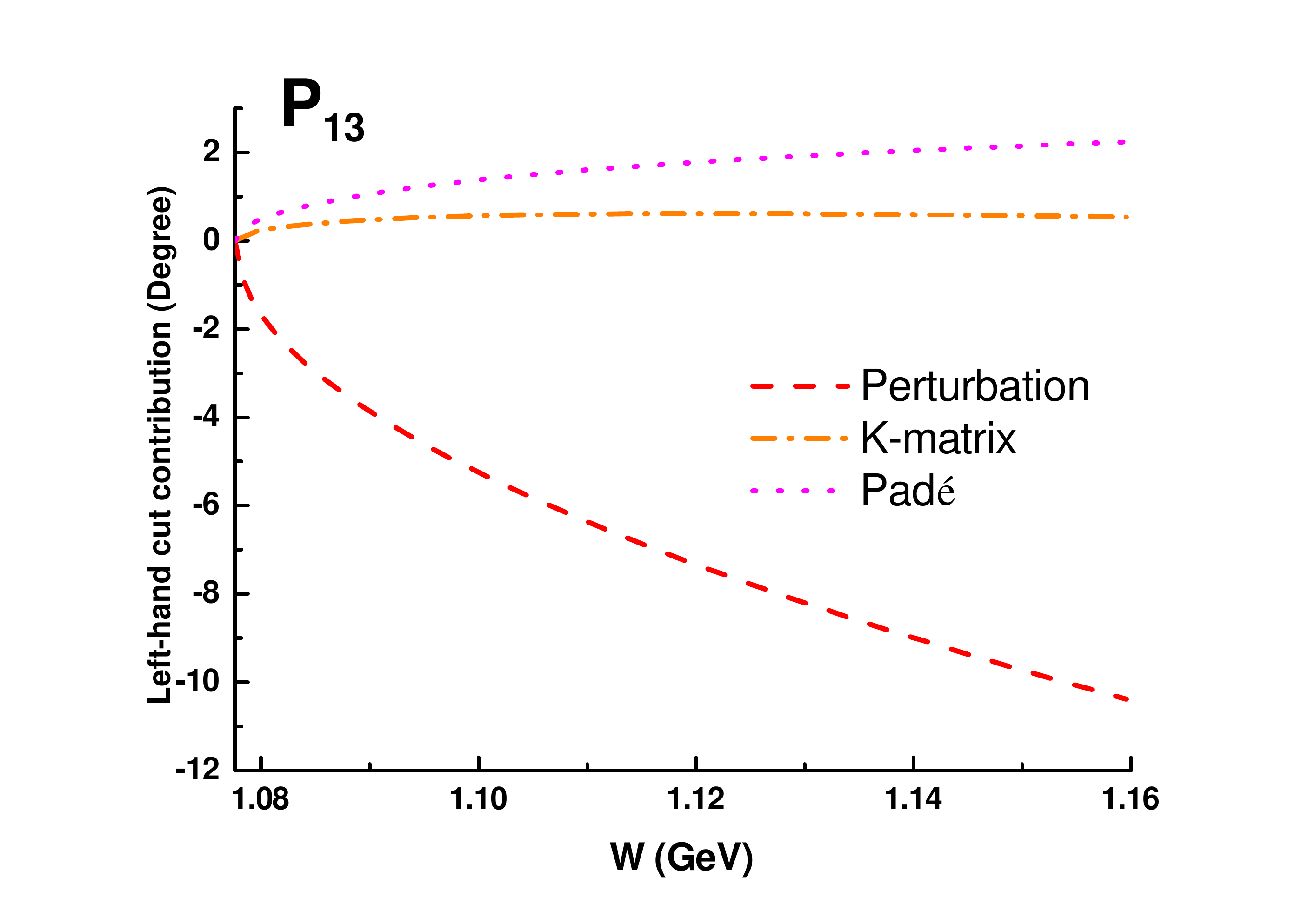}}}
\subfigure[]{
\label{p3uni:subfig:P33}
\scalebox{1.2}[1.2]{\includegraphics[width=0.4\textwidth]{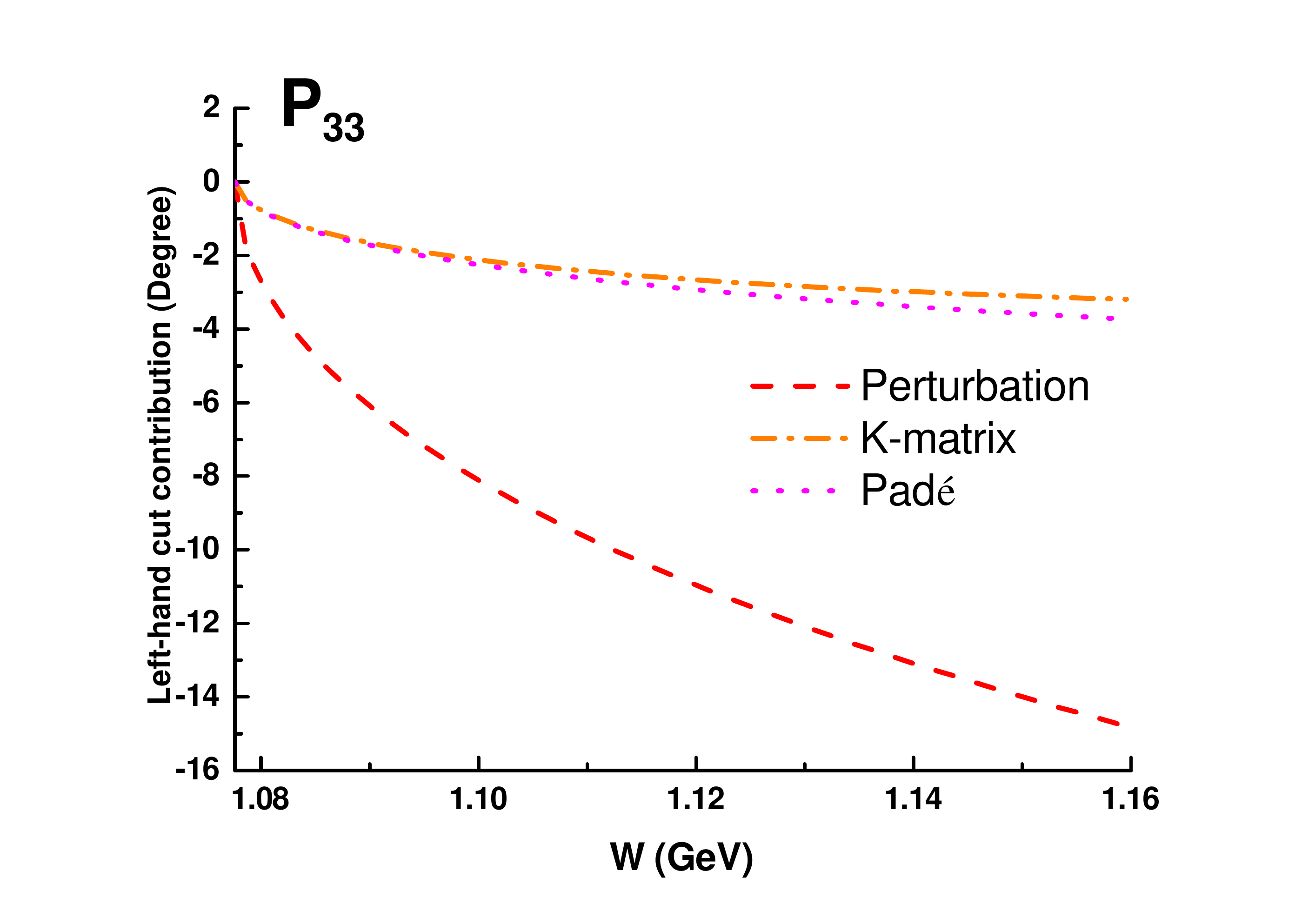}}}
\caption{Phase shifts from \textit{l.h.c.}s evaluated by perturbation theory and the two unitarization schemes in Eqs.~\eqref{Kdef} and~\eqref{Padedef}. The cut-off parameters are set as $s_c=-0.08$ GeV$^2$ and $\theta_c=1.18$ radian. }
\label{fig:p3uni}
\end{figure}

As already mentioned in the preceding paper~\cite{Wang:2017agd}, some conventional unitarization approaches like the $K$-matrix method, may give un-physical structures like spurious poles, but anyway at $\mathcal{O}(p^3)$ level the imaginary parts appears directly in the perturbation amplitudes, which would more or less improve the conventional unitarization approaches\footnote{In principle one can also figure out the locations of the spurious poles given by conventional unitarization approaches at $\mathcal{O}(p^3)$ level, just like what has been done in the preceding paper, but that requires much more efforts on numerical calculations since $\mathcal{O}(p^3)$ perturbation amplitudes have no analytic expressions, hence is not done in this paper. }. Roughly speaking, the differences between perturbation and unitarized results provide one possible point of views to estimate the uncertainties of \textit{l.h.c.}s calculation.

At $\mathcal{O}(p^3)$ level there are two traditional schemes to unitarize the amplitude, i.e. $K$-matrix approach in Eq.~\eqref{Kdef}, and Pad\'{e} approximation in Eq.~\eqref{Padedef}
\begin{align}
&T_K=\frac{T_1+T_2+\text{Re}T_3}{1-i\rho(T_1+T_2+\text{Re}T_3)}\ \mbox{, }\label{Kdef}\\
&T_P=\frac{T_1^2}{T_1-T_2-T_3}\ \mbox{, }\label{Padedef}
\end{align}
where $T_i$ represents the perturbation amplitude at $\mathcal{O}(p^i)$ level in each channel; $\text{Re}T_3$ stands for the real part of $\mathcal{O}(p^3)$ amplitude in physical region, and generally should be regarded as $T_3-i\rho T_1^2$. The two methods both give manifestly unitary $T$, i.e. $\text{Im}T=\rho|T|^2$. The contributions of \textit{l.h.c.}s given by perturbation calculation and the two unitarization methods are shown in Fig.~\ref{fig:p3uni}. It is found that the uncertainties in $P_{31}$ and $P_{13}$ channels may be larger since the two unitarization methods both give positive results in these two channels.

{In the end we emphasize that,  as claimed in the preceding paper~\cite{Wang:2017agd}, conventional unitarization methods may contain spurious poles that violate causality, thus they are not trustworthy in analyzing pole structures. }


\begin{thebibliography}{99}


\bibitem{Bransden:1973}  B.H. Bransden  and R.G. Moorhouse,
    {\it The pion-nucleon system},  Princeton University Press, 1973;

\bibitem{Hohler:1983}  G. H\"ohler, {\it Landolt-B\"ornstein, Vol. 9b2},
    edited by H. Schopper, Springer, Berlin, 1983.

\bibitem{Koch:1980ay}
  R.~Koch and E.~Pietarinen,
  Nucl.\ Phys.\ A {\bf 336}, 331 (1980).


\bibitem{Koch:1985bn}
  R.~Koch,
  Nucl.\ Phys.\ A {\bf 448}, 707 (1986).


\bibitem{Matsinos:2006sw}
  E.~Matsinos, W.~S.~Woolcock, G.~C.~Oades, G.~Rasche and A.~Gashi,
  Nucl.\ Phys.\ A {\bf 778}, 95 (2006)
  [hep-ph/0607080].


\bibitem{SAID}
Computer code SAID, online program at http://gwdac.phys.gwu.edu/ (solution WI08), and
R. A. Arndt et al., Phys. Rev. C {\bf 74}, 045205 (2006) (solution SM01).

\bibitem{Roper:1964zza}
  L.~D.~Roper,
  Phys.\ Rev.\ Lett.\  {\bf 12}, 340 (1964).


\bibitem{Bottino:1999ei}
  A.~Bottino, F.~Donato, N.~Fornengo and S.~Scopel,
  Astropart.\ Phys.\  {\bf 13}, 215 (2000)
  [hep-ph/9909228].


\bibitem{Bottino:2001dj}
  A.~Bottino, F.~Donato, N.~Fornengo and S.~Scopel,
  Astropart.\ Phys.\  {\bf 18}, 205 (2002)
  [hep-ph/0111229].


\bibitem{Gasser:1987rb}
  J.~Gasser, M.~E.~Sainio and A.~Svarc,
  Nucl.\ Phys.\ B {\bf 307}, 779 (1988).


\bibitem{Jenkins:1990jv}
  E.~E.~Jenkins and A.~V.~Manohar,
  Phys.\ Lett.\ B {\bf 255}, 558 (1991).


\bibitem{Bernard:1992qa}
  V.~Bernard, N.~Kaiser, J.~Kambor and U.~G.~Mei{\ss}ner,
  Nucl.\ Phys.\ B {\bf 388}, 315 (1992).


\bibitem{Ellis:1997kc}
  P.~J.~Ellis and H.~B.~Tang,
  Phys.\ Rev.\ C {\bf 57}, 3356 (1998)
  [hep-ph/9709354].


\bibitem{Becher:1999he}
  T.~Becher and H.~Leutwyler,
  Eur.\ Phys.\ J.\ C {\bf 9}, 643 (1999)
  [hep-ph/9901384].


\bibitem{Gegelia:1999qt}
  J.~Gegelia, G.~Japaridze and X.~Q.~Wang,
  J.\ Phys.\ G {\bf 29}, 2303 (2003)
  [hep-ph/9910260].


\bibitem{Gegelia:1999gf}
  J.~Gegelia and G.~Japaridze,
  Phys.\ Rev.\ D {\bf 60}, 114038 (1999)
  [hep-ph/9908377].


\bibitem{Fuchs:2003qc}
  T.~Fuchs, J.~Gegelia, G.~Japaridze and S.~Scherer,
  Phys.\ Rev.\ D {\bf 68}, 056005 (2003)
  [hep-ph/0302117].

\bibitem{Epelbaum:2015vea}
  E.~Epelbaum, J.~Gegelia, U.~G.~Mei{\ss}ner and D.~L.~Yao,
  Eur.\ Phys.\ J.\ C {\bf 75}, no. 10, 499 (2015)
  doi:10.1140/epjc/s10052-015-3728-7
  [arXiv:1510.02388 [hep-ph]].

\bibitem{Fettes:1998ud}
  N.~Fettes, U.~G.~Mei{\ss}ner and S.~Steininger,
  Nucl.\ Phys.\ A {\bf 640}, 199 (1998)
  [hep-ph/9803266].


\bibitem{Fettes:2000xg}
  N.~Fettes and U.~G.~Mei{\ss}ner,
  Nucl.\ Phys.\ A {\bf 676}, 311 (2000)
  [hep-ph/0002162].


\bibitem{Fettes:2000bb}
  N.~Fettes and U.~G.~Mei{\ss}ner,
  Nucl.\ Phys.\ A {\bf 679}, 629 (2001)
  [hep-ph/0006299].


\bibitem{Becher:2001hv}
  T.~Becher and H.~Leutwyler,
  JHEP {\bf 0106}, 017 (2001).
  [hep-ph/0103263].

\bibitem{Mai:2009ce}
  M.~Mai, P.~C.~Bruns, B.~Kubis and Ulf-G.~Mei{\ss}ner,
  Phys.\ Rev.\ D {\bf 80}, 094006 (2009).
  [hep-ph/0905.2810].

\bibitem{Bruns:2010sv}
  P.~C.~Bruns, M.~Mai and Ulf-G.~Mei{\ss}ner,
  Phys.\ Lett.\ B {\bf 697}, 254 (2011).
  [nucl-th/1012.2233].

\bibitem{Alarcon:2011kh}
  J.~M.~Alarcon, J.~Martin Camalich, J.~A.~Oller and L.~Alvarez-Ruso,
  Phys.\ Rev.\ C {\bf 83}, 055205 (2011)
  Erratum: [Phys.\ Rev.\ C {\bf 87}, no. 5, 059901 (2013)]
  [arXiv:1102.1537 [nucl-th]].


\bibitem{Chen:2012nx}
  Y.~H.~Chen, D.~L.~Yao and H.~Q.~Zheng,
  Phys.\ Rev.\ D {\bf 87}, 054019 (2013)
  [arXiv:1212.1893 [hep-ph]].


\bibitem{Yao:2016vbz}
  D.~L.~Yao, D.~Siemens, V.~Bernard, E.~Epelbaum, A.~M.~Gasparyan, J.~Gegelia, H.~Krebs and U.~G.~Mei{\ss}ner,
  JHEP {\bf 1605}, 038 (2016)
  [arXiv:1603.03638 [hep-ph]].


\bibitem{Siemens:2016hdi}
  D.~Siemens, V.~Bernard, E.~Epelbaum, A.~Gasparyan, H.~Krebs and U.~G.~Mei{\ss}ner,
  Phys.\ Rev.\ C {\bf 94}, no. 1, 014620 (2016)
  [arXiv:1602.02640 [nucl-th]].

\bibitem{Lu:2018zof}
  J.~X.~Lu, L.~S.~Geng, X.~L.~Ren, and M.~L.~Du,
  [arXiv:1812.03799 [nucl-th]].

\bibitem{Pascalutsa:2004ga}
  V.~Pascalutsa, B.~R.~Holstein and M.~Vanderhaeghen,
  Phys.\ Lett.\ B {\bf 600}, 239 (2004)
  [hep-ph/0407313].


\bibitem{Pascalutsa:2011fp}
  V.~Pascalutsa,
  AIP Conf.\ Proc.\  {\bf 1388}, 60 (2011)
  [arXiv:1105.2509 [hep-ph]].


\bibitem{Alarcon:2012kn}
  J.~M.~Alarcon, J.~Martin Camalich and J.~A.~Oller,
  Annals Phys.\  {\bf 336}, 413 (2013)
  [arXiv:1210.4450 [hep-ph]].


\bibitem{Chew:1957zz}
  G.~F.~Chew, M.~L.~Goldberger, F.~E.~Low and Y.~Nambu,
  Phys.\ Rev.\  {\bf 106}, 1337 (1957).


\bibitem{Hamilton:1963zz}
  J.~Hamilton and W.~S.~Woolcock,
  Rev.\ Mod.\ Phys.\  {\bf 35}, 737 (1963).


\bibitem{Steiner:1970mh}
  F.~Steiner,
  Fortsch.\ Phys.\  {\bf 18}, 43 (1970).


\bibitem{Gasparyan:2010xz}
  A.~Gasparyan and M.~F.~M.~Lutz,
  Nucl.\ Phys.\ A {\bf 848}, 126 (2010)
  [arXiv:1003.3426 [hep-ph]].


\bibitem{Ditsche:2012fv}
  C.~Ditsche, M.~Hoferichter, B.~Kubis and U.-G.~Mei{\ss}ner,
  JHEP {\bf 1206}, 043 (2012)
  [arXiv:1203.4758 [hep-ph]].


\bibitem{Hoferichter:2015hva}
  M.~Hoferichter, J.~Ruiz de Elvira, B.~Kubis and U.~G.~Mei{\ss}ner,
  Phys.\ Rept.\  {\bf 625}, 1 (2016)
  [arXiv:1510.06039 [hep-ph]].

\bibitem{Wang:2017agd}
  Y.~F.~Wang, D.~L.~Yao and H.~Q.~Zheng,
  Eur.\ Phys.\ J.\ C {\bf 78}, no. 7, 543 (2018)
  [arXiv:1712.09257 [hep-ph]].

\bibitem{Xiao:2000kx}
  Z.~Xiao and H.~Q.~Zheng,
  Nucl.\ Phys.\ A {\bf 695}, 273 (2001)
  [hep-ph/0011260].


\bibitem{Zheng:2003rw}
  H.~Q.~Zheng, Z.~Y.~Zhou, G.~Y.~Qin, Z.~Xiao, J.~J.~Wang and N.~Wu,
  Nucl.\ Phys.\ A {\bf 733}, 235 (2004)
  [hep-ph/0310293].


\bibitem{Zhou:2006wm}
  Z.~Y.~Zhou and H.~Q.~Zheng,
  Nucl.\ Phys.\ A {\bf 775}, 212 (2006)
  [hep-ph/0603062].


\bibitem{Zhou:2004ms}
  Z.~Y.~Zhou, G.~Y.~Qin, P.~Zhang, Z.~Xiao, H.~Q.~Zheng and N.~Wu,
  JHEP {\bf 0502}, 043 (2005)
  [hep-ph/0406271].


\bibitem{Qin:2002hk}
  G.~Y.~Qin, W.~Z.~Deng, Z.~Xiao and H.~Q.~Zheng,
  Phys.\ Lett.\ B {\bf 542}, 89 (2002).
  [hep-ph/0205214].

\bibitem{Guo:2007ff}
  Z.~H.~Guo, J.~J.~Sanz Cillero and H.~Q.~Zheng,
  JHEP {\bf 06}, 030 (2007).
  [hep-ph/0701232].

\bibitem{Guo:2007hm}
  Z.~H.~Guo, J.~J.~Sanz Cillero and H.~Q.~Zheng,
  Phys.\ Lett.\ B {\bf 661}, 342 (2008).
  [hep-ph/0710.2163].

\bibitem{MacDowell:1959zza}
  S.~W.~MacDowell,
  Phys.\ Rev.\  {\bf 116}, 774 (1959).


\bibitem{Kennedy:1961}
  J.~Kennedy and T.~D.~Spearman,
  Phys.\ Rev.\  {\bf 126}, 1596 (1961).

\bibitem{Fettes:2000gb}
  N.~Fettes, U.~G.~Mei{\ss}ner, M.~Mojzis and S.~Steininger,
  Annals Phys.\  {\bf 283}, 273 (2000).
  Erratum: [Annals Phys.\  {\bf 288}, 249 (2001)]
  [hep-ph/0001308].


\bibitem{Gasser:1983yg}
  J.~Gasser and H.~Leutwyler,
  Annals Phys.\  {\bf 158}, 142 (1984).


\bibitem{Siemens:2016jwj}
  D.~Siemens, J.~Ruiz de Elvira, E.~Epelbaum, M.~Hoferichter, H.~Krebs, B.~Kubis and U.-G.~Mei{\ss}ner,
  Phys.\ Lett.\ B {\bf 770}, 27 (2017).
  [arXiv:1610.08978 [nucl-th]].

\bibitem{Hu:1948zz}
  N.~Hu,
  Phys.\ Rev.\  {\bf 74}, 131 (1948).

\bibitem{Regge:1958ft}
  T.~Regge,
  Nuovo Cim.\  {\bf 8}, 671 (1958).

\bibitem{Wang:2018gul}
  Y.~F.~Wang, D.~L.~Yao and H.~Q.~Zheng,
  Front.\ Phys.\ (Beijing) {\bf 14}, no. 2, 24501 (2019).

\bibitem{Caprini:2005zr}
      I.~Caprini, G.~Colangelo and H.~Leutwyler,
      Phys.\ Rev.\ Lett.\ {\bf 96}, 132001 (2006).

\bibitem{DescotesGenon:2006uk}
      S.~Descotes-Genon and B.~Moussallam,
      Eur.\ Phys.\ J.\ C {\bf 48}, 553 (2006).

\bibitem{Passarino:1978jh}
  G.~Passarino and M.~J.~G.~Veltman,
  Nucl.\ Phys.\ B {\bf 160}, 151 (1979).
\end{thebibliography}
\end{document}